\documentclass[article,12pt]{article}

\usepackage{lineno}
\usepackage{hyperref}
\modulolinenumbers[5]

\usepackage{amscd,amsmath,amssymb,amsfonts,latexsym,mathrsfs,amsthm}
\usepackage{graphicx} 
\usepackage{setspace}
\usepackage{sectsty}
\usepackage[english]{babel}
\usepackage{bm}
\usepackage{mathrsfs}
\usepackage{subfigure}
\usepackage{rotating}
\usepackage{multirow}
\usepackage[utf8]{inputenc} 
\usepackage{color}
\usepackage{verbatim}
\usepackage{appendix}

\usepackage{colortbl}

\definecolor{MYCOLOR0}{rgb}{0.92,0.92,0.92}
\definecolor{MYCOLOR}{rgb}{1,1,0}
\definecolor{MYCOLOR2}{rgb}{0.5,1,0.5}

\definecolor{MYCOLOR3}{rgb}{0.88,1,1}

\def\x{{\mathbf x}}

\newtheorem{rem}{{\bf Remark}}

\title{A Review of Multiple Try MCMC algorithms for Signal Processing}

\author{Luca Martino \\ 
{\small Image Processing Lab., Universitat de Val\`encia (Spain)} \\
{\small Universidad Carlos III de Madrid, Leganes (Spain)} }

\date{}

\begin{document}



\maketitle



\begin{abstract}
Many applications in signal processing require the estimation of some parameters of interest given a set of observed data. More specifically, Bayesian inference needs the computation of {\it a-posteriori} estimators which are often expressed as complicated multi-dimensional integrals. Unfortunately, analytical expressions for these estimators cannot be found in most real-world applications, and Monte Carlo methods are the only feasible approach. A very powerful class of Monte Carlo techniques is formed by the Markov Chain Monte Carlo (MCMC) algorithms. They generate a Markov chain such that its stationary distribution coincides with the target posterior density. In this work, we perform a thorough review of MCMC methods using multiple candidates in order to select the next state of the chain, at each iteration. With respect to the classical Metropolis-Hastings method, the use of multiple try techniques foster the exploration of the sample space. We present different Multiple Try Metropolis schemes, Ensemble MCMC methods, Particle Metropolis-Hastings algorithms and the Delayed Rejection Metropolis technique. { We highlight limitations, benefits, connections and differences among the different methods, and compare them by numerical simulations.}

\end{abstract}



\linenumbers


\section{Introduction}
\label{sec:intro}



Bayesian methods have become very popular in signal processing over the last years \cite{Doucet:MCSignalProcessing2005,Fitzgerald01,Martino09,FUSS}. 
They require the application of sophisticated Monte Carlo techniques, such as Markov chain Monte Carlo (MCMC) and particle filters, for the efficient computation of {\it a-posteriori} estimators \cite{Bugallo07,Djuric03,Fitzgerald01}. More  specifically, the MCMC algorithms generate a Markov chain such that its stationary distribution coincides with the posterior probability density function (pdf) \cite{Liang10,Liu04b,Robert04}. Typically, the only requirement is to be able to evaluate the target function, where the knowledge of the normalizing constant is usually not needed. 

The most popular MCMC method is undoubtedly the Metropolis-Hastings (MH) algorithm  \cite{Hastings70,Metropolis53}. The MH technique is a very simple method, easy to be applied: this is the reason of its success. In MH, at each iteration, one new candidate is generated  from a proposal pdf and then is properly compared with the previous state of the chain, in order to decide the next state. 
However, the performance of MH are often not satisfactory. For instance, when the posterior is multimodal, or when the dimension of the space increases, the correlation among the generated samples is usually high and, as a consequence, the variance of the resulting estimators grows. To speed up the convergence and reduce the ``burn-in'' period of the MH chain, several extensions have been proposed in literature.

 In this work, we provide an exhaustive review of more sophisticated MCMC methods that, at each iteration, consider different candidates as possible new state of the chain.  More specifically, at each iteration different samples are compared by certain weights and then one of them is selected as possible future state.
 The main advantage of these algorithms is that they foster the exploration of a larger portion of the sample space, decreasing the correlation among the states of the generated chain. In this work, we describe different algorithms of this family, independently introduced in literature. 
The main contribution is to present them under the same the framework and notation, remarking differences, relationships, limitations and strengths. 
  All the discussed techniques yield an ergodic chain converging to the posterior density of interest (in the following, referred also as target pdf).

The first scheme of this MCMC class, called Orientational Bias Monte Carlo (OBMC) \cite[Chapter 13]{Frenkel96}, was proposed in the context of molecular simulation. Later on a more general algorithm, called Multiple Try Metropolis (MTM), was introduced \cite{Liu00}.\footnote{MTM includes OBMC as a special case (see Section \ref{SpecialCaseStandMTM}).} 
The MTM algorithm has been extensively studied and generalized in different ways \cite{LucaJesse1,Bedard12,Craiu07,Casarin13,Pandolfi10}. Other techniques, alternative to the MTM schemes, are the so-called the Ensemble MCMC (EnMCMC) methods \cite{Neal11,Calderhead14,LWMCMC,Austad07}. They follow a similar approach to MTM but employ a different acceptance function for selecting the next state of the chain. With respect to (w.r.t.) a generic MTM scheme, EnMCMC does not require any generation of auxiliary samples (as in a MTM scheme employing a generic proposal pdf) and hence, in this sense, EnMCMC are less costly. 

In all the previous techniques, the candidates are drawn in a batch way and compared jointly. In the Delayed Rejection Metropolis  (DRM) algorithm \cite{Haario06,Mira01,Tierney1999}, in case of rejection of the novel possible state, the authors suggest to perform an additional acceptance test considering a new candidate. If this candidate is again rejected, the procedure can be iterated until reaching a desired number of attempts. The main benefit of DRM is that the proposal pdf can be improved at each intermediate stage. However, the acceptance function progressively becomes more complex so that the implementation of DRM for a great number of attempts is not straightforward (compared to the implementation of a MTM  scheme with a generic number of tries).
  
 In the last years, other Monte Carlo methods which combine particle filtering and MCMC have become very popular in the signal processing community. For instance, this is the case of the Particle Metropolis Hastings (PMH) and the Particle Marginal Metropolis Hastings (PMMH) algorithms, which have been widely used in signal processing in order to make inference and smoothing about dynamical and static parameters in state space models \cite{PMCMC,PMH_DSP}. PMH can be interpreted as a MTM scheme where the different candidates are generated and weighted by the use of a particle filter \cite{MTM_PMH14,GIS17}. In this work, we present PMH and PMMH and discuss their connections and differences with the classical MTM approach.  Furthermore, we describe a suitable procedure for recycling some candidates in the final Monte Carlo estimators, called Group Metropolis Sampling (GMS) \cite{GIS17,GMSeusipco}. The GMS scheme can be also seen as a way of generating a chain of sets of weighted samples. Finally, note that other similar and related techniques can be found within the so-called data augmentation approach \cite{Leman09,Storvik11}.
 
The remaining of the paper is organized as follows. Section \ref{PS_section} recalls the problem statement and some background material, introducing also the required notation. The basis of MCMC and the Metropolis-Hastings (MH) algorithm are presented in Section \ref{MCMCsect}. Section \ref{MainSect} is the core of the work, which describes the different MCMC using multiple candidates. Section \ref{SimuSect} provides some numerical results, applying different techniques in a hyperparameter tuning problem for a Gaussian Process regression model, and in a localization problem considering a wireless sensor network. Some conclusions are given in Section \ref{conclSect}.

\section{Problem statement and preliminaries}
\label{PS_section}

In many signal processing applications, the goal consists in inferring a variable of interest, ${\bm \theta}=[\theta_1,\ldots,\theta_{D}]\in \mathcal{D}\subseteq \mathbb{R}^{D}$, given a set of observations or measurements, ${\bf y}\in \mathbb{R}^{P}$. In the Bayesian framework, the total knowledge about the parameters, after the data have been observed, is represented by the posterior probability density function (pdf) \cite{Liu04b,Robert04}, i.e.,
\begin{eqnarray}
	\bar{\pi}({\bm \theta})&=&p({\bm \theta}| {\bf y})= \frac{\ell({\bf y}|{\bm \theta}) g({\bm \theta})}{Z({\bf y})}, \nonumber \\ 
&=& \frac{1}{Z} \pi({\bm \theta}),
\label{eq:posterior}
\end{eqnarray}
where $\ell({\bf y}|{\bm \theta})$ denotes the likelihood function (i.e., the observation model), $g({\bm \theta})$ is the prior probability density function (pdf) and $Z=Z({\bf y})$ is the marginal likelihood (a.k.a., Bayesian evidence) \cite{Bernardo94,Robert07} and $\pi({\bm \theta})=\ell({\bf y}|{\bm \theta}) g({\bm \theta})$.
In general $Z$ is unknown, and it is possible to evaluate only the unnormalized target function,
$\pi({\bm \theta}) \propto \bar{\pi}({\bm \theta})$.

The analytical study of the posterior density $\bar{\pi}({\bm \theta})$ is often unfeasible and integrals involving $\bar{\pi}({\bm \theta})$ are typically intractable \cite{Bernardo94,Box73,Robert07}. For instance, one might be interested in the estimation of
\begin{eqnarray}
\label{MainInt}
I=E_\pi[f({\bm \theta})]&=&\int_{ \mathcal{D}} f({\bm \theta})\bar{\pi}({\bm \theta}) d{\bm \theta}, \\
  &=&\frac{1}{Z}\int_{ \mathcal{D}} f({\bm \theta})\pi({\bm \theta}) d{\bm \theta},
\end{eqnarray}
where $f({\bm \theta})$ is a generic integrable function w.r.t. $\bar{\pi}$. 
\newline
{\bf Dynamic and static parameters.}
In some specific application, the variable of interest ${\bm \theta}$ can be split in two disjoint parts, ${\bm \theta}=[{\bf x},{\bm \lambda}]$, where one, ${\bf x}$, is involved into a dynamical system (for instance, $\bf x$ is the hidden state in a state-space model) and the other, ${\bm \lambda}$, is a static parameter (for instance, an unknown parameter of the model). The strategies for making inference about ${\bf x}$ and ${\bm \lambda}$ should take into account the different nature of the two parameters (e.g., see Section \ref{PMMH}). The main notation and acronyms are summarized in Tables \ref{tab:notation}-\ref{tab:acro}.

\begin{table}[!ht]
\centering
\caption{Summary of the main notation. }
\vspace{0.1cm}
	\begin{tabular}{|c|l||c|l|}
	 \hline
      \cellcolor{MYCOLOR0} ${\bm \theta}=[\theta_1,\ldots,\theta_D]^{\top}$ & \multicolumn{3}{l|}{Variable of interest, ${\bm \theta} \in\mathcal{D}\subseteq\mathbb{R}^{D}$. } \\
      \cellcolor{MYCOLOR0} ${\bf y}$ & \multicolumn{3}{l|}{Observed measurements/data. } \\      
      \cellcolor{MYCOLOR0} ${\bm \theta}=[\x,{\bm \lambda}]^{\top}$ & \multicolumn{3}{l|}{$\x:$ dynamic parameters; ${\bm \lambda}:$ static parameters. } \\
  \cellcolor{MYCOLOR0} $\bar{\pi}({\bm \theta})$ & \multicolumn{3}{l|}{Normalized posterior pdf, $\bar{\pi}({\bm \theta})= p({\bm \theta}| {\bf y})$.} \\
  \cellcolor{MYCOLOR0} $\pi({\bm \theta})$ & \multicolumn{3}{l|}{Unnormalized posterior function,  $\pi({\bm \theta}) \propto \bar{\pi}({\bm \theta})$.} \\
   \cellcolor{MYCOLOR0} $\widehat{\pi}({\bm \theta})$ &  \multicolumn{3}{l|}{Particle approximation of $\bar{\pi}({\bm \theta})$.} \\        
     \hline
        \cellcolor{MYCOLOR0}  $I$ &  \multicolumn{3}{l|}{Integral of interest, in Eq. \eqref{MainInt}.} \\ 
         \cellcolor{MYCOLOR0}  $\widehat{I},\widetilde{I}$ &  \multicolumn{3}{l|}{Estimators of $I$.} \\ 
       \cellcolor{MYCOLOR0}  $Z$ &  \multicolumn{3}{l|}{Marginal likelihood; normalizing constant of $\pi({\bm \theta})$.} \\ 
              \cellcolor{MYCOLOR0}  $\widehat{Z}$, $\widetilde{Z}$ &  \multicolumn{3}{l|}{Estimators of the marginal likelihood $Z$.} \\ 
     \hline
\end{tabular}
\label{tab:notation}
\end{table}

\begin{table}[!h]
\centering
\caption{Summary of the main acronyms. }
\vspace{0.1cm}
	\begin{tabular}{|c|l|}
	        	\hline          
	     \cellcolor{MYCOLOR0} MCMC &  Markov  Chain Monte Carlo\\
  \cellcolor{MYCOLOR0} MH & Metropolis-Hastings \\
    \cellcolor{MYCOLOR0} I-MH &  Independent Metropolis-Hastings \\
    \cellcolor{MYCOLOR0} MTM & Multiple Try Metropolis \\
      \cellcolor{MYCOLOR0} I-MTM & Independent Multiple Try Metropolis\\
        \cellcolor{MYCOLOR0} I-MTM2 & Independent Multiple Try Metropolis (version 2)\\
       \cellcolor{MYCOLOR0} PMH & Particle Metropolis-Hastings \\ 
          \cellcolor{MYCOLOR0} PMMH & Particle Marginal Metropolis-Hastings \\ 
           \cellcolor{MYCOLOR0} GMS & Group Metropolis Sampling \\    
        \cellcolor{MYCOLOR0} EnMCMC & Ensemble MCMC \\ 
         \cellcolor{MYCOLOR0} I-EnMCMC & Independent  Ensemble MCMC \\    
          \cellcolor{MYCOLOR0} DRM  & Delayed Rejection Metropolis \\    
          	 \hline
	   \cellcolor{MYCOLOR0} IS &  Importance Sampling\\
	      \cellcolor{MYCOLOR0} SIS & Sequential Importance Sampling\\
	            \cellcolor{MYCOLOR0} SIR & Sequential Importance Resampling\\
     \hline
\end{tabular}
\label{tab:acro}
\end{table}

\subsection{Monte Carlo integration}
In many practical scenarios, the integral $I$ cannot be computed in a closed form, and numerical approximations are typically required. Many deterministic quadrature methods are available in the literature \cite{Burden00,Kythe04}. However, as the dimension $D$ of the inference problem grows (${\bm \theta}\in  \mathbb{R}^{D}$), the deterministic quadrature schemes become less efficient. In this case, a common approach consists of approximating the integral $I$ in Eq. \eqref{MainInt} by using Monte Carlo (MC) quadrature \cite{Liu04b,Robert04}. Namely, considering $T$ independent and identically distributed (i.i.d.) samples drawn from the posterior target pdf, i.e. ${\bm \theta}_1,\ldots,{\bm \theta}_T \sim \bar{\pi}({\bm \theta})$,\footnote{In this work, for simplicity, we use the same notation for denoting a random variable or one realization of a random variable.} we can build the consistent estimator
\begin{equation}
\label{IdealMC}
 {\widehat I}_T=\frac{1}{T} \sum_{t=1}^T f({\bm \theta}_t).
\end{equation}
${\widehat I}_T$ converges in probability to $I$ due to the weak law of large numbers. The approximation above ${\widehat I}_T$ is known as a direct (or ideal) Monte Carlo estimator if the samples ${\bm \theta}_t$ are independent and identically distributed  (i.i.d.) from $\bar \pi$. Unfortunately, in many practical applications, direct methods for drawing independent samples from $\bar{\pi}({\bm \theta})$ are not available. Therefore, different approaches are required, such as the Markov chain Monte Carlo (MCMC) techniques.

\section{Markov chain Monte Carlo (MCMC) methods}
\label{MCMCsect}
A MCMC algorithm generates an ergodic Markov chain with invariant (a.k.a., stationary) density given by the posterior pdf $ \bar{\pi}({\bm \theta})$ \cite{Liang10,Robert04}. Specifically, given a starting state ${\bm \theta}_0$, a sequence of {\it correlated} samples is generated, 
${\bm \theta}_0\rightarrow{\bm \theta}_1\rightarrow {\bm \theta}_2\rightarrow ....  \rightarrow{\bm \theta}_T$. Even if the samples are now correlated, the estimator
\begin{equation}
\label{corrEst}
{\widetilde I}_T=\frac{1}{T} \sum_{t=1}^T f({\bm \theta}_t)
\end{equation}
is consistent, regardless the starting vector ${\bm \theta}^{(0)}$ \cite{Robert04}.\footnote{Recall we are assuming that the Markov chain is ergodic and hence the starting value is forgotten.} With respect to the direct Monte Carlo approach using i.i.d. samples, the application of an MCMC algorithm entails a loss of efficiency of the estimator ${\widetilde I}_T$, since the samples are {\it positively} correlated, in general. In other words, to achieve a given variance obtained with the direct Monte Carlo estimator, it is necessary to generate more samples.
 Thus, in order to improve the performance of an MCMC technique we have to decrease the correlation among the states of the chain.\footnote{For the sake of simplicity, we use all the generated states in the final estimators, without removing any burn-in period \cite{Robert04}.}

\subsection{The Metropolis-Hastings (MH) algorithm}
One of the most popular and widely applied MCMC algorithm is  the Metropolis-Hastings (MH) method \cite{Metropolis53,Liang10,Robert04}. Recall that  we are able to evaluate point-wise a function proportional to the target, i.e., $\pi({\bm \theta})\propto {\bar\pi}({\bm \theta})$.  A proposal density (a pdf which is easy to draw from) is denoted as $q({\bm \theta}|{\bm \theta}_{t-1})>0$, with ${\bm \theta},{\bm \theta}_{t-1}\in\mathbb{R}^D$.  In Table \ref{alg:MH}, we describe the standard MH algorithm in detail.

\begin{table}[!h]
	\centering
	\caption{{\bf The MH algorithm}}
	    \begin{tabular}{|p{0.95\columnwidth}|}
		\hline			
\begin{enumerate}
 \item \texttt{Initialization:}  Choose an initial state ${\bm \theta}_0$.
   \item \texttt{FOR} $t = 1, \ldots, T$:
\begin{enumerate}
\item  Draw a sample ${\bm \theta}'\sim q({\bm \theta}|{\bm \theta}_{t-1})$.
\item Accept the new state, ${\bm \theta}_t={\bm \theta}'$, with probability 
\begin{eqnarray}
\label{AlfaMH}
\alpha({\bm \theta}_{t-1},{\bm \theta}') &=& \min\left[1, \frac{\pi({\bm \theta}')  q({\bm \theta}_{t-1}|{\bm \theta}')}{\pi({\bm \theta}_{t-1})q({\bm \theta}'|{\bm \theta}_{t-1})}\right],  
\end{eqnarray}
Otherwise, set ${\bm \theta}_t={\bm \theta}_{t-1}$.
\end{enumerate}
\item \texttt{Return:}  $\{{\bm \theta}_t\}_{t=1}^T$. 
\end{enumerate}

\\
\hline
	\end{tabular}		
	\label{alg:MH}
\end{table}
The algorithm returns the sequence of states $\{{\bm \theta}_1, {\bm \theta}_2, \ldots, {\bm \theta}_t, \ldots, {\bm \theta}_T\}$ (or a subset of them removing the burn-in period if an estimation of its length is available). We can see that the next state ${\bm \theta}_t$ can be the proposed sample ${\bm \theta}'$ (with probability $\alpha$) or the previous state ${\bm \theta}_{t-1}$ (with probability $1-\alpha$).  Under some mild regularity conditions, when $t$ grows, the pdf of the current state ${\bm \theta}_t$ converges to the target density ${\bar \pi}({\bm \theta})$  \cite{Robert04}. The MH algorithm satisfies the so-called detailed balance condition which is sufficient to guarantee that the output chain is ergodic and has ${\bar \pi}$ as stationary distribution \cite{Liang10,Robert04}. Note that the acceptance probability $\alpha$ can be rewritten as 
\begin{eqnarray}
\alpha({\bm \theta}_{t-1},{\bm \theta}') = \min\left[1, \frac{\pi({\bm \theta}')  q({\bm \theta}_{t-1}|{\bm \theta}')}{\pi({\bm \theta}_{t-1})q({\bm \theta}'|{\bm \theta}_{t-1})}\right]= \min\left[1,\frac{w({\bm \theta}'|{\bm \theta}_{t-1})}{w({\bm \theta}_{t-1}|{\bm \theta}')}\right]. \label{AlfaMH2} 
\end{eqnarray}
where we have denoted $w({\bm \theta}'|{\bm \theta}_{t-1})=\frac{\pi({\bm \theta}')}{q({\bm \theta}'|{\bm \theta}_{t-1})}$ and $w({\bm \theta}_{t-1}|{\bm \theta}')=\frac{\pi({\bm \theta}_{t-1})}{q({\bm \theta}_{t-1}|{\bm \theta}')}$ in a similar fashion of the importance sampling weights of ${\bm \theta}'$ and ${\bm \theta}_{t-1}$ \cite{Robert04}. If the  proposal pdf is independent from the previous state, i.e., $q({\bm \theta}|{\bm \theta}_{t-1})=q({\bm \theta})$, the acceptance function depends on the ratio of the importance weights $w({\bm \theta}')=\frac{\pi({\bm \theta}')}{q({\bm \theta}')}$ and $w({\bm \theta}_{t-1})=\frac{\pi({\bm \theta}_{t-1})}{q({\bm \theta}_{t-1})}$, as shown in Table \ref{I_MH_table}. We refer to this special MH case as the Independent MH (I-MH) algorithm. It is strictly related to other techniques described in the following (e.g., see Section \ref{superSection}).

\begin{table}[!h]
	\centering
	\caption{{\bf The Independent MH (I-MH) algorithm}}
	   \begin{tabular}{|p{0.95\columnwidth}|}
		\hline			
\begin{enumerate}
 \item \texttt{Initialization:}  Choose an initial state ${\bm \theta}_0$.
   \item \texttt{FOR} $t = 1, \ldots, T$:
\begin{enumerate}
\item  Draw a sample ${\bm \theta}'\sim q({\bm \theta})$.
\item Accept the new state, ${\bm \theta}_t={\bm \theta}'$, with probability 
\begin{eqnarray}
\label{Alfa_I-MH}
\alpha({\bm \theta}_{t-1},{\bm \theta}') &=& \min\left[1, \frac{w({\bm \theta}')}{w({\bm \theta}_{t-1})}\right],  
\end{eqnarray}
Otherwise, set ${\bm \theta}_t={\bm \theta}_{t-1}$.
\end{enumerate}
\item \texttt{Return:}  $\{{\bm \theta}_t\}_{t=1}^T$. 
\end{enumerate}	  	   
\\
\hline
	\end{tabular}		
	\label{I_MH_table}
\end{table}

\section{MCMC using multiple candidates}
\label{MainSect}
\begin{figure}[h!]
\centering
\centerline{
\subfigure[Standard MH]{\includegraphics[height=3cm]{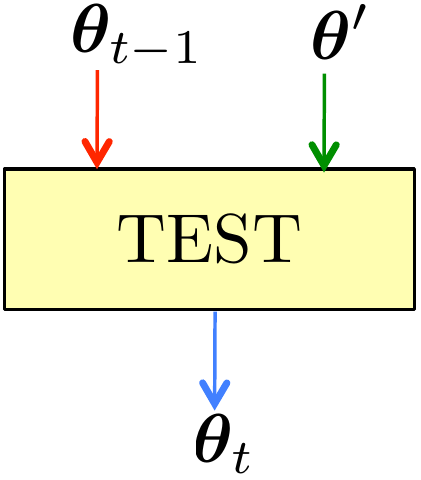} }
\hspace{1cm}
\subfigure[MCMC using multiple tries]{\includegraphics[height=3cm]{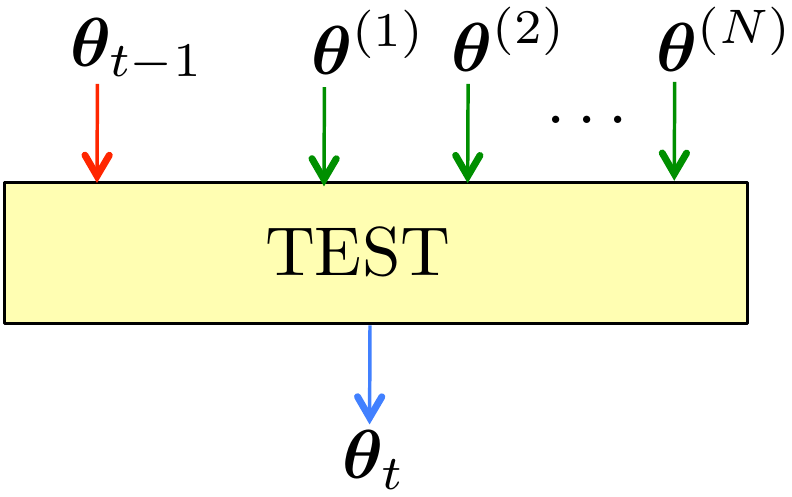} }
}
\caption{Graphical representation of the classical MH method and the MCMC schemes using different candidates at each iteration. 
 }
\label{FigMH_MCMC}
\end{figure}

In the standard MH technique described above, at each iteration one new sample ${\bm \theta}'$ is generated to be tested with the previous state ${\bm \theta}_{t-1}$ by the acceptance probability $\alpha({\bm \theta}_{t-1},{\bm \theta}')$. Other generalized MH schemes generate several candidates at each iteration to be tested as new possible state. In all these schemes, an extended acceptance probability $\alpha$ is properly designed in order to guarantee the ergodicity of the chain. Figure \ref{FigMH_MCMC} provides a graphical representation of the difference between MH and the techniques using several candidates.

Below, we describe the most important examples of this class of MCMC algorithms. In most of them, a single MH-type test is performed at each iteration whereas in other methods a sequence of tests is employed. 
Furthermore, most of these techniques use an Importance Sampling (IS) approximation of the target density \cite{Liu04b,Robert04} in order to improve the proposal procedure employed within a MH-type algorithm. Namely, they build an IS approximation, and then draw one sample from this approximation (resampling step). Finally, the selected sample is compared with the previous state of the chain, ${\bm \theta}_{t-1}$, according to a suitable generalized acceptance probability $\alpha$. It can be proved that all the methodologies presented in this work yield a ergodic chain with the posterior ${\bar \pi}$ as invariant density.

\subsection{The Multiple Try Metropolis (MTM) algorithm}
\label{MTMsect}




The  Multiple Try Metropolis (MTM) algorithms are examples of this class of methods, where $N$ samples ${\bm \theta}^{(1)},{\bm \theta}^{(2)},\ldots, {\bm \theta}^{(N)}$ (called also ``tries'' or ``candidates'') are drawn from the proposal pdf $q({\bm \theta})$, at each iteration \cite{Liu00,LucaJesse1,Bedard12,Craiu07,Casarin13,LucaJesse2,MTMissue17}. Then, one of them is selected according to some suitable weights. Finally, the selected candidate is accepted or rejected as new state according to a generalized probability function $\alpha$. 

The MTM algorithm is given in Table \ref{alg:MTM}. For the sake of simplicity, we have considered the use of the importance weights $w({\bm \theta}|{\bm \theta}_{t-1})=\frac{\pi({\bm \theta})}{q({\bm \theta}|{\bm \theta}_{t-1})}$, but there is not a unique possibility, as also shown below  \cite{Liu00,LucaJesse2}. In its general form, when the proposal depends on the previous state of the chain $q({\bm \theta}|{\bm \theta}_{t-1})$, the MTM requires the generation of $N-1$ auxiliary samples, ${\bf v}^{(i)}$, which are employed in the computation of the acceptance function $\alpha$. They are needed in order to guarantee the ergodicity. Indeed,  the resulting MTM kernel satisfies the detailed balance condition, so that the chain is reversible \cite{Liu00,LucaJesse2}. Note that for $N=1$, we have ${\bm \theta}^{(j)}={\bm \theta}^{(1)}$, ${\bf v}^{(1)}={\bm \theta}_{t-1}$ and the acceptance probability of the MTM method becomes
\begin{eqnarray}
   \alpha({\bm \theta}_{t-1},{\bm \theta}^{(1)})&=& \min\left(1,\frac{w({\bm \theta}^{(1)}|{\bm \theta}_{t-1})}{w({\bf v}^{(1)}|{\bm \theta}^{(1)})}\right), \nonumber \\
   &=& \min\left(1,\frac{w({\bm \theta}^{(1)}|{\bm \theta}_{t-1})}{w({\bm \theta}_{t-1}|{\bm \theta}^{(1)})}\right)=\min\left(1,\frac{\pi({\bm \theta}^{(1)}) q({\bm \theta}_{t-1}|{\bm \theta}^{(1)})}{\pi({\bm \theta}_{t-1}) q({\bm \theta}^{(1)}|{\bm \theta}_{t-1})}\right),  
     \end{eqnarray}
that is the  acceptance probability of the classical MH technique. Several variants have been studied, for instance, with correlated tries and considering the use of different proposal pdfs \cite{LucaJesse2,cascralei:2012}.

\begin{table}[!h]
	\centering
	\caption{{\bf The MTM algorithm with importance sampling weights.}}
	    \begin{tabular}{|p{0.95\columnwidth}|}
		\hline
\begin{enumerate}
	\item \texttt{Initialization:}  Choose an initial state ${\bm \theta}_0$.
    \item \texttt{FOR} $t = 1, \ldots, T$:
    	\begin{enumerate}
        \item\label{1step} Draw ${\bm \theta}^{(1)},{\bm \theta}^{(2)},\ldots, {\bm \theta}^{(N)}  \sim q({\bm \theta}|{\bm \theta}_{t-1})$.
       \item Compute the importance weights
       \begin{equation}   
       w({\bm \theta}^{(n)}|{\bm \theta}_{t-1})=\frac{\pi({\bm \theta}^{(n)})}{q({\bm \theta}^{(n)}|{\bm \theta}_{t-1})}, \quad \mbox{ with } \quad  n=1,\ldots,N.
        \end{equation}
     \item Select one sample ${\bm \theta}^{(j)}\in\{{\bm \theta}^{(1)},\ldots,{\bm \theta}^{(N)}\}$, according to the probability mass function $\bar{w}_n=\frac{w({\bm \theta}^{(n)}|{\bm \theta}_{t-1})}{\sum_{i=1}^N w({\bm \theta}^{(i)}|{\bm \theta}_{t-1})}$. 
  \item\label{AuxPoints} Draw $N-1$ auxiliary samples from $q({\bm \theta}|{\bm \theta}^{(j)})$, denoted as 
  ${\bf v}^{(1)},\ldots, {\bf v}^{(j-1)}, {\bf v}^{(j+1)}, \ldots, {\bf v}^{(N)} \sim q({\bm \theta}|{\bm \theta}^{(j)})$,
   and set ${\bf v}^{(j)}={\bm \theta}_{t-1}$. 
 \item\label{AuxPoints2}  Compute the weights of the auxiliary samples
        \begin{equation}             w({\bf v}^{(n)}|{\bm \theta}^{(j)})=\frac{\pi({\bf v}^{(n)})}{q({\bf v}^{(n)}|{\bm \theta}^{(j)})}, \quad \mbox{ with } \quad  n=1,\ldots,N.
  \end{equation}
\item Set ${\bm \theta}_t={\bm \theta}^{(j)}$ with probability
  \begin{eqnarray}
   \alpha({\bm \theta}_{t-1},{\bm \theta}^{(j)})
                    	&=& \min\left(1,\frac{\sum_{n=1}^N w({\bm \theta}^{(n)}|{\bm \theta}_{t-1})}{\sum_{n=1}^N w({\bf v}^{(n)}|{\bm \theta}^{(j)})}\right),           
                \label{eq:alphaMTM}
  \end{eqnarray}
  otherwise, set ${\bm \theta}_t={\bm \theta}_{t-1}$.
         \end{enumerate}
         \item \texttt{Return:}  $\{{\bm \theta}_t\}_{t=1}^T$. 
   \end{enumerate}	
\\
\hline
	\end{tabular}		
	\label{alg:MTM}
\end{table}

{\rem  The MTM method in Table \ref{alg:MTM} needs at step \ref{AuxPoints} the generation of $N-1$ auxiliary samples and at step \ref{AuxPoints2} the computation of their weights (and, as a consequence, $N-1$ additional evaluation of the target pdf are required), that are only employed in the computation of the acceptance function $\alpha$. }

\subsubsection{Generic form of the weights}
\label{SpecialCaseStandMTM}

The importance weights are not the unique possible choice.
It is possible to show that the MTM algorithm generates an ergodic chain with invariant density $\bar{\pi}$, if the weight function $w({\bm \theta}|{\bm \theta}_{t-1})$ is chosen with the form
\begin{equation}
\label{Weightforms}
w({\bm \theta}|{\bm \theta}_{t-1})=\pi({\bm \theta})q({\bm \theta}_{t-1}|{\bm \theta}) \xi({\bm \theta}_{t-1},{\bm \theta}),
\end{equation}
where
$$
\xi({\bm \theta}_{t-1},{\bm \theta})= \xi({\bm \theta},{\bm \theta}_{t-1}), \qquad \forall {\bm \theta},{\bm \theta}_{t-1} \in \mathcal{D}.
$$
For instance, choosing $\xi({\bm \theta}_{t-1},{\bm \theta})=\frac{1}{q({\bm \theta}|{\bm \theta}_{t-1}) q({\bm \theta}_{t-1}|{\bm \theta})}$, we obtain the importance weights $w({\bm \theta}|{\bm \theta}_{t-1})=\frac{\pi({\bm \theta})}{q({\bm \theta}|{\bm \theta}_{t-1})}$ used above. If we set $\xi({\bm \theta}_{t-1},{\bm \theta})=1$, we have $w({\bm \theta}|{\bm \theta}_{t-1})=\pi({\bm \theta})q({\bm \theta}_{t-1}|{\bm \theta})$. Another interesting example can be employed if the proposal is symmetric, i.e., $q({\bm \theta}|{\bm \theta}_{t-1})=q({\bm \theta}_{t-1}|{\bm \theta})$. In this case, we can choose $\xi({\bm \theta}_{t-1},{\bm \theta})=\frac{1}{q({\bm \theta}_{t-1}|{\bm \theta})}$ and then $w({\bm \theta}|{\bm \theta}_{t-1})=w({\bm \theta})=\pi({\bm \theta})$, i.e., the weights only depend on the value of the target density at ${\bm \theta}$. Thus, 
MTM contains the Orientational Bias Monte Carlo (OBMC) scheme \cite[Chapter 13]{Frenkel96} as a special case, when a symmetric proposal pdf is employed, and then one candidate is chosen with weights proportional to the target density, i.e., $w({\bm \theta}|{\bm \theta}_{t-1})=\pi({\bm \theta})$.

\subsubsection{Independent Multiple Try Metropolis (I-MTM) schemes}
The MTM method described in Table \ref{alg:MTM} requires to draw $2N-1$ samples at each iteration ($N$ candidates and $N-1$ auxiliary samples) and $N-1$ are only used in the acceptance probability function. The generation of the auxiliary points
$$
{\bf v}^{(1)},\ldots, {\bf v}^{(j-1)}, {\bf v}^{(j+1)}, \ldots, {\bf v}^{(N)} \sim q({\bm \theta}|{\bm \theta}^{(j)}), 
$$
can be avoided if the proposal pdf is independent from the previous state, i.e., $q({\bm \theta}|{\bm \theta}_{t-1})=q({\bm \theta})$. Indeed, in this case, we should draw $N-1$ samples again from $q({\bm \theta})$ at the step \ref{AuxPoints} of Table \ref{alg:MTM}. Since we have already drawn $N$ samples from $q({\bm \theta})$ at step \ref{1step} of Table \ref{alg:MTM}, we can set 
\begin{equation}
\label{AuxP_2}
{\bf v}^{(1)}={\bm \theta}^{(1)},\ldots,{\bf v}^{(j-1)}={\bm \theta}^{(j-1)}, {\bf v}^{(j)}={\bm \theta}^{(j+1)}\dots {\bf v}^{(N-1)}={\bm \theta}^{(N)},
\end{equation}
 without jeopardizing the ergodicity of the chain (recall that ${\bf v}^{(j)}={\bm \theta}_{t-1}$). Hence, we can avoid step \ref{AuxPoints} and  the acceptance function can be rewritten as  
   \begin{eqnarray}
   \alpha({\bm \theta}_{t-1},{\bm \theta}^{(j)})
                    	&=& \min\left(1,\frac{w({\bm \theta}^{(j)})+\sum_{n=1,n\neq j}^N w({\bm \theta}^{(n)})}{w({\bm \theta}_{t-1})+\sum_{n=1,n\neq j}^N w({\bm \theta}^{(n)})}\right).
                \label{eq:alphaMTM_fora}
  \end{eqnarray}   
The I-MTM algorithm is provided in Table \ref{alg:I-MTM}.  

{\rem An I-MTM method requires only $N$ new evaluations of the target pdf at each iteration, instead of $2N-1$ new evaluations in the generic MTM scheme in Table \ref{alg:MTM}.}
\newline
Note that we can also write $\alpha({\bm \theta}_{t-1},{\bm \theta}^{(j)})$ as 
   \begin{eqnarray}
   \alpha({\bm \theta}_{t-1},{\bm \theta}^{(j)})&=& \min\left(1,\frac{\widehat{Z}_1}{\widehat{Z}_2}\right),   
                \label{eq:alphaMTM_fora2}
  \end{eqnarray} 
 where we have denoted 
   \begin{eqnarray} 
 \widehat{Z}_1=\frac{1}{N}\sum_{n=1}^N w({\bm \theta}^{(n)}), \qquad \widehat{Z}_2=\frac{1}{N} \left(w({\bm \theta}_{t-1})+\sum_{n=1,n\neq j}^N w({\bm \theta}^{(n)})\right).
   \end{eqnarray} 
\begin{table}[!h]
	\centering
	\caption{{\bf The Independent Multiple Try Metropolis (I-MTM) algorithm.}}
	    \begin{tabular}{|p{0.95\columnwidth}|}
\hline
\begin{enumerate}
	\item \texttt{Initialization:}  Choose an initial state ${\bm \theta}_0$.
    \item \texttt{FOR} $t = 1, \ldots, T$:
    	\begin{enumerate}
        \item Draw ${\bm \theta}^{(1)},{\bm \theta}^{(2)},\ldots, {\bm \theta}^{(N)}  \sim q({\bm \theta})$.
       \item Compute the importance weights
       \begin{equation}           
       w({\bm \theta}^{(n)})=\frac{\pi({\bm \theta}^{(n)})}{q({\bm \theta}^{(n)})}, \quad \mbox{ with } \quad  n=1,\ldots,N.
        \end{equation}
     \item Select one sample ${\bm \theta}^{(j)}\in\{{\bm \theta}^{(1)},\ldots,{\bm \theta}^{(N)}\}$,  according to the probability mass function $\bar{w}_n=\frac{w({\bm \theta}^{(n)})}{\sum_{i=1}^N w({\bm \theta}^{(i)})}$. 
\item Set ${\bm \theta}_t={\bm \theta}^{(j)}$ with probability
  \begin{eqnarray}
   \alpha({\bm \theta}_{t-1},{\bm \theta}^{(j)})
                    	&=& \min\left(1,\frac{w({\bm \theta}^{(j)})+\sum_{n=1,n\neq j}^N w({\bm \theta}^{(n)})}{w({\bm \theta}_{t-1})+\sum_{n=1,n\neq j}^N w({\bm \theta}^{(n)})}\right), \nonumber\\
                 &=& \min\left(1,\frac{\widehat{Z}_1}{\widehat{Z}_2}\right),   
                \label{eq:alphaI_MTM}
  \end{eqnarray}
  otherwise, set ${\bm \theta}_t={\bm \theta}_{t-1}$.  
         \end{enumerate}
         \item \texttt{Return:}  $\{{\bm \theta}_t\}_{t=1}^T$. 
   \end{enumerate}
   \\
\hline
   	\end{tabular}		
	\label{alg:I-MTM}
\end{table}

\begin{figure}[h!]
\centering
\centerline{
\includegraphics[width=16cm]{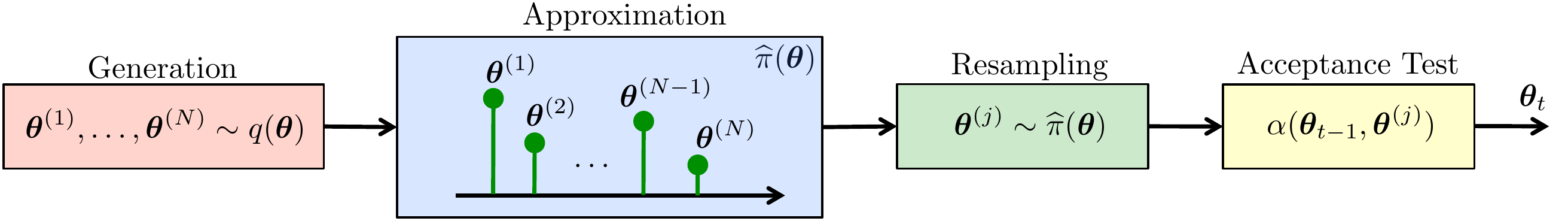} 
}
\caption{Graphical representation of the I-MTM schemes.
 }
\label{FigMTM}
\end{figure}

{\bf Alternative version (I-MTM2).} From the IS theory, we know that $ \widehat{Z}_1=\frac{1}{N}\sum_{n=1}^N w({\bm \theta}^{(n)})$ is an unbiased estimator of the normalizing constant $Z$ of the target $\pi$ (a.k.a, Bayesian evidence or marginal likelihood). It suggests to replace $\widehat{Z}_2$ with other unbiased estimators of $Z$ (without jeopardizing the ergodicity of the chain). For instance, instead of recycling the samples generated in the same iteration as auxiliary points as in Eq. \eqref{AuxP_2},  we could reuse samples generated in the previous iteration $t-1$. This alternative version of I-MTM method (I-MTM2) is given in Table \ref{alg:I-MTM2}. Note that, in both cases I-MTM and I-MTM2, the selected candidate ${\bm \theta}^{(j)}$ is drawn from the following particle approximation of the target $\bar{\pi}$,
 \begin{equation}
 \label{ApproxPI_IMP}
 \widehat{\pi}({\bm \theta}|{\bm \theta}^{(1:N)})= \sum_{i=1}^N \bar{w}({\bm \theta}^{(i)})\delta({\bm \theta}-{\bm \theta}^{(i)}), \qquad \bar{w}_i=\bar{w}({\bm \theta}^{(i)})=\frac{w({\bm \theta}^{(i)})}{\sum_{n=1}^N w({\bm \theta}^{(n)})},
\end{equation}
 i.e., ${\bm \theta}^{(j)}\sim  \widehat{\pi}({\bm \theta}|{\bm \theta}^{(1:N)})$.  The acceptance probability $\alpha$ used in I-MTM2 can be also justified considering a proper IS weighting of a resampled particle \cite{GISssp16} and using the expression \eqref{AlfaMH2} related to the standard MH method, as discussed in \cite{GIS17}.  Figure \ref{FigMTM} provides a graphical representation of the I-MTM schemes.



\begin{table}[!h]
	\centering
	\caption{{\bf Alternative version of I-MTM method (I-MTM2).}}
	    \begin{tabular}{|p{0.95\columnwidth}|}
\hline
\begin{enumerate}
	\item \texttt{Initialization:}  Choose an initial state ${\bm \theta}_0$, and obtain an initial approximation $\widehat{Z}_0\approx Z$.   
	 \item \texttt{FOR $t = 1, \ldots, T$:}
    	\begin{enumerate}
        \item Draw ${\bm \theta}^{(1)},{\bm \theta}^{(2)},\ldots, {\bm \theta}^{(N)}  \sim q({\bm \theta})$.
       \item Compute the importance weights
       \begin{equation}        
        w({\bm \theta}^{(n)})=\frac{\pi({\bm \theta}^{(n)})}{q({\bm \theta}^{(n)})}, \quad \mbox{ with } \quad  n=1,\ldots,N.
        \end{equation}
     \item Select one sample ${\bm \theta}^{(j)}\in\{{\bm \theta}^{(1)},\ldots,{\bm \theta}^{(N)}\}$, according to the probability mass function $\bar{w}_n=\frac{1}{N\widehat{Z}^*}w({\bm \theta}^{(n)})$ where $\widehat{Z}^*=\frac{1}{N}\sum_{i=1}^N w({\bm \theta}^{(i)})$. 
\item Set ${\bm \theta}_t={\bm \theta}^{(j)}$ and $\widehat{Z}_t=\widehat{Z}^*$ with probability
  \begin{equation}
   \alpha({\bm \theta}_{t-1},{\bm \theta}^{(j)})
                    	= \min\left(1,\frac{\widehat{Z}^*}{\widehat{Z}_{t-1}}\right),
                \label{eq:alphaIMTM2}
  \end{equation}
  otherwise, set ${\bm \theta}_t={\bm \theta}_{t-1}$ and $\widehat{Z}_t=\widehat{Z}_{t-1}$.
         \end{enumerate}
   \end{enumerate}
  \\
\hline
   	\end{tabular}		
	\label{alg:I-MTM2}
\end{table}


\subsubsection{Reusing candidates in parallel I-MTM chains}
\label{Par_MTM}

Let us consider to run $C$ independent parallel chains yielded by an I-MTM scheme. In this case, we have $NC$ evaluations of the target function $\pi$ and $C$ resampling steps performed at each iteration (so that we have $NCT$ total target evaluations and $CT$ total resampling steps). 

In literature, different authors have suggested to recycle the $N$ candidates, ${\bm \theta}^{(1)},{\bm \theta}^{(2)},\ldots, {\bm \theta}^{(N)} \sim q({\bm \theta})$, in order to reduce the number of evaluations of the target pdf \cite{OMCMC_DSP} . The idea is to performs $C$-times the resampling procedure considering the same set of candidates, ${\bm \theta}^{(1)},{\bm \theta}^{(2)},\ldots, {\bm \theta}^{(N)}$ (a similar approach was proposed in \cite{Calderhead14}). Each resampled candidate is then tested as possible future state of one chain.
 In this scenario, The number of target evaluations per iteration is only $N$ (hence, the total number of evaluation of $\pi$ is $NT$). However, the resulting $C$ parallel chains are no longer independent, and there is a lose of performance w.r.t. the independent chains. There exists also the possibility of reducing the total number of resampling steps, as suggested in the Block Independent MTM scheme \cite{OMCMC_DSP} (but  the dependence among the chains grows even more).

\subsection{Particle Metropolis-Hastings (PMH) method}
\label{PMHsection}

Assume that the variable of interest is formed by only a dynamical variable, i.e., ${\bm \theta}={\bf x}=x_{1:D}=[x_1\dots,x_D]^{\top}$ (see Section \ref{PS_section}). This is the case of inferring a hidden state in state-space model, for instance. More generally, let assume that we are able to factorize the target density as
\begin{eqnarray}
\bar{\pi}({\bf x})\propto \pi({\bf x})&=& \gamma_1(x_1) \gamma_2(x_2|x_1) \cdots \gamma_D(x_D|x_{1:D-1}), \\
&=&\gamma_1(x_1) \prod_{d=2}^D  \gamma_d(x_d|x_{d-1}).
\end{eqnarray}
The Particle Metropolis Hastings (PMH) method \cite{Andrieu10} is an efficient MCMC technique, proposed independently from the MTM algorithm, specifically designed for being applied in this framework. Indeed, we can take advantage of the factorization of the target pdf and consider a proposal pdf decomposed in the same fashion
$$
q({\bf x})=q_1(x_1) q_2(x_2|x_1) \cdots  q_D(x_D|x_{1:D-1})=q_1(x_1) \prod_{d=2}^D  q_d(x_d|x_{d-1}).
$$
Then, as in a batch IS scheme, given an $n$-th sample ${\bf x}^{(n)}=x_{1:D}^{(n)}\sim q({\bf x})$ with $x_{d}^{(n)}\sim q_d(x_d|x_{d-1})$, we assign the importance weight  
\begin{equation}
\label{EqFinRecW_now}
w({\bf x}^{(n)})=w_D^{(n)}=\frac{\pi({\bf x}^{(n)})}{q({\bf x}^{(n)})}=\frac{ \gamma_1(x_1^{(n)}) \gamma_2(x_2^{(n)}|x_1^{(n)}) \cdots  \gamma_D(x_D^{(n)}|x_{1:D-1}^{(n)})}{q_1(x_1^{(n)}) q_2(x_2^{(n)}|x_1^{(n)}) \cdots  q_D(x_D^{(n)}|x_{1:D-1}^{(n)})}.
\end{equation}
 The previous expression suggests a recursive procedure for computing the importance weights: starting with $w_1^{(n)}=\frac{\pi(x_1^{(n)})}{q(x_1^{(n)})}$ and then
 \begin{gather}
 \label{RecWeights_now}
 \begin{split}
  w_d^{(n)}=w_{d-1}^{(n)} \beta_d^{(n)}=\prod_{j=1}^{d} \beta_j^{(n)},   \quad \quad d=1,\ldots,D,   
 \end{split}
 \end{gather}
 where we have set
 \begin{equation}
 \beta_1^{(n)}=w_1^{(n)} \quad \mbox{and} \quad \beta_d^{(n)}=\frac{\gamma_d(x_d^{(n)}|x_{1:d-1}^{(n)})}{q_d(x_d^{(n)}|x_{1:d-1}^{(n)})},
\end{equation} 
for $d=2,\ldots,D$. This method is usually referred as Sequential Importance Sampling (SIS).  If resampling steps are also employed at some iteration, the method is called Sequential Importance Resampling (SIR), a.k.a., {\it particle filtering} (PF) (see Appendix \ref{AppPF}). PMH uses a SIR approach for providing the particle approximation $
 \widehat{\pi}({\bf x}|{\bf x}^{(1:N)})= \sum_{i=1}^N \bar{w}_D^{(i)}\delta({\bf x}-{\bf x}^{(i)})$ where $ \bar{w}_D^{(i)}=\frac{w_D^{(i)}}{\sum_{n=1}^Nw_D^{(n)}}$ and $w_D^{(i)}=w({\bf x}^{(i)})$, obtained using Eq. \eqref{RecWeights_now} (with a   proper weighting of a resampled particle \cite{GISssp16,GIS17}). Then, one particle is drawn from this approximation, i.e., with a probability proportional to the corresponding normalized weight. 
\newline
\newline
{\bf Estimation of the marginal likelihood $Z$ in particle filtering.} SIR combines the SIS approach with the application of resampling procedures. In SIR, a consistent estimator of $Z$ is given by 
\begin{equation}
\label{EstZ2}
\widetilde{Z}=\prod_{d=1}^D\left[\sum_{n=1}^N{\bar w}_{d-1}^{(n)}\beta_d^{(n)}\right], \quad \mbox{where} \quad \bar{w}_{d-1}^{(i)}=\frac{w_{d-1}^{(i)}}{\sum_{n=1}^Nw_{d-1}^{(n)}}.
\end{equation}
Due to the application of the resampling, in SIR the standard estimator 
\begin{equation}
\widehat{Z}=\frac{1}{N}\sum_{n=1}^N w_{D}^{(n)}=\frac{1}{N}\sum_{n=1}^N w({\bf x}^{(n)}),
\end{equation}
is a possible alternative {\it only if} a proper weighting of the resampled particles is applied \cite{GIS17,GISssp16} (otherwise, it is not an estimator of $Z$).  If a proper weighting of a resampled particle is employed, both $\widetilde{Z}$ and $\widehat{Z}$ are equivalent estimators of $Z$ \cite{GISssp16,GIS17,MTM_PMH14}. Without the use of resampling steps (i.e., in SIS),  $\widetilde{Z}$ and $\widehat{Z}$ are always equivalent estimators \cite{GIS17}. See also Appendix \ref{AppPF}.
\newline
\newline
 The complete description of PMH is provided in Table \ref{tab:PMH} considering the use of $\widetilde{Z}$. At each iteration, a particle filter is run in order to provide an approximation by $N$ weighted samples of the measure of the target. Then, a sample among the $N$ weighted particles is chosen by one resampling step. This selected sample is then accepted or rejected as next state of the chain according to an MH-type acceptance probability, which involves two estimators of marginal likelihood $Z$. PMH is also related to other popular method in molecular simulation called Configurational Bias Monte Carlo (CBMC) \cite{Siepmann92}. 
 




\begin{table}[!t]
\caption{\textbf{Particle Metropolis-Hastings (PMH) algorithm.}}
\begin{tabular}{|p{0.95\columnwidth}|}
   \hline
\begin{enumerate}
\item \texttt{Initialization:} Choose a initial state ${\bf x}_0$, obtain an initial estimation $\widetilde{Z}_{0}\approx Z$.
\item \texttt{For $t=1,\ldots,T$:}
\begin{enumerate}
\item\label{PMHahora} We employ a SIR approach for drawing with $N$ particles and weighting them, $\{{\bf x}^{(i)},w_D^{(i)}\}_{i=1}^N$, i.e., we obtain sequentially a particle approximation
$\widehat{\pi}({\bf x})= \sum_{i=1}^N \bar{w}_D^{(i)}\delta({\bf x}-{\bf x}^{(i)})$ where ${\bf x}^{(i)}=[x_{1}^{(i)},\ldots,x_{D}^{(i)}]^{\top}$. Furthermore, we also obtain $\widetilde{Z}^*$ as in Eq. \eqref{EstZ2}.
\item Draw ${\bf x}^* \sim  \widehat{\pi}({\bf x}|{\bf x}^{(1:N)})$, i.e., choose a particle ${\bf x}^*=\{{\bf x}^{(1)},\ldots,{\bf x}^{(N)}\}$ with probability  $\bar{w}_D^{(i)}$, $i=1,...,N$.	 
\item Set  ${\bf x}_t={\bf x}^*$ and $\widetilde{Z}_{t}=\widetilde{Z}^*$ with probability
\begin{equation}
\label{A1pmh}
\alpha=\min\left[1, \frac{{\widetilde Z}^*}{{\widetilde Z}_{t-1}}\right],
\end{equation}
otherwise set ${\bf x}_t={\bf x}_{t-1}$ and $\widetilde{Z}_{t}=\widetilde{Z}_{t-1}$.
\item \texttt{Return:} $\{{\bf x}_t\}_{t=1}^{T}$ where ${\bf x}_t=[x_{1,t},\ldots,x_{D,t}]^{\top}$.
\end{enumerate}
\end{enumerate} \\ \\
\hline 
\end{tabular}
\label{tab:PMH}
\end{table}

\subsubsection{Relationship among I-MTM2, PMH and I-MH}
\label{superSection}
\begin{figure*}[h!]
\begin{center}
\centerline{
  \subfigure[Batch-IS or SIS.]{\includegraphics[width=0.43\textwidth]{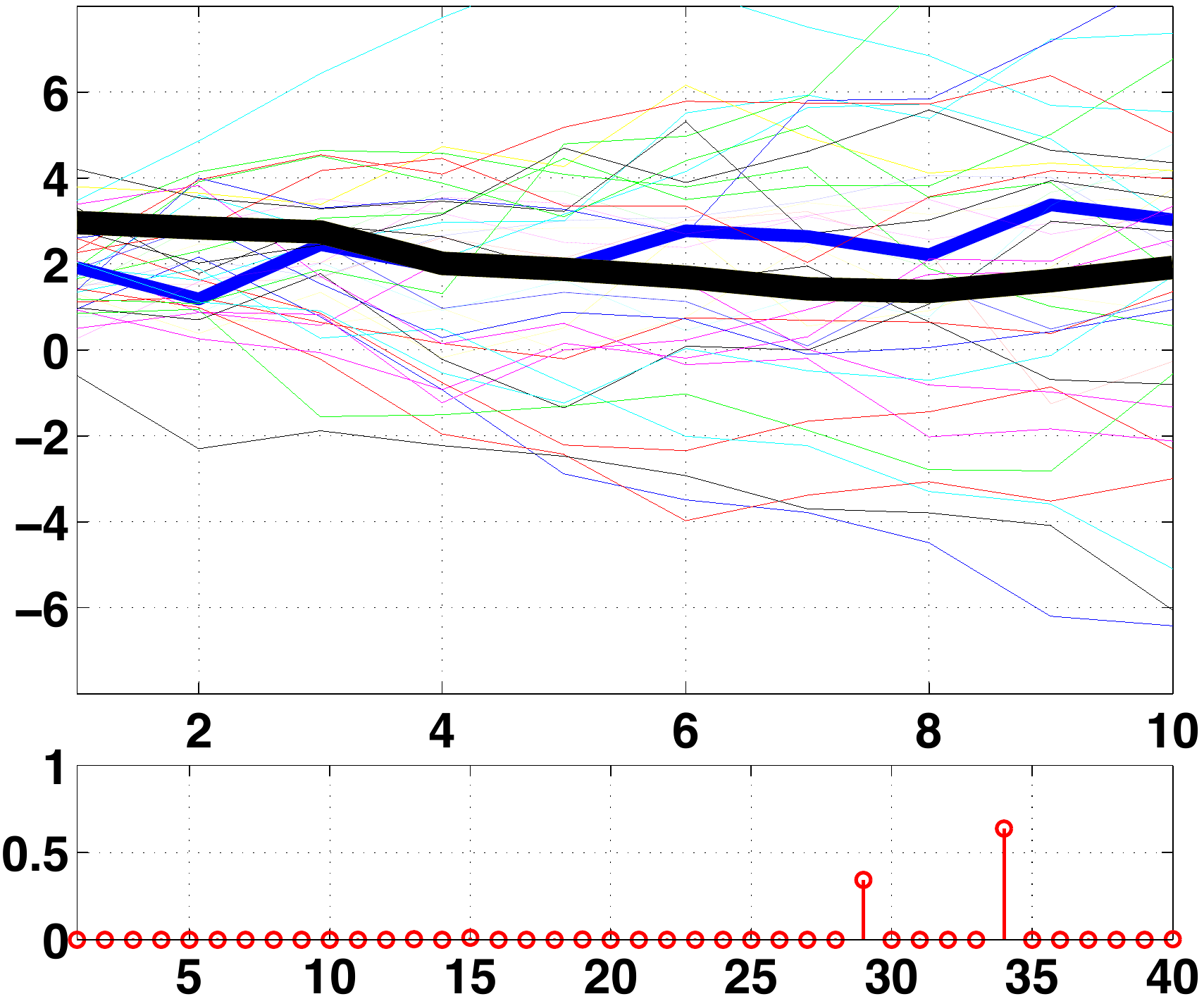}}
  \hspace{0.2cm}
   \subfigure[SIR with resampling at $d=4,8$.]{\includegraphics[width=0.43\textwidth]{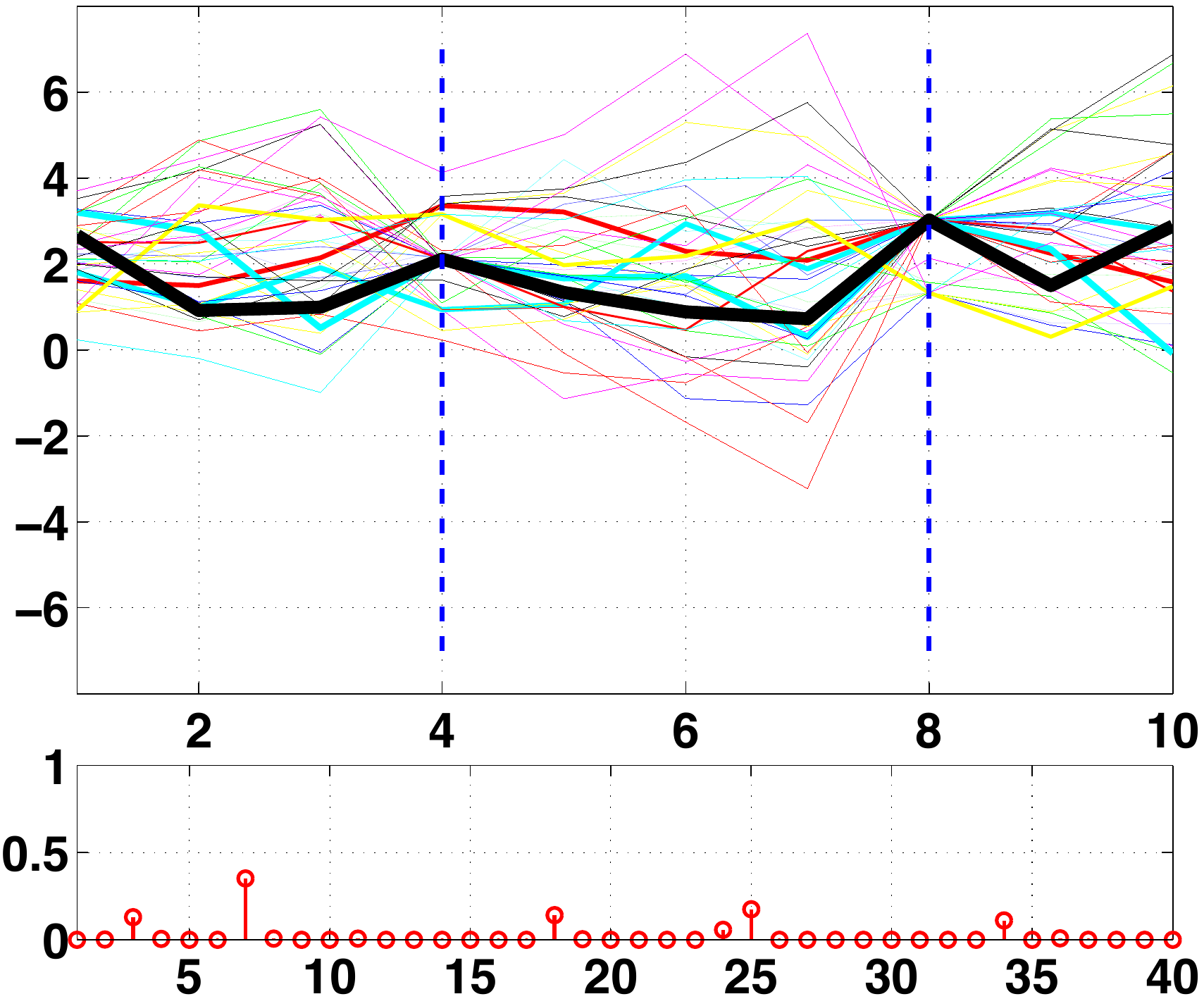}}
}
\caption{
{\footnotesize 
Graphical representation of SIS and SIR. We consider as target density a multivariate Gaussian pdf, ${\bar \pi}({\bf x})=\prod_{d=1}^{10} \mathcal{N}(x_d|2,\frac{1}{2})$. In each figure, every component of different particles are represented, so that each particle ${\bf x}^{(i)}=x_{1:D}^{(i)}$ forms a {\it path} (with $D=10$). We set $N=40$. The normalized weights ${\bar w}_D^{(i)}=w({\bf x}^{(i)})$ corresponding to each figure are also shown in the bottom. The line-width of each path is proportional to the corresponding weight ${\bar w}_D^{(i)}$. The particle corresponding to the greatest weight  is always depicted in black. The proposal pdfs used are $q_1(x_1)=\mathcal{N}(x_1|2,1)$ and $q(x_d|x_{d-1})=\mathcal{N}(x_d|x_{d-1},1)$ for $d\geq 2$. {\bf (a)} Batch IS or SIS. {\bf (b)} SIR resampling steps at the iterations $d=4,8$.
 } 
  }
\label{FigSIS_SIR}
\end{center}
\end{figure*}

A simple look at I-MTM2 and PMH shows that they are strictly related \cite{MTM_PMH14}. Indeed, the structure of the two algorithms coincides. The main difference lies that the candidates in PMH are generated sequentially, using a SIR scheme. If no resampling steps are applied, then I-MTM2 and PMH are {\it exactly} the same algorithm, where the candidates are drawn in a {\it batch} setting or {\it sequential} way. Hence, the application of resampling steps is the main difference between the generation procedures of PMH and I-MTM2. Owing to the use of resampling, the candidates $\{{\bf x}^{(1)},\ldots,{\bf x}^{(N)}\}$ proposed by PMH are not independent (differently from I-MTM2). As an example, Figure \ref{FigSIS_SIR} shows $N=40$ particles (with $D=10$) generated and weighted by SIS and SIR procedures (each path is a generated particle ${\bf x}^{(i)}=x_{1:10}^{(i)}$). The generation of correlated samples can be also considered in MTM methods without jeopardizing  the ergodicity of the chain, as simply shown for instance in \cite{Craiu07}, for instance. Another difference is the use of ${\widetilde Z}$ or ${\widehat Z}$. However, if a proper weighting of a resampled particle is employed, both estimators coincide  \cite{GIS17,GISssp16,MTM_PMH14}. Furthermore, both I-MTM2 and PMH can be considered as I-MH schemes where a proper importance sampling weighting of a resampled particle is employed \cite{GISssp16}. Namely, I-MTM2 and PMH are equivalent to an I-MH technique using the following complete proposal pdf, 
\begin{equation}
\label{CompleteProp}
\widetilde{q}({\bm \theta})=\int_{\mathcal{D}^N} \widehat{\pi}({\bm \theta}|{\bm \theta}^{(1:N)}) \left[\prod_{i=1}^N q({\bm \theta}^{(i)})\right]d{\bm \theta}^{(1:N)},
\end{equation}
 where $\widetilde{\pi}$ is given in Eq. \eqref{ApproxPI_IMP}, i.e., ${\bm \theta}^{(j)}\sim \widetilde{q}({\bm \theta})$, and then considering the generalized (proper) IS weighting, $w({\bm \theta}^{(j)})=\widehat{Z}^*$, $w({\bm \theta}_{t-1})=\widehat{Z}_{t-1}$  \cite{GIS17,GISssp16}. For further details see Appendix \ref{AfterRES}.

\subsubsection{Particle Marginal Metropolis-Hastings (PMMH) method}
\label{PMMH}

Assume now that the variable of interest if formed by both dynamical and static variables, i.e., ${\bm \theta}=[{\bf x}, {\bm \lambda}]^{\top}$. For instance, this is the case of inferring both, an hidden state ${\bf x}$ in state-space model, and static parameters ${\bm \lambda}$ of the model. The Particle Marginal Metropolis-Hastings (PMMH) technique is a extension of PMH which addresses this problem.

Let us consider ${\bf x}=x_{1:D}=[x_1,x_2,\ldots,x_D]\in\mathbb{R}^{d_x}$, and an additional model parameter ${\bm \lambda} \in \mathbb{R}^{d_\lambda}$ to be inferred as well (${\bm \theta}=[{\bf x}, {\bm \lambda}]^{\top}\in \mathbb{R}^D$, with $D=d_x+d_\lambda$). Assuming a prior pdf $g_\lambda({\bm \lambda})$ over ${\bm \lambda}$, and a factorized complete posterior pdf $\bar{\pi}({\bm \theta})=\bar{\pi}({\bf x},{\bm \lambda})$, 
\begin{eqnarray*}
\bar{\pi}({\bf x},{\bm  \lambda})\propto \pi({\bf x},{\bm  \lambda})&=&g_\lambda({\bm \lambda}) \pi({\bf x}|{\bm \lambda}),
\end{eqnarray*}
where $\pi({\bf x}|{\bm \lambda})=\gamma_1(x_1|{\bm \lambda}) \prod_{d=2}^D\gamma_d(x_d|x_{1:d-1},{\bm \lambda})$. For a specific value of ${\bm \lambda}$, we can use a particle filter approach, obtaining the approximation $\widehat{\pi}(\x|{\bm \lambda})=\sum_{n=1}^N {\bar w}_D^{(n)}\delta(\x-\x^{(n)})$ and the estimator $\widetilde{Z}({\bm \lambda})$, as described above. The PMMH technique is then summarized in Table \ref{alg:PMMH}. The pdf $q_\lambda({\bm \lambda}|{\bm \lambda}_{t-1})$ denotes the proposal density for generating possible values of ${\bm \lambda}$.
 Observe that, with the specific choice $q_\lambda({\bm \lambda}|{\bm \lambda}_{t-1})=g_\lambda({\bm \lambda})$, then the acceptance function becomes
\begin{equation}
\alpha = \min\left[1, \frac{\widetilde{Z}({\bm \lambda}^*)}{\widetilde{Z}({\bm \lambda}_{t-1})}\right].
\end{equation}
Note also that PMMH  w.r.t. to ${\bm \lambda}$ can be interpreted as MH method where an the posterior cannot be evaluated point-wise. Indeed, $\widetilde{Z}({\bm \lambda})$ approximates the marginal likelihood $p({\bf y}|{\bm \lambda})$ \cite{andrieu2009}.

\begin{table}[!h]
	\centering
	\caption{{\bf Particle Marginal MH (PMMH) algorithm}}
	    \begin{tabular}{|p{0.95\columnwidth}|}
		\hline
		  \vspace{-0.5cm} 
  \begin{enumerate}
\item  \texttt{Initialization:} Choose the initial states $\x_0$, ${\bm \lambda}_0$, and an initial approximation $\widetilde{Z}_0({\bm \lambda})\approx Z({\bm \lambda}) \approx p({\bf y}|{\bm \lambda})$.
\item \texttt{For $t=1,\ldots,T$:}
\begin{enumerate}
\item Draw ${\bm \lambda}^*\sim q_\lambda({\bm \lambda}|{\bm \lambda}_{t-1})$.
\item Given ${\bm \lambda}^*$, run a particle filter obtaining $\widehat{\pi}(\x|{\bm \lambda}^*)=\sum_{n=1}^N {\bar w}_D^{(n)}\delta(\x-\x^{(n)})$ and $\widetilde{Z}({\bm \lambda}^*)$, as in Eq. \eqref{EstZ2}.
\item \item Draw ${\bf x}^* \sim  \widehat{\pi}({\bf x}|{\bm \lambda}^*, {\bf x}^{(1:N)})$, i.e., choose a particle ${\bf x}^*=\{{\bf x}^{(1)},\ldots,{\bf x}^{(N)}\}$ with probability  $\bar{w}_D^{(i)}$, $i=1,...,N$.	 
\item Set  ${\bm \lambda}_t={\bm \lambda}^*$, $\x_t={\bf x}^*$, with probability 
\begin{equation}
\label{AlfaPMMH}
\alpha = \min\left[1, \frac{\widetilde{Z}({\bm \lambda}^*) g_\lambda({\bm \lambda}^*) q_\lambda({\bm \lambda}_{t-1}|{\bm \lambda}^*)}{\widetilde{Z}({\bm \lambda}_{t-1})  g_\lambda({\bm \lambda}_{t-1}) q_\lambda({\bm \lambda}^*|{\bm \lambda}_{t-1})}\right].
\end{equation}
Otherwise, set ${\bm \lambda}_t={\bm \lambda}^*$ and $\x_t=\x_{t-1}$.
\end{enumerate}
  \item \texttt{ Return:} $\{\x_t\}_{t=1}^T$ and $\{{\bm \lambda}_t\}_{t=1}^T$.
  \vspace{-0.3cm}
    \end{enumerate} 	\\
	\hline
	\end{tabular}		
	\label{alg:PMMH}
\end{table}

\subsection{Group Metropolis Sampling}

The  auxiliary weighted samples in the I-MTM schemes (i.e., the $N-1$ samples drawn at each iteration that are not selected to be compared with the previous state ${\bm \theta}_{t-1}$) can be recycled providing a consistent and more efficient estimators \cite{GISssp16,GIS17}.

\begin{figure}[h!]
\centering
\centerline{
\includegraphics[width=10cm]{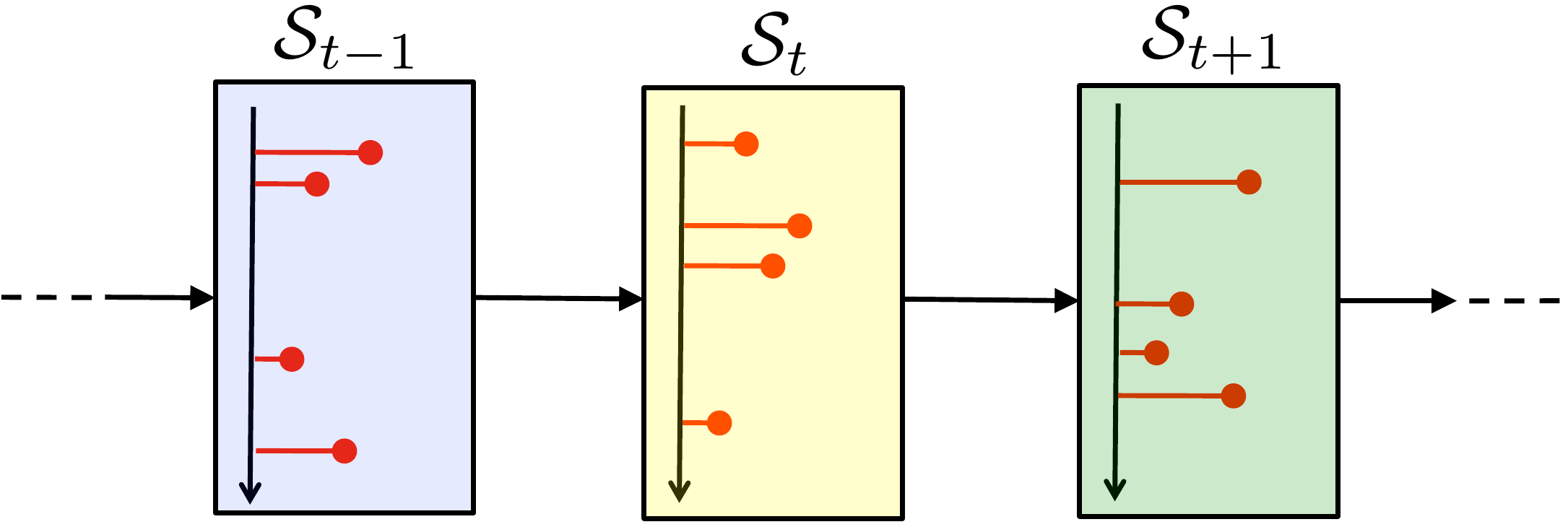} }
\caption{Chain of sets $\mathcal{S}_t=\{{\bm \theta}_{n,t},\rho_{n,t}\}_{n=1}^N$ generated by the GMS method (graphical representation with $N=4$). 
 }
\label{FigGMS}
\end{figure}

The so-called  Group Metropolis Sampling (GMS) method is shown in Table \ref{alg:GMS}.
 GMS yields a sequence of sets of weighted samples $\mathcal{S}_t=\{{\bm \theta}_{n,t},\rho_{n,t}\}_{n=1}^N$, for $t=1,\ldots,T$, where we have denoted with  $\rho_{n,t}$ the importance weights assigned to the samples ${\bm \theta}_{n,t}$ (see Figure \ref{FigGMS}). All the samples are then employed for a joint particle approximation of the target. Alternatively, GMS can directly provide an approximation of a specific moment of the target pdf (i.e., given a particular function $f$). The estimator of this specific moment provided by GMS is 
\begin{eqnarray}
\label{Echecazzo}
\widetilde{I}_{NT}=\frac{1}{T}\sum_{t=1}^T \sum_{n=1}^N \frac{\rho_{n,t}}{\sum_{i=1}^N \rho_{i,t}} f({\bm \theta}_{n,t})=\frac{1}{T}\sum_{t=1}^T \widetilde{I}_N^{(t)}.  
\end{eqnarray}

\begin{table}[!h]
	\centering
	\caption{{\bf Group Metropolis Sampling}}
	    \begin{tabular}{|p{0.95\columnwidth}|}
\hline
\begin{enumerate}
	\item  \texttt{Initialization:}  Choose an initial state ${\bm \theta}_0$ and an initial approximation $\widehat{Z}_0\approx Z$.
    \item \texttt{FOR} $t = 1, \ldots, T$:
    	\begin{enumerate}
        \item Draw ${\bm \theta}^{(1)},{\bm \theta}^{(2)},\ldots, {\bm \theta}^{(N)}  \sim q({\bm \theta})$.
       \item Compute the importance weights
       \begin{equation}        w({\bm \theta}^{(n)})=\frac{\pi({\bm \theta}^{(n)})}{q({\bm \theta}^{(n)})}, \quad \mbox{ with } \quad  n=1,\ldots,N,
        \end{equation}
  define $\mathcal{S}^*=\{{\bm \theta}^{(n)},w({\bm \theta}^{(n)})\}_{n=1}^N$; and compute $\widehat{Z}^*=\frac{1}{N}\sum_{n=1}^N w({\bm \theta}^{(n)})$.     
\item   Set  $\mathcal{S}_t=\mathcal{S}^*$, i.e.,
$$
\mathcal{S}_t=\left\{{\bm \theta}_{n,t}={\bm \theta}^{(n)},\rho_{n,t}=w({\bm \theta}^{(n)})\right\}_{n=1}^N,
$$
and $\widehat{Z}_{t}=\widehat{Z}^*$, with probability 
\begin{equation}
\label{AlfaGMS}
\alpha(\mathcal{S}_{t-1},\mathcal{S}^*) = \min\left[1, \frac{\widehat{Z}^*}{\widehat{Z}_{t-1}}\right].
\end{equation}
Otherwise, set $\mathcal{S}_t=\mathcal{S}_{t-1}$ and $\widehat{Z}_{t}=\widehat{Z}_{t-1}$.
\end{enumerate}
\item \texttt{Return:} All the sets $\{\mathcal{S}_t\}_{t=1}^T$, or $\{\widetilde{I}_N^{(t)}\}_{t=1}^T$ where
\begin{equation}
\widetilde{I}_N^{(t)}=\sum\limits_{n=1}^N\frac{\rho_{n,t}}{\sum_{i=1}^N \rho_{i,t}}g({\bm \theta}_{n,t}), 
\end{equation}
and  $\widetilde{I}_{NT}=\frac{1}{T}\sum_{t=1}^T \widetilde{I}_N^{(t)}$.
   \end{enumerate}
  \\
\hline
   	\end{tabular}		
	\label{alg:GMS}
\end{table}


%

%
 Unlike in the I-MTM schemes, no resampling steps are performed in GMS. However, we can recover an  I-MTM chain from the GMS output applying one resampling step when $\mathcal{S}_t\neq  \mathcal{S}_{t-1}$, i.e.,       
\begin{gather}
{\bm \theta}_{t}= \left\{
\begin{split}
\label{RecChain}
&{\bm \theta}_{t} \sim \sum_{n=1}^N \frac{\rho_{n,t}}{\sum_{i=1}^N \rho_{i,t}}  \delta({\bm \theta}-{\bm \theta}_{n,t}),   \quad \mbox{ if }  \quad \mathcal{S}_t\neq  \mathcal{S}_{t-1}, \\
&{\bm \theta}_{t-1}, \quad\quad\quad\quad\quad\quad\quad\quad\quad\quad\quad\mbox{ }\mbox{ }\mbox{ } \mbox{ if } \quad \mathcal{S}_t=  \mathcal{S}_{t-1},
\end{split}
\right. 
\end{gather}
for $t=1,\ldots,T$. More specifically, $\{{\bm \theta}_t\}_{t=1}^T$ is a Markov chain obtained by one run of an I-MTM2 technique. 
The consistency of the GMS estimators is discussed in Appendix \ref{ConGMS}.
GMS can be also interpreted as an iterative IS scheme where an IS approximation of $N$ samples is built at each iteration and compared with the previous IS approximation. This procedure is iterated $T$ times and all the accepted IS estimators $\widetilde{I}_N^{(t)}$ are finally combined to provide a unique global approximation of $NT$ samples. Note that the temporal combination of the IS estimators is obtained dynamically by the random repetitions due to the rejections in the acceptance test. 

{\rem The complete weighting procedure in GMS can be interpreted as the composition of two weighting schemes: (a) by an IS approach building $\{\rho_{n,t}\}_{n=1}^N$ and (b) by the possible random repetitions due to the rejections in the acceptance test.}
\newline
Figure \ref{FigGMS} depicts a graphical representation of the GMS outputs as chain of sets $\mathcal{S}_t=\{{\bm \theta}_{n,t},\rho_{n,t}\}_{n=1}^N$.

\subsection{Ensemble MCMC algorithms}

Another alternative procedure, often referred as Ensemble MCMC (EnMCMC) methods (a.k.a., called Locally weighted MCMC), involving several tries at each iteration \cite{Neal11,LWMCMC}. Related techniques has been proposed independently in different works \cite{Calderhead14,Austad07}. 
First, let us define the joint proposal density
\begin{equation}
q({\bm \theta}^{(1)},\ldots,{\bm \theta}^{(N)}|{\bm \theta}_{t}): \mathcal{D}^{N} \rightarrow \mathbb{R},
\end{equation}
and, considering $N+1$ possible elements, $\mathcal{S}=\{{\bm \theta}^{(1)}, \ldots, {\bm \theta}^{(N)},{\bm \theta}^{(N+1)}\}$, we define the $D\times N$ matrix
\begin{equation}
\label{MatrixTeta}
{\bf \Theta}_{\neg k}=[{\bm \theta}^{(1)},\ldots,{\bm \theta}^{(k-1)},{\bm \theta}^{(k+1)},\ldots,{\bm \theta}^{(N+1)}],
\end{equation}
with columns all the vectors in $\mathcal{S}$ with the exception of ${\bm \theta}^{(k)}$. For simplicity, in the followings we abuse of the notation writing $q({\bm \theta}^{(1)},\ldots,{\bm \theta}^{(N)}|{\bm \theta}_{t})=q({\bf \Theta}_{\neg N+1}|{\bm \theta}_{t})$, for instance. One simple example of joint proposal pdf is
\begin{equation}
q({\bm \theta}^{(1)},\ldots,{\bm \theta}^{(N)}|{\bm \theta}_{t})=\prod_{n=1}^N q({\bm \theta}^{(n)}|{\bm \theta}_{t}),
\end{equation}
i.e., considering independence among ${\bm \theta}^{(n)}$'s (and having the same marginal proposal pdf $q$). More sophisticated joint proposal densities can be employed. A generic EnMCMC algorithm is outlined in Table \ref{alg:En-MCMCgen}.

\begin{table}[!h]
	\centering
	\caption{{\bf Generic EnMCMC algorithm}}
	    \begin{tabular}{|p{0.95\columnwidth}|}
\hline
\begin{enumerate}
	\item \texttt{Initialization:}  Choose an initial state ${\bm \theta}_0$.
    \item \texttt{FOR} $t = 1, \ldots, T$:
    	\begin{enumerate}
        \item Draw $\widetilde{{\bm \theta}}^{(1)},\ldots, \widetilde{{\bm \theta}}^{(N)}  \sim q({\bm \theta}^{(1)},\ldots,{\bm \theta}^{(N)}|{\bm \theta}_{t-1})$.
        \item Set $\widetilde{{\bm \theta}}^{(N+1)}={\bm \theta}_{t-1}$.
     \item Set ${\bm \theta}_t=\widetilde{{\bm \theta}}^{(j)}$, resampling $\widetilde{{\bm \theta}}^{(j)}$  within the set 
     $$\{\widetilde{{\bm \theta}}^{(1)},\ldots,\widetilde{{\bm \theta}}^{(N)},\widetilde{{\bm \theta}}^{(N+1)}={\bm \theta}_{t-1}\},
     $$
      formed by $N+1$ samples, according to the probability mass function 
     $$
     \alpha({\bm \theta}_{t-1},\widetilde{{\bm \theta}}^{(j)})=\frac{\pi(\widetilde{{\bm \theta}}^{(j)})q({\bf \Theta}_{\neg j}|\widetilde{{\bm \theta}}^{(j)})}{\sum_{\ell=1}^{N+1} \pi(\widetilde{{\bm \theta}}^{(\ell)})q({\bf \Theta}_{\neg \ell}|\widetilde{{\bm \theta}}^{(\ell)}) },
     $$
     for $j=1,\ldots,N+1$ and ${\bf \Theta}_{\neg j}$ is defined in Eq. \eqref{MatrixTeta}.
         \end{enumerate}
         \item \texttt{Return:}  $\{{\bm \theta}_t\}_{t= 1}^T$. 
   \end{enumerate}
  \\
\hline
   	\end{tabular}		
	\label{alg:En-MCMCgen}
\end{table}

{\rem Note that with respect to the generic MTM method, EnMCMC does not require to draw auxiliary samples and weights. Therefore, in EnMCMC a smaller number of evaluation of target is required w.r.t.  a generic MTM scheme. }

\subsubsection{Independent Ensemble MCMC}

In this section, we present an interesting special case, which employs a single proposal pdf $q({\bm \theta})$ independent on the previous state of the chain, i.e.,
\begin{equation}
q({\bm \theta}^{(1)},\ldots,{\bm \theta}^{(N)}|{\bm \theta}_{t})=q({\bm \theta}^{(1)},\ldots,{\bm \theta}^{(N)})=
\prod_{n=1}^N q({\bm \theta}^{(n)}).
\end{equation}
In this case, the technique can be simplified as shown below. At each iteration, the algorithm described in Table \ref{alg:En-MCMCind} generates $N$ new samples ${\bm \theta}^{(1)},{\bm \theta}^{(2)},\ldots, {\bm \theta}^{(N)}$  and then resample the new state ${\bm \theta}_t$ within a set of $N+1$ samples, $\{{\bm \theta}^{(1)},\ldots,{\bm \theta}^{(N)},{\bm \theta}_{t+1}\}$ (which includes the previous state), according to the probabilities
 \begin{equation}
 \label{A_enMCMC}
     \alpha({\bm \theta}_{t-1},{\bm \theta}^{(j)})=\bar{w}_j=\frac{w({\bm \theta}^{(j)})}{\sum_{i=1}^N w({\bm \theta}^{(i)})+ w({\bm \theta}_{t-1})}, \qquad j=1,\ldots,N+1,
\end{equation}
where $w({\bm \theta})=\frac{\pi({\bm \theta})}{q({\bm \theta})}$ denotes the importance sampling weight. Note that Eq. \eqref{A_enMCMC} for $N=1$ becomes 
\begin{eqnarray}
     \alpha({\bm \theta}_{t-1},{\bm \theta}^{(j)})&=&\frac{w({\bm \theta}^{(j)})}{ w({\bm \theta}^{(i)})+ w({\bm \theta}_{t-1})}, \nonumber \\
     &=&\frac{\frac{\pi({\bm \theta}^{(j)})}{q({\bm \theta}^{(j)})}}{ \frac{\pi({\bm \theta}^{(j)})}{q({\bm \theta}^{(j)})}+ \frac{\pi({\bm \theta}_{t-1})}{q({\bm \theta}_{t-1})}},\nonumber \\
     &=&\frac{\pi({\bm \theta}^{(j)})q({\bm \theta}_{t-1})}{\pi({\bm \theta}^{(j)})q({\bm \theta}_{t-1})+ \pi({\bm \theta}_{t-1})q({\bm \theta}^{(j)})},  
 \label{A_enMCMC_2}
\end{eqnarray}
that is the Barker's acceptance function (see \cite{Barker65,MHWiley17}).

\begin{table}[!h]
	\centering
	\caption{{\bf EnMCMC with an independent proposal pdf (I-EnMCMC).}}
	    \begin{tabular}{|p{0.95\columnwidth}|}
\hline
\begin{enumerate}
	\item \texttt{Initialization:}  Choose an initial state ${\bm \theta}_0$.
    \item \texttt{FOR} $t = 1, \ldots, T$:
    	\begin{enumerate}
        \item Draw ${\bm \theta}^{(1)},{\bm \theta}^{(2)},\ldots, {\bm \theta}^{(N)}  \sim q({\bm \theta})$, and set ${\bm \theta}^{(N+1)}={\bm \theta}_{t-1}$.
       \item Compute the importance weights
       \begin{equation}           
       w({\bm \theta}^{(n)})=\frac{\pi({\bm \theta}^{(n)})}{q({\bm \theta}^{(n)})}, \quad \mbox{ with } \quad  n=1,\ldots,N+1.
        \end{equation}
     \item Set ${\bm \theta}_t={\bm \theta}^{(j)}$, resampling  ${\bm \theta}^{(j)}$  within the set $\{{\bm \theta}^{(1)},\ldots,{\bm \theta}^{(N)},{\bm \theta}_{t+1}\}$ formed by $N+1$ samples, according to the probability mass function 
     $$
     \alpha({\bm \theta}_{t-1},{\bm \theta}^{(j)})=\bar{w}_j=\frac{w({\bm \theta}^{(j)})}{\sum_{i=1}^N w({\bm \theta}^{(i)})+ w({\bm \theta}_{t-1})}.
     $$
         \end{enumerate}
         \item \texttt{Return:}  $\{{\bm \theta}_t\}_{t= 1}^T$. 
   \end{enumerate}
  \\
\hline
   	\end{tabular}		
	\label{alg:En-MCMCind}
\end{table}

\begin{figure}[h!]
\centering
\centerline{
\includegraphics[width=16cm]{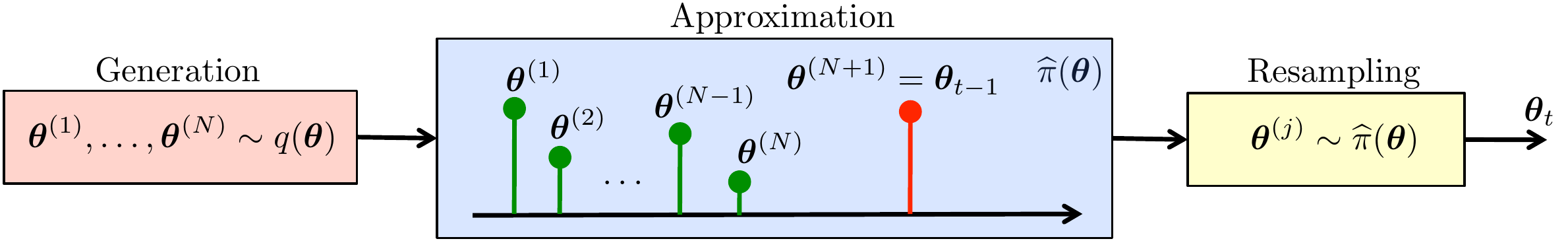} 
}
\caption{Graphical representation of the steps of the I-EnMCMC scheme.
 }
\label{FigEnMCMC}
\end{figure}

As discussed in \cite[Appendix B]{OMCMC_DSP}, \cite[Appendix C]{LAIS17}, \cite{GISssp16}, the density of a resampled candidate becomes closer and closer to $\bar{\pi}$ as $N$ grows, i.e., $N\rightarrow \infty$.
Hence, the performance of  I-EnMCMC clearly improves with $N\rightarrow \infty$ (see Appendix \ref{AfterRES}).
 The I-EnMCMC algorithm produces an ergodic chain with invariant density $\bar{\pi}$, by resampling $N+1$ samples  at each iteration ($N$ new samples ${\bm \theta}^{(1)},\ldots, {\bm \theta}^{(N)}$ from $q$ and setting ${\bm \theta}^{(N+1)}={\bm \theta}_{t-1}$). Figure \ref{FigEnMCMC} summarizes the steps of I-EnMCMC.

%

\subsection{Delayed Rejection Metropolis (DRM) Sampling}

\begin{figure}[h!]
\centering
\centerline{
\includegraphics[width=16cm]{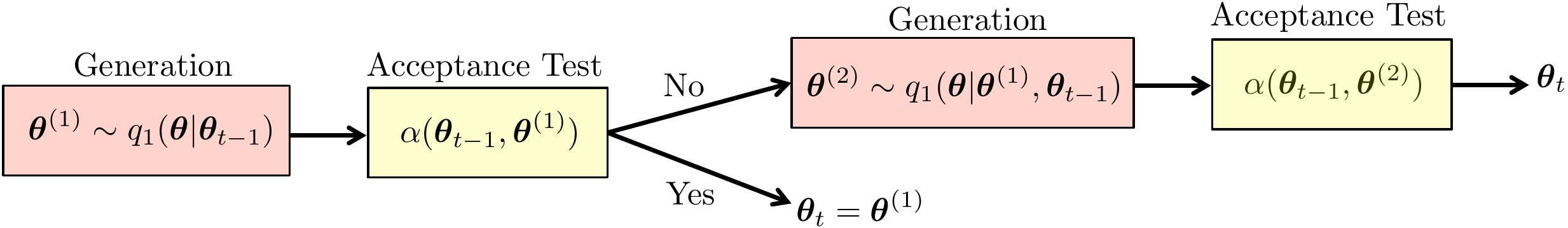} 
}
\caption{Graphical representation of the DRM scheme with two acceptance tests.
 }
\label{FigDRM}
\end{figure}

An alternative use of different candidates in one iteration of a Metropolis-type method is given in \cite{Haario06,Mira01,Tierney1999}. 
The idea behind the proposed algorithm, called Delayed Rejection Metropolis (DRM) algorithm, is the following. As in a standard MH method, at each iteration, one sample is proposed ${\bm \theta}^{(1)}\sim q_1({\bm \theta}|{\bm \theta}_{t-1})$ and accepted with probability
$$
\alpha_1({\bm \theta}_{t-1},{\bm \theta}^{(1)})= \min\left(1,\frac{\pi({\bm \theta}^{(1)})q_1({\bm \theta}_{t-1}|{\bm \theta}^{(1)})}{\pi({\bm \theta}_{t-1}) q_1({\bm \theta}^{(1)}|{\bm \theta}_{t-1})}\right). 
$$
If ${\bm \theta}^{(1)}$ is accepted then ${\bm \theta}_t={\bm \theta}^{(1)}$ and the chain is moved forward. If  ${\bm \theta}^{(1)}$ is rejected, the DRM method suggests of drawing another samples  ${\bm \theta}^{(2)}\sim q_2({\bm \theta}|{\bm \theta}^{(1)},{\bm \theta}_{t-1})$ (considering a different proposal pdf $q_2$ taking into account possibly the previous candidate ${\bm \theta}^{(1)}$) and accepted with a suitable acceptance probability
{\footnotesize
         \begin{eqnarray*}
        &&\alpha_2({\bm \theta}_{t-1},{\bm \theta}^{(2)})= \\
      &&  =\min\left(1,\frac{ \pi({\bm \theta}^{(2)}) q_1({\bm \theta}^{(1)}|{\bm \theta}^{(2)})q_2({\bm \theta}_{t-1}|{\bm \theta}^{(1)},{\bm \theta}^{(2)})(1-\alpha_1({\bm \theta}^{(2)},{\bm \theta}^{(1)}))  }{  \pi({\bm \theta}_{t-1}) q_1({\bm \theta}^{(1)}|{\bm \theta}_{t-1}) q_2({\bm \theta}^{(2)}|{\bm \theta}^{(1)},{\bm \theta}_{t-1})(1-\alpha_1({\bm \theta}_{t-1},{\bm \theta}^{(1)})) }\right).       
      \end{eqnarray*} 
      }
 The acceptance function $\alpha_2({\bm \theta}_{t-1},{\bm \theta}^{(2)})$ is designed in order to ensure the ergodicity of the chain. If ${\bm \theta}^{(2)}$ is rejected we can set ${\bm \theta}_t={\bm \theta}_{t-1}$ and perform another iteration of the algorithm, or continue with this iterative strategy  drawing ${\bm \theta}^{(3)}\sim q_3({\bm \theta}|{\bm \theta}^{(2)},{\bm \theta}^{(1)},{\bm \theta}_{t-1})$ and test it with a proper probability $\alpha_3({\bm \theta}_{t-1},{\bm \theta}^{(3)})$.
  The DRM algorithm with only 2 acceptance stages is outlined in Table \ref{alg:DelayedRejection} and summarized in Figure \ref{FigDRM}.
   
   {\rem Note that the proposal pdf can be improved at each intermediate stage (${\bm \theta}^{(1)}\sim q_1({\bm \theta}|{\bm \theta}_{t-1})$, ${\bm \theta}^{(2)}\sim q_2({\bm \theta}|{\bm \theta}^{(1)},{\bm \theta}_{t-1})$ etc.), using the information provided by the previous generated samples and the corresponding target evaluations.}

The idea behind DRM of creating a path of intermediate points, then improving the proposal pdf, and hence fostering larger jumps have been also considered in other works \cite{LucaJesse1,ModeJump}.   

\begin{table}[!h]
	\centering
	\caption{{\bf Delayed Rejection Metropolis algorithm with 2 acceptance steps.}}
	    \begin{tabular}{|p{0.95\columnwidth}|}
\hline
\begin{enumerate}
	\item \texttt{Initialization:} Choose an initial state ${\bm \theta}_0$. 
	     \item \texttt{FOR} $t = 1, \ldots, T$:
    	\begin{enumerate}
         \item Draw ${\bm \theta}^{(1)}\sim q_1({\bm \theta}|{\bm \theta}_{t-1})$ and $u_1\sim \mathcal{U}([0,1])$.
      \item Define the probability
       \begin{eqnarray}
          \alpha_1({\bm \theta}_{t-1},{\bm \theta}^{(1)})&=& \min\left(1,\frac{\pi({\bm \theta}^{(1)})q_1({\bm \theta}_{t-1}|{\bm \theta}^{(1)})}{\pi({\bm \theta}_{t-1}) q_1({\bm \theta}^{(1)}|{\bm \theta}_{t-1})}\right), 
                \label{eq:alphaDelay1} 
      \end{eqnarray}
    \item  If $u_1\leq  \alpha_1({\bm \theta}_{t-1},{\bm \theta}^{(1)})$,  set ${\bm \theta}_t={\bm \theta}^{(1)}$. 
    \item   Otherwise,  if $ u_1>  \alpha_1({\bm \theta}_{t-1},{\bm \theta}^{(1)})$,  do: 
      \newline
              (d1)    Draw ${\bm \theta}^{(2)}\sim q_2({\bm \theta}|{\bm \theta}^{(1)},{\bm \theta}_{t-1})$ and  $u_2\sim \mathcal{U}([0,1])$.
        \newline    
              (d2)    Given the function
              \begin{eqnarray*}       
                \psi({\bm \theta}_{t-1},{\bm \theta}^{(2)}|{\bm \theta}^{(1)})&=&\pi({\bm \theta}_{t-1}) q_1({\bm \theta}^{(1)}|{\bm \theta}_{t-1}) \times\\
  &&q_2({\bm \theta}^{(2)}|{\bm \theta}^{(1)},{\bm \theta}_{t-1}) (1-\alpha_1({\bm \theta}_{t-1},{\bm \theta}^{(1)})),
               \end{eqnarray*} 
       and the  probability
                \begin{eqnarray}
        \alpha_2({\bm \theta}_{t-1},{\bm \theta}^{(2)})=\min\left(1,\frac{\psi({\bm \theta}^{(2)},{\bm \theta}_{t-1}|{\bm \theta}^{(1)})}{ \psi({\bm \theta}_{t-1},{\bm \theta}^{(2)}|{\bm \theta}^{(1)}) }\right).   
                \label{eq:alphaDelay2}
      \end{eqnarray} 
        \newline  
        (d3) If $u_2\leq  \alpha_2({\bm \theta}_{t-1},{\bm \theta}^{(2)})$,  set ${\bm \theta}_t={\bm \theta}^{(2)}$. 
        \newline
         (d4)   Otherwise,  if $ u_2>  \alpha_1({\bm \theta}_{t-1},{\bm \theta}^{(2)})$, set ${\bm \theta}_t={\bm \theta}_{t-1}$.
      \end{enumerate}
      \item \texttt{Return:}  $\{{\bm \theta}_t\}_{t= 1}^T$. 
  \end{enumerate}
  \\
\hline
   	\end{tabular}		
	\label{alg:DelayedRejection}
\end{table}

\section{Summary: computational cost, differences and connections}

The performance of the algorithms described above improves as $N$ grows, in general: the correlation among samples vanishes to zero, and the acceptance rate of new state approaches one (see Section \ref{First_EX}).\footnote{Generally, an acceptance rate close to $1$ is not an evidence of good performance for an MCMC algorithm. However, for the techniques tackled in this work, the situation is different: as $N$ grows, the procedure used for proposing a novel possible state (involving $N$ tries, resampling steps etc.) becomes better and better, yielding a better approximation of the target pdf. See Appendix \ref{AfterRES} for further details.} As $N$ increases, they become similar and similar to an  exact sampler drawing independent samples directly from the target density (for MTM, PMH and EnMCMC schemes the explanation is given in Appendix \ref{AfterRES}). However, this occurs at the expense of an additional computational cost. 
 
In Table \ref{SummEvalTable}, we summarize the total number of target evaluations, $E$, and the total number of samples 
used in the final estimators, $Q$ (without considering to remove any burn-in period). The generic MTM algorithm has the greatest number of target evaluations. However, a random-walk proposal pdf can be used in a generic MTM algorithm  and, in general, it fosters the exploration of the state space.  In this sense, the generic EnMCMC seems to be preferable w.r.t. MTM, since $E=NT$ and the random-walk proposal can be applied. A disadvantage of the EnMCMC schemes is that their acceptance function seems worse in terms of Peskun's ordering \cite{Peskun73} (see numerical results in Section \ref{First_EX}). Namely,  fixing the number of $N$ tries, the target $\pi$ the proposal $q$ pdfs, the MTM schemes seem to provide greater acceptance rates than the corresponding EnMCMC techniques. This is theoretically proved for $N=1$ \cite{Peskun73}, and the difference vanishes to zero as $N$ grows.
The GMS technique, like other strategies \cite{Calderhead14,OMCMC_DSP}, has been proposed to recycle samples or re-use target evaluations, in order to increase $Q$ (see also Section \ref{Par_MTM}).

\begin{table}[h!]
\begin{center}
\caption{{\bf Total number of target evaluations, $E$, and total number of samples, $Q$, used in the final estimators.}}
\vspace{0.1cm}
\begin{tabular}{|c|c|c|c|c|}
\hline 
{\bf Algorithm}  &  {\bf MTM} &  {\bf I-MTM}   &  {\bf I-MTM2} &  {\bf PMH}     \\ 
\hline
$E$   &  $2NT-1$  & $NT$  & $NT$ & $NT$      \\ 
\hline
$Q$   &  $T$ & $T$ & $T$ &   $T$  \\ 
\hline
\hline
{\bf Algorithm}  &  {\bf GMS} & {\bf EnMCMC} & {\bf I-EnMCMC} & {\bf DMR}  \\ 
\hline
$E$   &  $NT$  & $NT$   & $NT$ &  $NT$    \\ 
\hline
$Q$   &  $NT$  &  $T$ & $T$ &  $T$   \\ 
\hline
\end{tabular}
\label{SummEvalTable}
\end{center}
\end{table}

 In PMH, the components of the different tries are drawn sequentially and they are correlated due to the application of the resampling steps. In DMR, each candidate is drawn in a batch way (all the components jointly) but the different candidates are drawn in a sequential manner (see Figure \ref{FigDRM}), ${\bm \theta}^{(1)}$ then ${\bm \theta}^{(2)}$ etc.    
The benefit of this strategy is that the proposal pdf can be improved considering the previous generated tries. Hence, if the proposal takes into account the previous samples, DMR generates correlated candidates as well. The main disadvantage  of DRM is that the implementation for a generic $N>2$ is not straightforward.

 I-MTM and  I-MTM2 differs for the acceptance function employed.
Furthermore, The main difference between the I-MTM2 and PMH schemes is the use of resampling steps during the generation the different tries. For this reason, the candidates of PMH are correlated (unlike in I-MTM2).
I-MTM2 and PMH can be interpreted as I-MH methods using a sophisticated proposal density $\widehat{q}({\bm \theta})$ in Eq. \eqref{CompleteProp}, and an extended IS weighting procedure is employed. Note that, indeed, $\widehat{q}$ cannot be evaluated pointwise, hence a standard IS weighting strategy cannot be employed.

\section{Numerical Experiments}
\label{SimuSect}
We test different MCMC using multiple candidates in different numerical experiments. In the first example, an exhaustive  comparison among several techniques with an independent proposal is given. We have considered different number of tries, length of the chain, parameters of the proposal pdfs and also different dimension of the inference problem. In the second numerical simulation,  
we compare different particle methods. The third one regards the hyperparameter selection for a Gaussian Process (GP) regression model. The last two examples are  localization problems in a wireless sensor network (WSN): in the fourth one some parameters of the WSN are also tuned, whereas in last example a real data analysis is performed.  

\subsection{A first comparison of efficiency}
\label{First_EX}
In order to compare the performance of different techniques, in this section we consider a multi-modal, multidimensional Gaussian target density. More specifically, we have 
 \begin{equation}
{\bar \pi}({\bm \theta})=\frac{1}{3}\sum_{i=1}^3\mathcal{N}({\bm \theta}|{\bm \mu}_i,{\bm \Sigma}_i), \qquad {\bm \theta}\in \mathbb{R}^D,
\end{equation}
where ${\bm \mu}_1=[\mu_{1,1},\ldots,\mu_{1,D}]^{\top}$, ${\bm \mu}_2=[\mu_{2,1},\ldots,\mu_{2,D}]^{\top}$, ${\bm \mu}_3=[\mu_{3,1},\ldots,\mu_{3,D}]^{\top}$, with $\mu_{1,d}=-3$, $\mu_{2,d}=0$, $\mu_{3,d}=2$ for all $d=1,\ldots,D$. Moreover, the covariance matrices are diagonal, ${\bm \Sigma}_i=\delta_i {\bf I}_D$ (where ${\bf I}_D$ is the $D\times D$ identity matrix), with $\delta_i=0.5$ for $i=1,2,3$. Hence, given a random variable ${\bf \Theta} \sim {\bar \pi}({\bm \theta})$, we know analytically that $E[{\bm \Theta}]=[{\bar \theta}_{1},\ldots,{\bar \theta}_{D}] ^{\top} $ with ${\bar \theta}_{d}= -\frac{1}{3}$ for all $d$, and $\mbox{diag}\{\mbox{Cov}[{\bm \Theta}]\}=[ \xi_{1},\ldots, \xi_{D}] ^{\top}$ with $\xi_{d}= \frac{85}{18}$ for all $d=1,\ldots,D$.

 We apply I-MTM, IMTM2 and I-EnMCMC in order to estimate  all the expected values and all the variances of the marginal target pdfs. Namely, for a given dimension $D$,  we have to estimate all $\{{\bar \theta}_{d}\}_{d=1}^D$ and  
 $\{\xi_{d}\}_{d=1}^D$, hence $2D$ values. The results are averaged over $3000$ independent runs. At each run, we compute an averaged square error obtained in the estimation of the $2D$ values and then calculate the Mean Square Error (MSE) averaged over the $3000$ runs. For all the techniques, we consider a Gaussian proposal density $q({\bm \theta})= \mathcal{N}({\bm \theta}|{\bm \mu},\sigma^2{\bf I}_D)$ with ${\bm \mu}=[\mu_1=0,\ldots,\mu_D=0]^{\top}$ (independent from the previous state) and different values of $\sigma$ are considered.  

\subsubsection{Description of the experiments}
We perform several experiments varying the number of tries, $N$, the length of the generated chain, $T$,  the dimension of the inference problem, $D$, and the scale parameter of the proposal pdf, $\sigma$. In Figures \ref{FigEx1_1}(a)-(b)-(c)-(d)-(e), we show the MSE (obtained by different techniques)  as function of $N$, $T$, $D$ and $\sigma$, respectively. In Figure \ref{FigEx1_1}(d), we only consider I-MTM with $N\in\{1,5,100,1000\}$ in order to show the effect of using different tries $N$ in different dimensions $D$. Note that I-MTM with $N=1$ coincides with I-MH.  

Let us denote as $\phi(\tau)$ the auto-correlation function of the states of the generated chain. Figures \ref{FigEx1_2}(a)-(b)-(c) depicts the normalized auto-correlation function ${\bar \phi}(\tau)=\frac{\phi(\tau)}{\phi(0)}$ (recall that $\phi(0)\geq \phi(\tau)$ for $\tau \geq 0$) at different lags $\tau=1,2,3$, respectively. Furthermore, given the definition of the Effective Sample Size (ESS) \cite[Chapter 4]{Gamerman97bo}, 
\begin{equation}
ESS=\frac{T}{1+2\sum_{\tau=1}^\infty {\bar \phi}(\tau)},
\end{equation}
 in Figure \ref{FigEx1_2}(d), we show of the ratio $\frac{ESS}{T}$ (approximated; cutting off the series in the denominator at lag $\tau=10$), as function of $N$.\footnote{Since the MCMC algorithms yield {\it positive} correlated sequences of states, we have $ESS< T$ in general.}
 Finally, in Figures \ref{FigEx1_3}(a)-(b), we provide the Acceptance Rate (AR) of a new state (i.e., the expected number of accepted jumps to a novel state), as function of $N$ and $D$, respectively. 

\subsubsection{Comment on the results}

\begin{figure*}[h!]
\begin{center}
\centerline{
\subfigure[MSE versus $N$ ($T=2000$, $\sigma^2=2$, $D=1$)]{\includegraphics[width=0.38\textwidth]{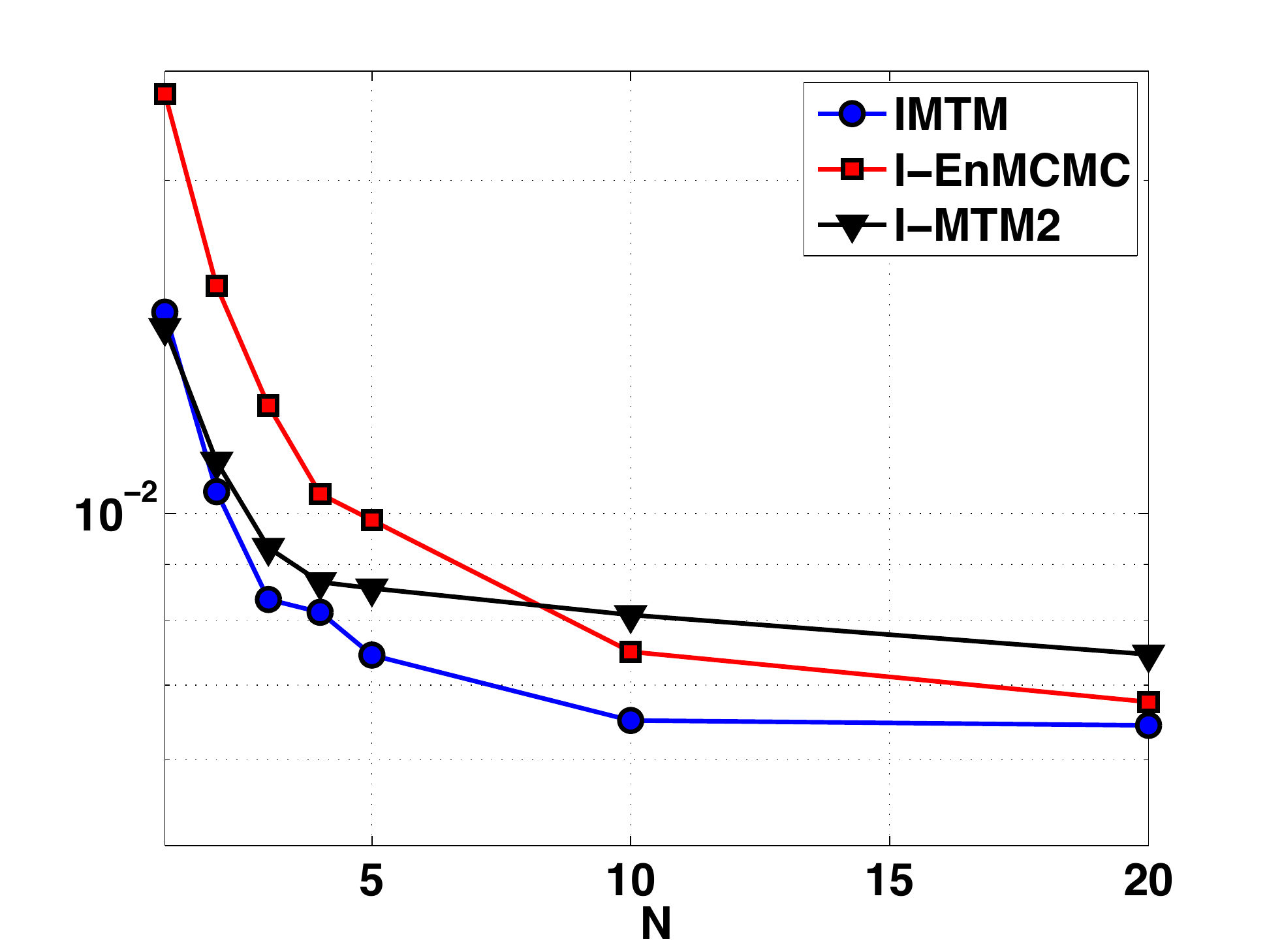}}
\subfigure[MSE versus $N$ ($T=500$, $\sigma^2=2$, $D=10$)]{\includegraphics[width=0.38\textwidth]{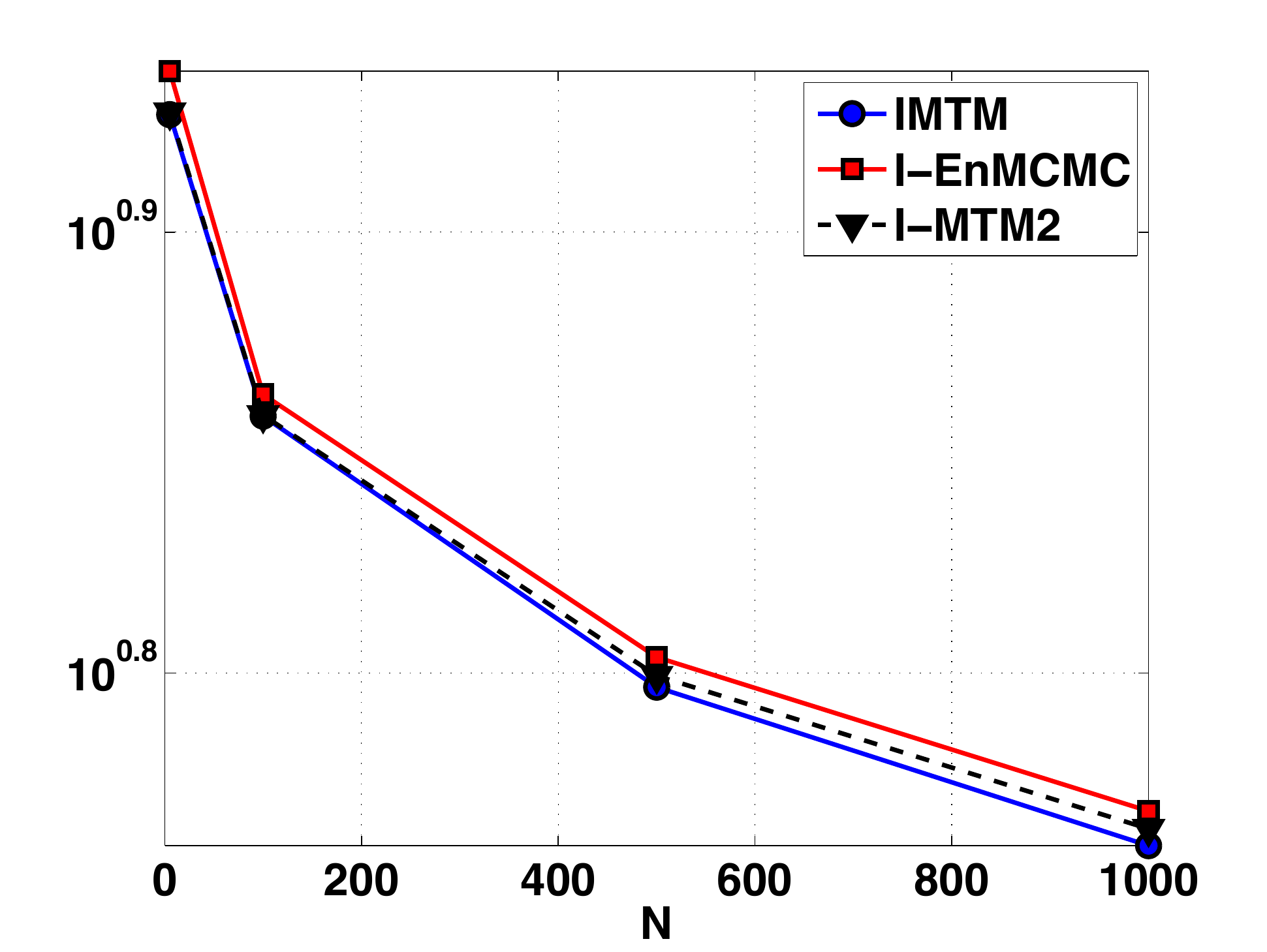}}
\subfigure[MSE versus $T$ ($N=5$, $\sigma^2=2$, $D=1$)]{\includegraphics[width=0.38\textwidth]{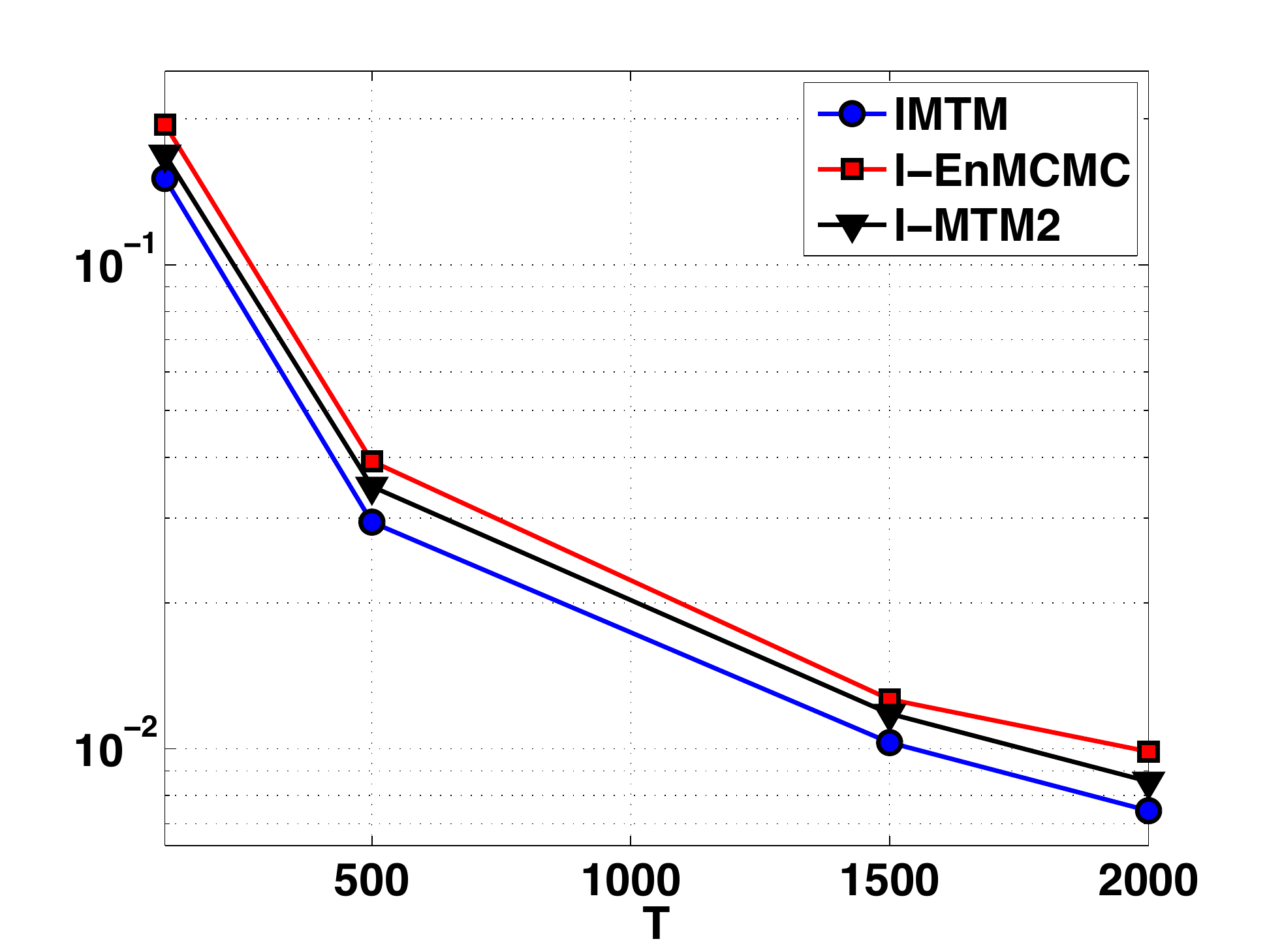}} 
}
\centerline{  
\subfigure[MSE versus $D$ ($T=500$, $\sigma^2=2$)]{\includegraphics[width=0.5\textwidth]{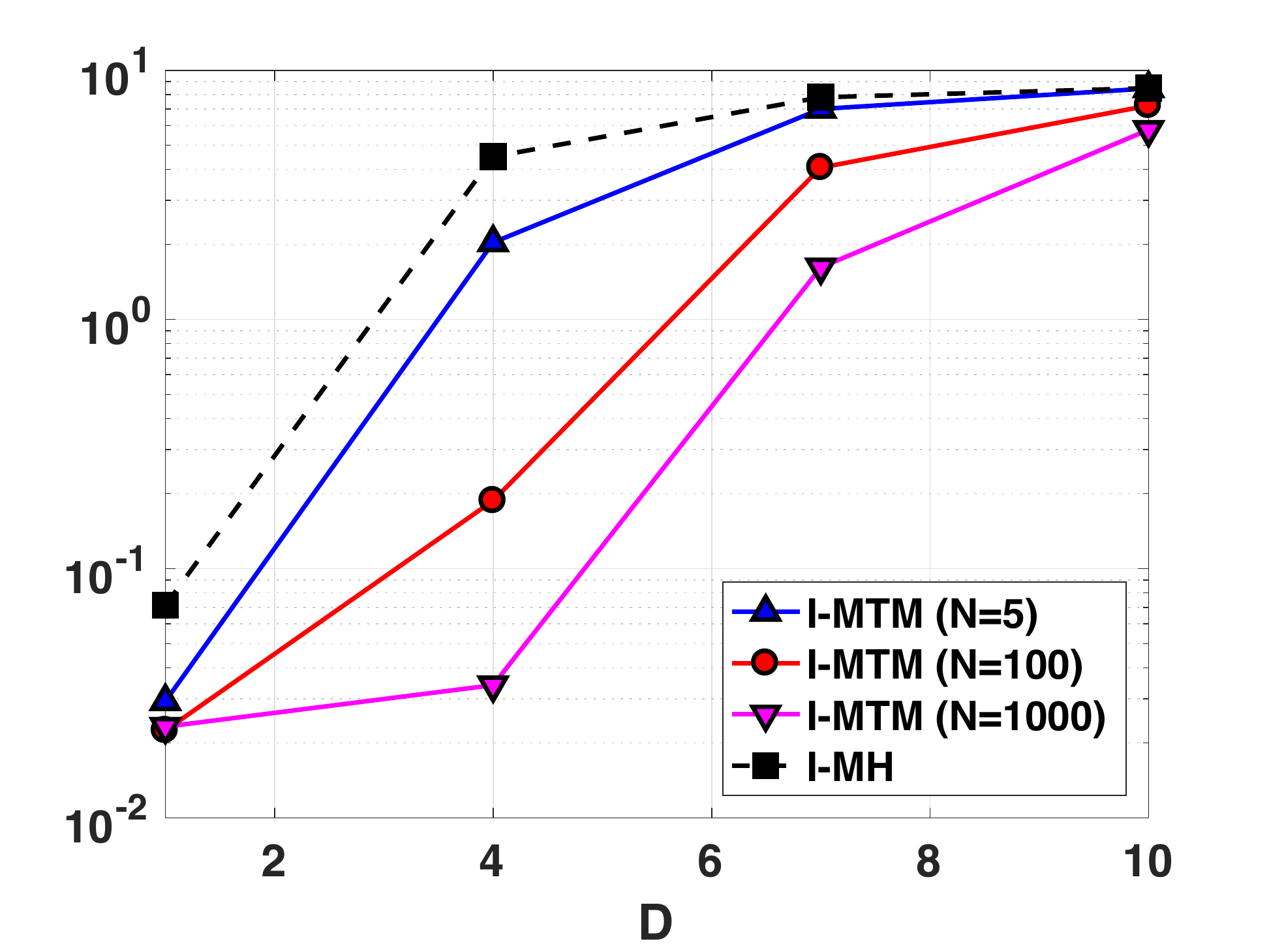}}
\subfigure[MSE versus $\sigma$ ($N=5$, $T=100$, $D=1$)]{\includegraphics[width=0.5\textwidth]{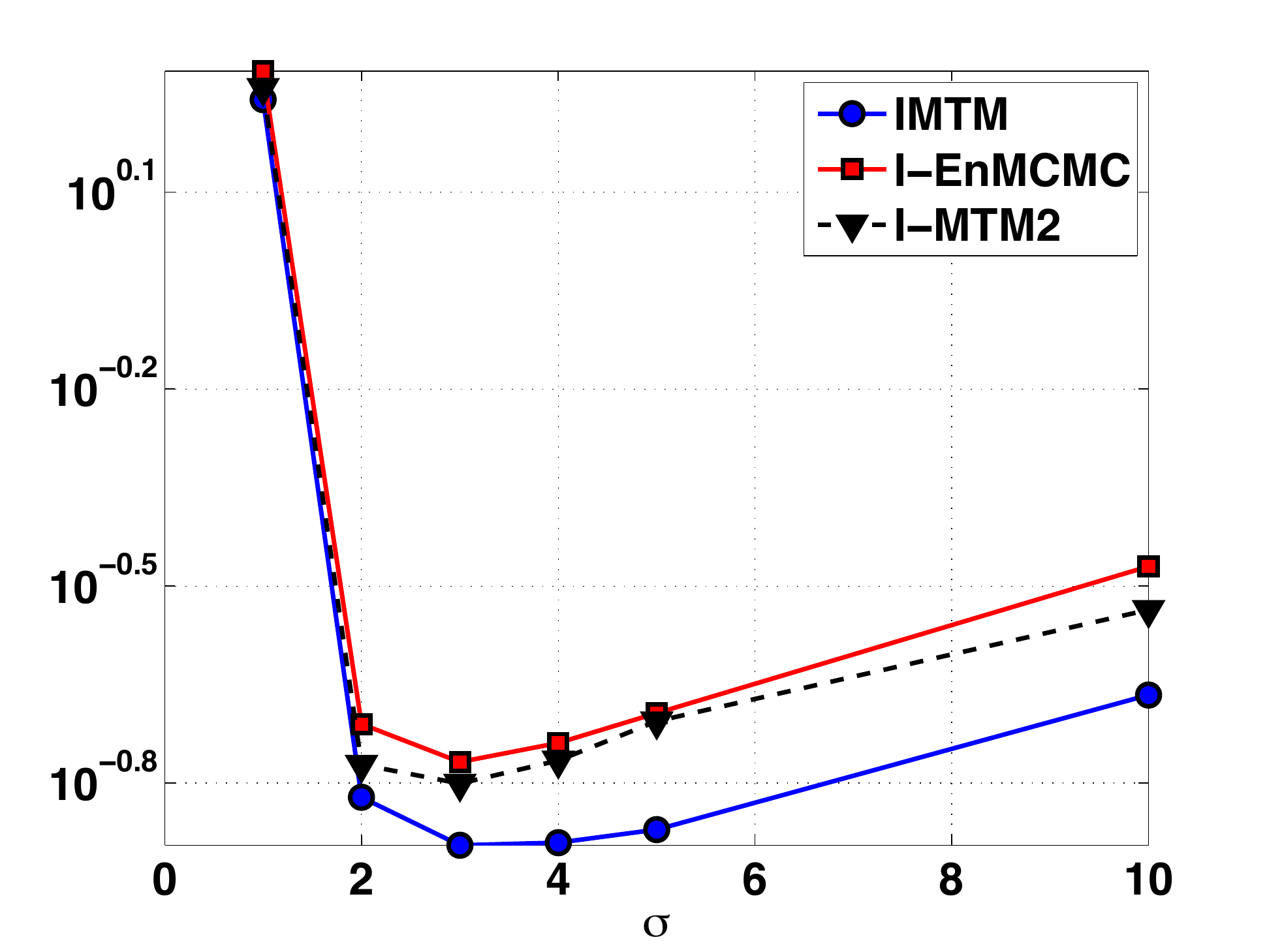}}
}
\caption{MSE as functions of different parameters (log-scale in the vertical axis). {\bf (a)}-{\bf (b)} MSE versus $N$ with $T\in\{2000,500\}$, $\sigma^2=2$, $D\in\{1,10\}$. {\bf (c)} MSE versus $T$ with $N=5$, $\sigma^2=2$, $D=1$. {\bf (d)} MSE versus $D$ with $N\in\{1,5,10^2,10^3\}$, $T=2000$, $\sigma^2=2$ testing only for I-MTM (for $N=1$, it coincides with I-MH). {\bf (e)} MSE versus $\sigma$ with $N=5$, $T=100$, $D=1$. }
\label{FigEx1_1}
\end{center}
\end{figure*}

\begin{figure*}[h!]
\begin{center}
\centerline{
\subfigure[${\bar \phi}(1)$ versus $N$ ($T=2000$, $\sigma=2$, $D=1$).]{\includegraphics[width=0.5\textwidth]{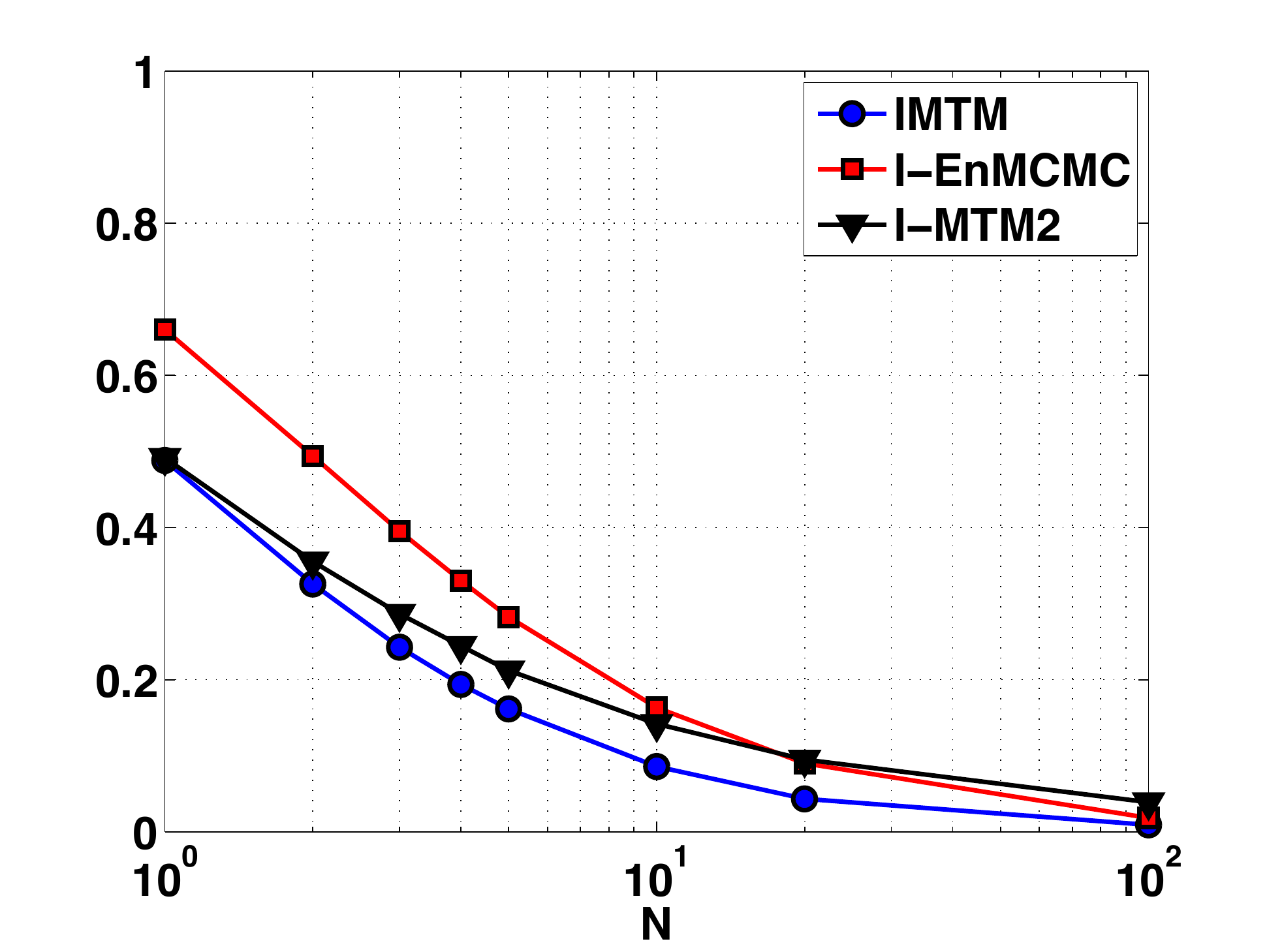}}
\subfigure[${\bar \phi}(2)$ versus $N$ ($T=2000$, $\sigma=2$, $D=1$).]{\includegraphics[width=0.5\textwidth]{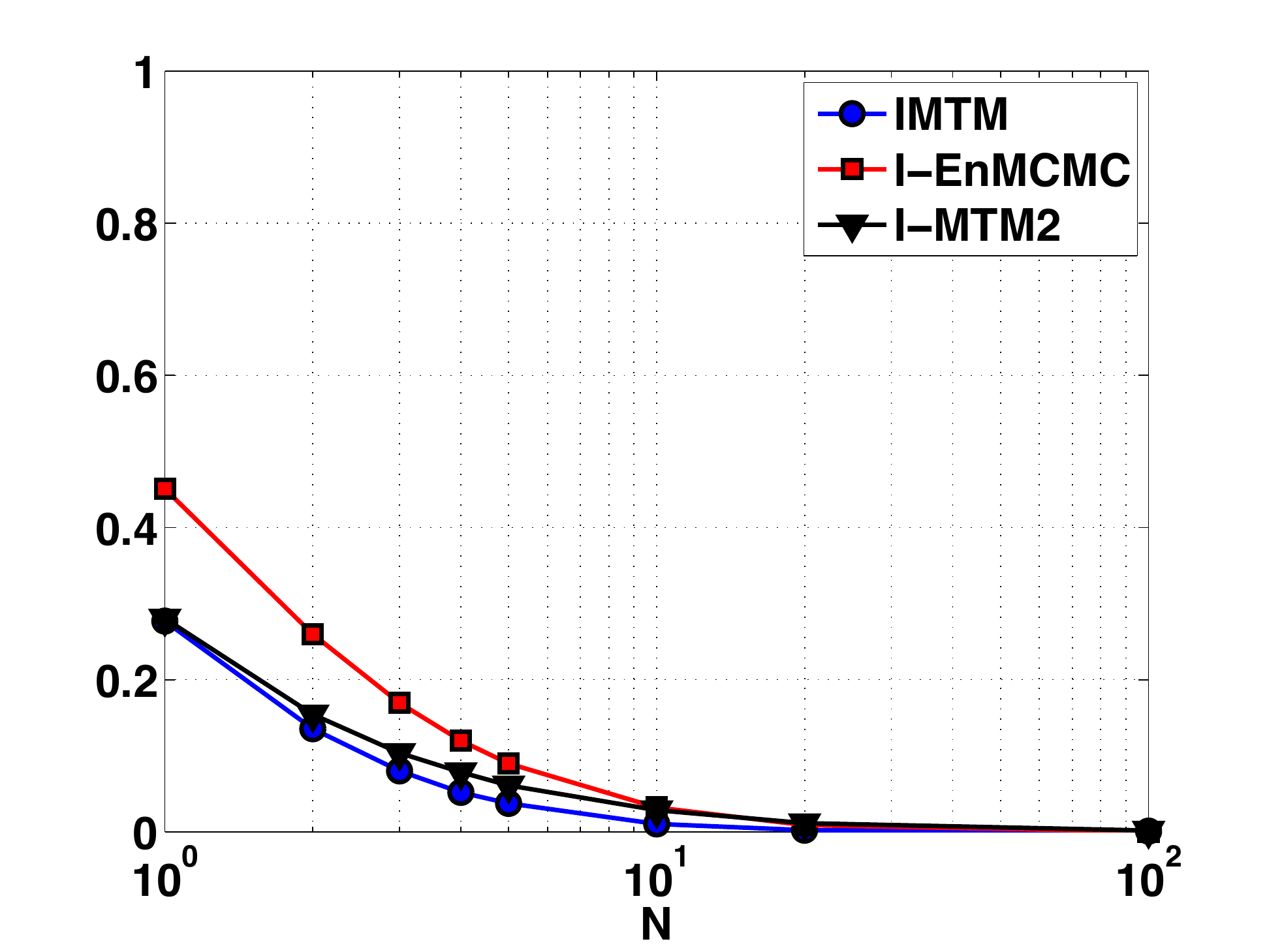}}
}
\centerline{
\subfigure[${\bar \phi}(3)$ versus $N$ ($T=2000$, $\sigma=2$, $D=1$).]{\includegraphics[width=0.5\textwidth]{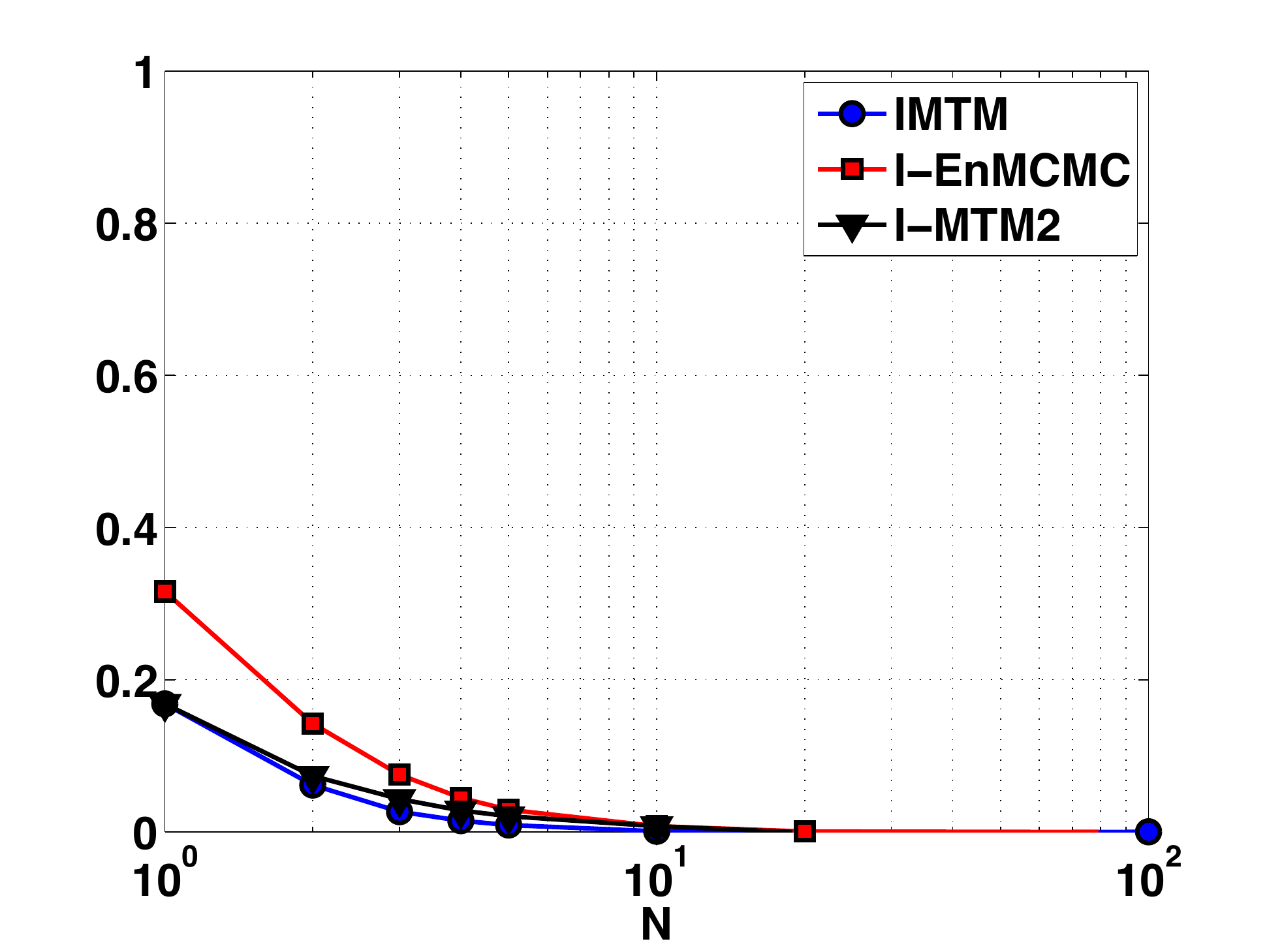}}
\subfigure[$\frac{ESS}{T}$ versus $N$ ($T=2000$, $\sigma=2$, $D=1$). ]{\includegraphics[width=0.5\textwidth]{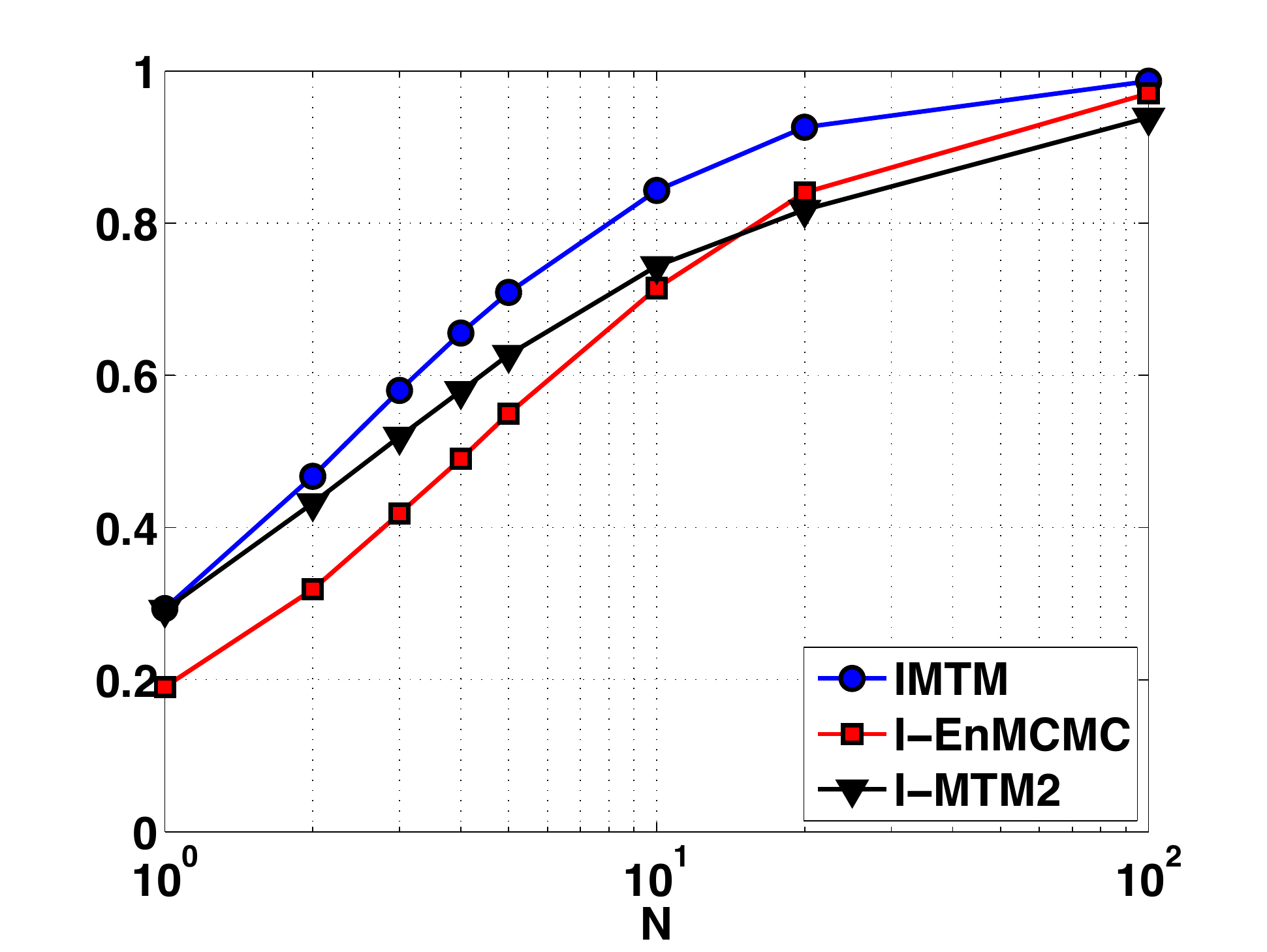}}
}
\caption{ (log-scale in the horizontal axis) {(\bf a)}-{(\bf b)}-{(\bf c)}  Auto-correlation function ${\bar \phi}(\tau)$ of the states of generated chain as function of $N$, at different lags $\tau=1,2,3$, respectively ($T=2000$, $\sigma=2$, $D=1$). {(\bf d)} The Effective Sample Size rate $\frac{ESS}{T}$ as function of $N$, keeping fixed $T=2000$, $\sigma=2$, $D=1$. }
\label{FigEx1_2}
\end{center}
\end{figure*}

\begin{figure*}[h!]
\begin{center}
\centerline{
\subfigure[AR versus $N$ ($T=2000$, $\sigma^2=2$, $D=1$).]{\includegraphics[width=0.5\textwidth]{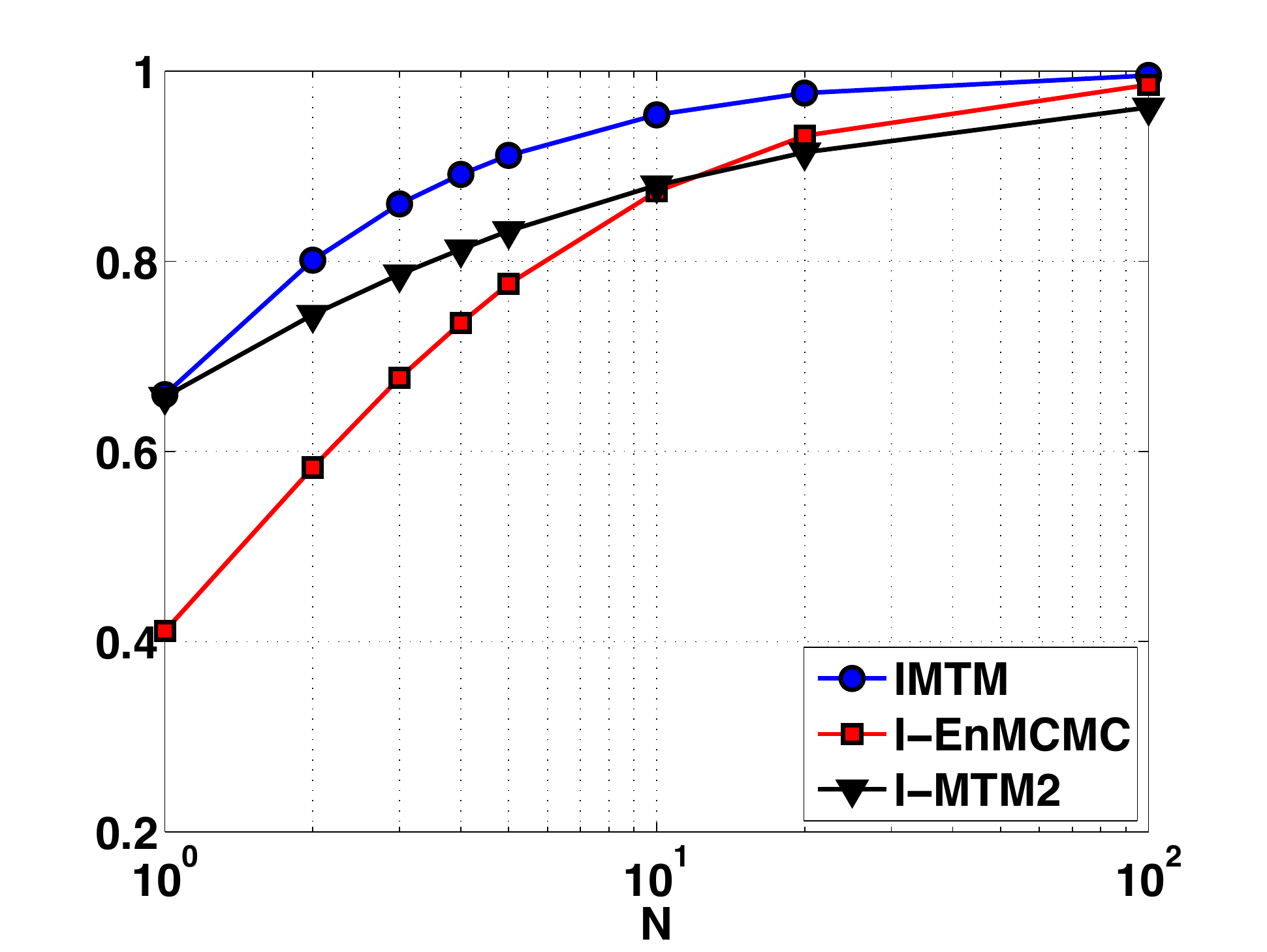}}
\subfigure[AR versus $D$ ($N=100$, $T=2000$, $\sigma^2=2$).]{\includegraphics[width=0.5\textwidth]{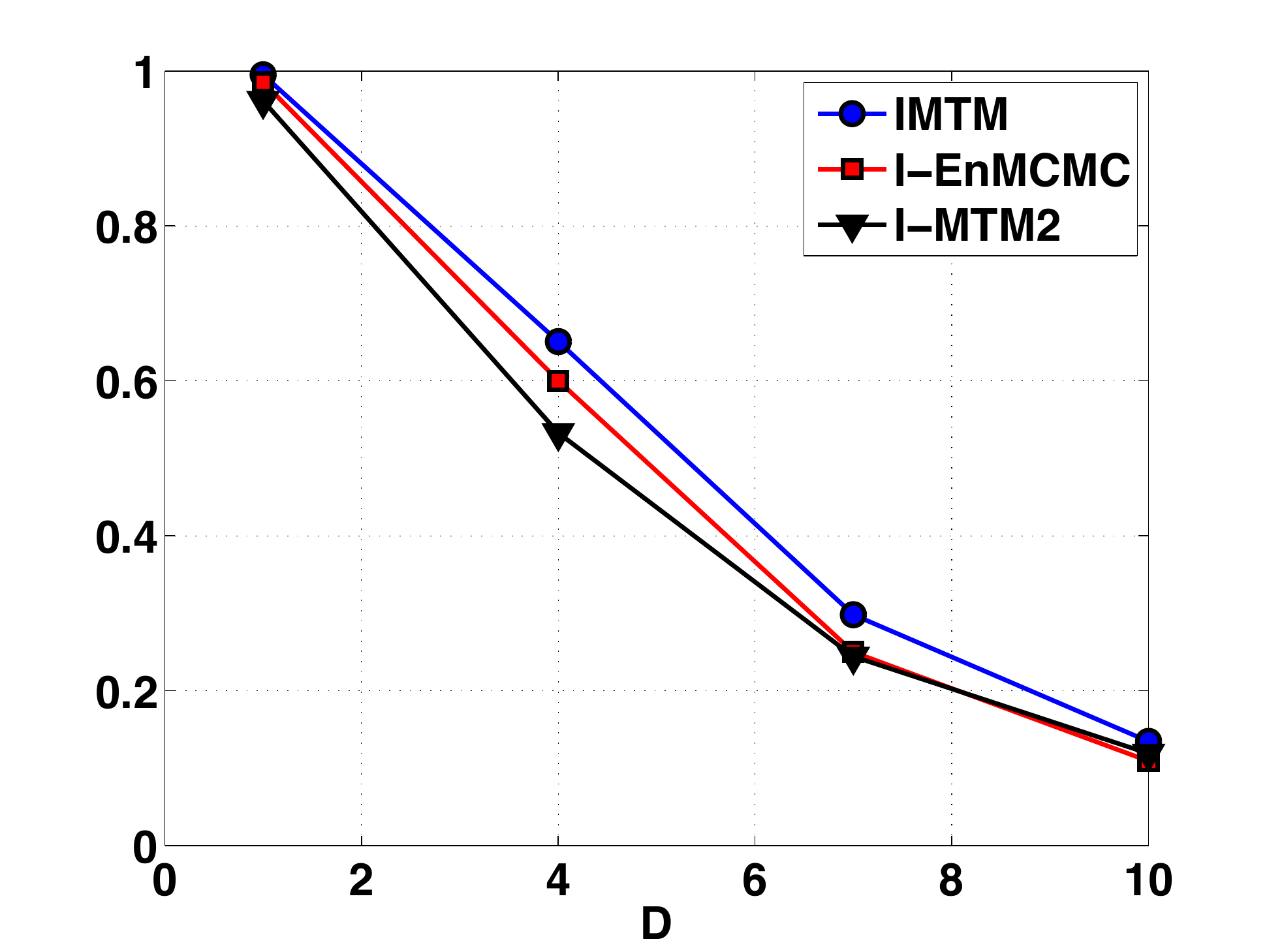}}
}
\caption{{\bf (a)} Acceptance Rate (AR) as function of $N$, with $T=2000$, $\sigma^2=2$, $D=1$ (log-scale in the horizontal axis). {\bf (b)} AR as function of $N$, with $N=100$, $T=2000$, $\sigma^2=2$.}
\label{FigEx1_3}
\end{center}
\end{figure*}

Figures \ref{FigEx1_1}(a)-(b), \ref{FigEx1_2} and \ref{FigEx1_3}(a), clearly show that the performance improves as $N$ grows, for all the algorithms.  
The MSE values and the correlation decrease, and  the ESS and the AR grow.
I-MTM seems to provide the best performance. Recall that for $N=1$,  I-EnMCMC becomes an I-MH with Baker's acceptance function and I-MTM becomes the I-MH in Table \ref{I_MH_table} \cite{MHWiley17}. For $N=1$, the results confirm the Peskun's ordering about the acceptance function for a MH method \cite{Peskun73,MHWiley17}. Observing the results, the Peskun's ordering appears valid also for the multiple try case, $N>1$. 
I-MTM2 seems to have worse performance than I-MTM for all $N$. With respect to  I-EnMCMC, I-MTM2 performs better for smaller $N$. The difference among the MSE values obtained by the samplers becomes smaller as $N$ grows, as shown in Figure \ref{FigEx1_1}(a)-(b) (note that in the first one $D=1$, in the other $D=10$, and the range of $N$ is different).
 The comparison among I-MTM, I-MTM2 and I-EnMCMC  seems not to be affected by changing $T$ and $\sigma$, as depicted in Figures \ref{FigEx1_1}(c)-(e). Namely, the MSE values change but the ordering of the methods (e.g., best and worst) seems to depend mainly on $N$. 
Obviously, for greater $D$, more tries are required in order to obtain good performance (see Figures \ref{FigEx1_1}(d) and\ref{FigEx1_3}(b), for instance). Note that for $N\rightarrow \infty$,   I-MTM, I-MTM2 and I-EnMCMC perform similarly to an exact sampler drawing $T$ independent samples from ${\bar \pi}({\bm \theta})$: the correlation $\phi(\tau)$ among the samples approaches zero (for all $\tau$), ESS approaches $T$ and AR approaches $1$.

\subsection{Numerical experiment comparing particle schemes}

\begin{figure*}[h!]
\begin{center}
\centerline{
 \subfigure[MSE as function of $T$.]{\includegraphics[width=0.5\textwidth]{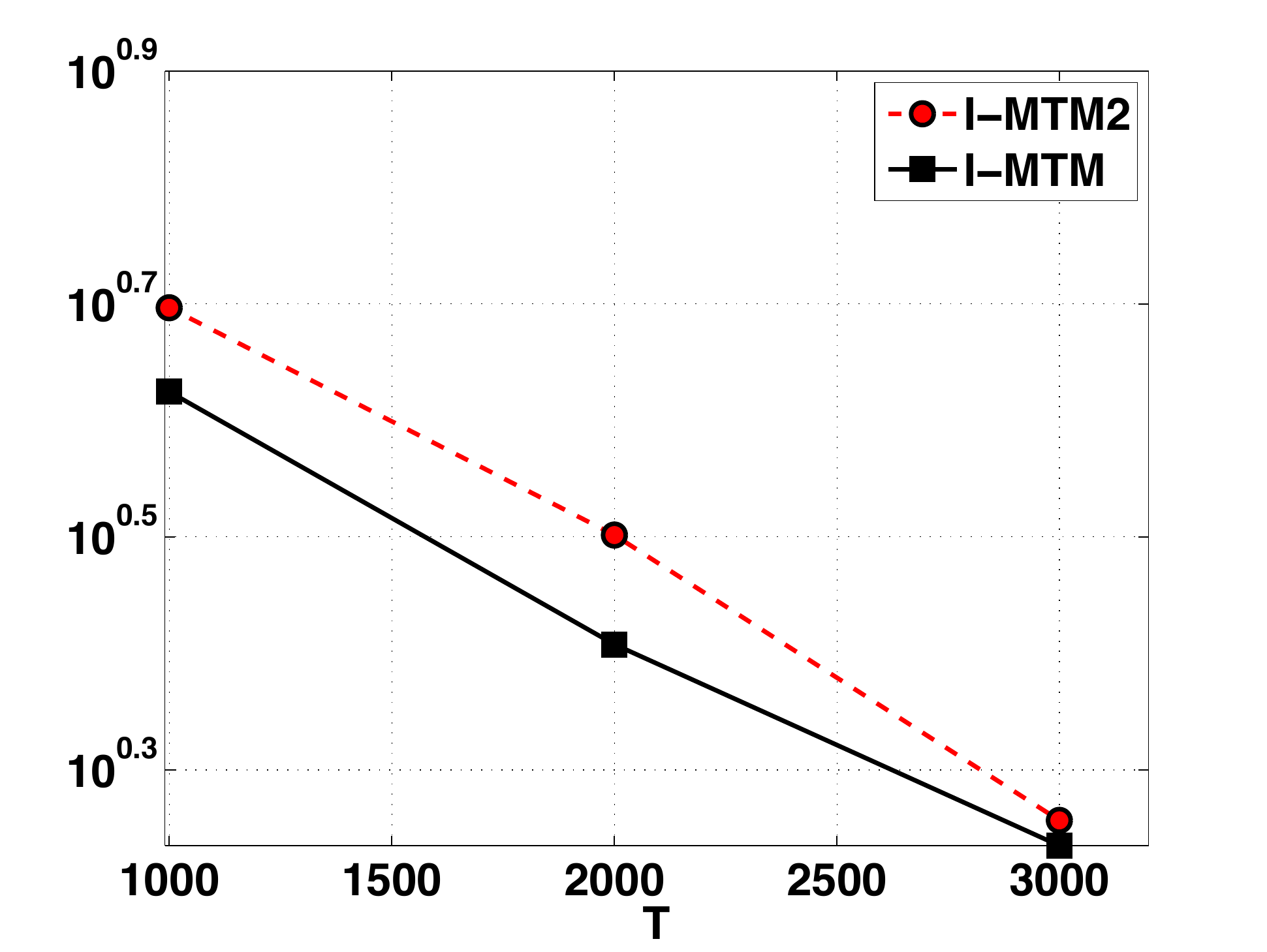}}
 \subfigure[MSE as function of $T$.]{\includegraphics[width=0.5\textwidth]{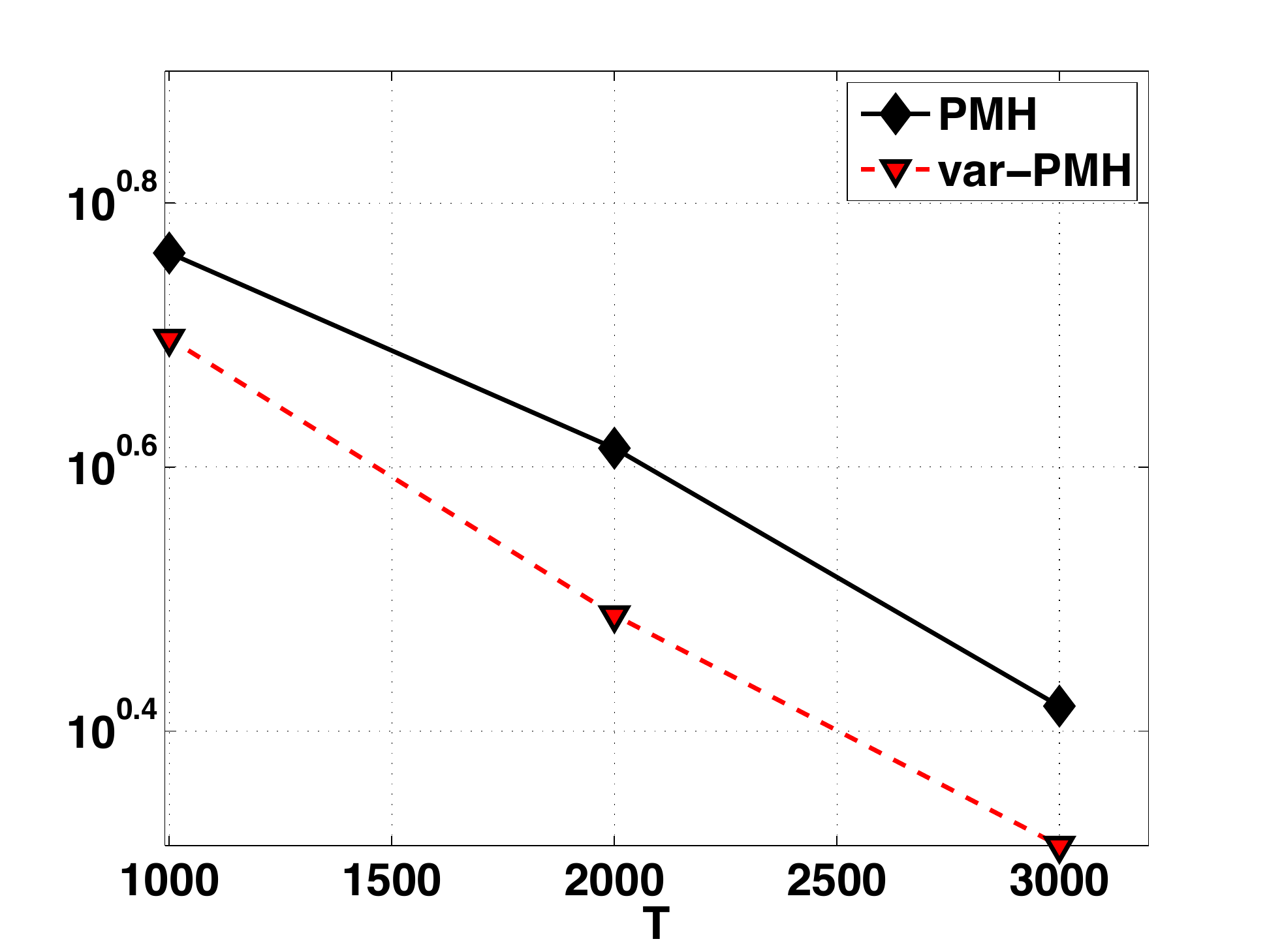}}
 }
 \centerline{
 \subfigure[MSE as function of $N$.]{\includegraphics[width=0.5\textwidth]{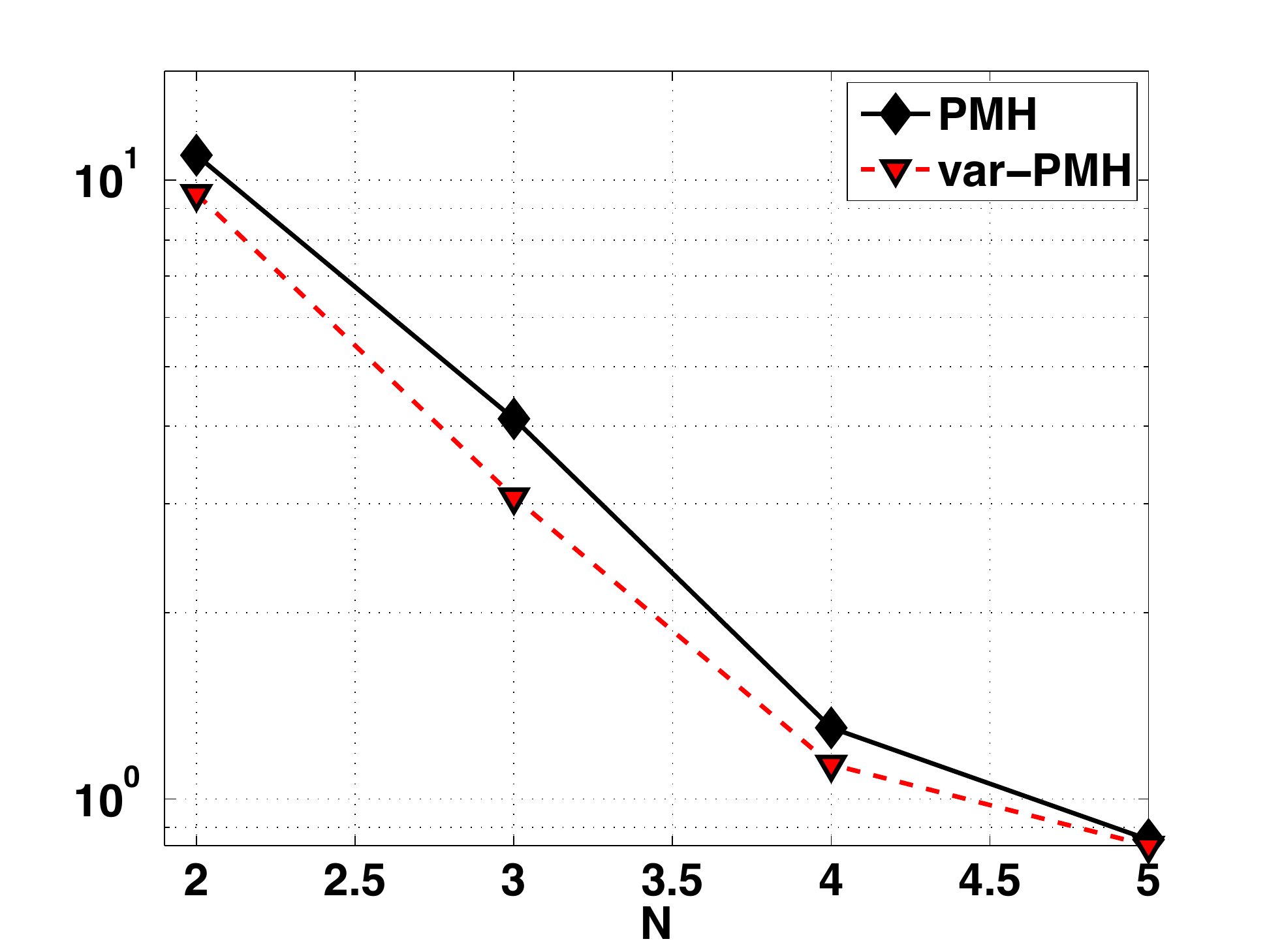}}
 \subfigure[Different states of the var-PMH chain.]{\includegraphics[width=0.5\textwidth]{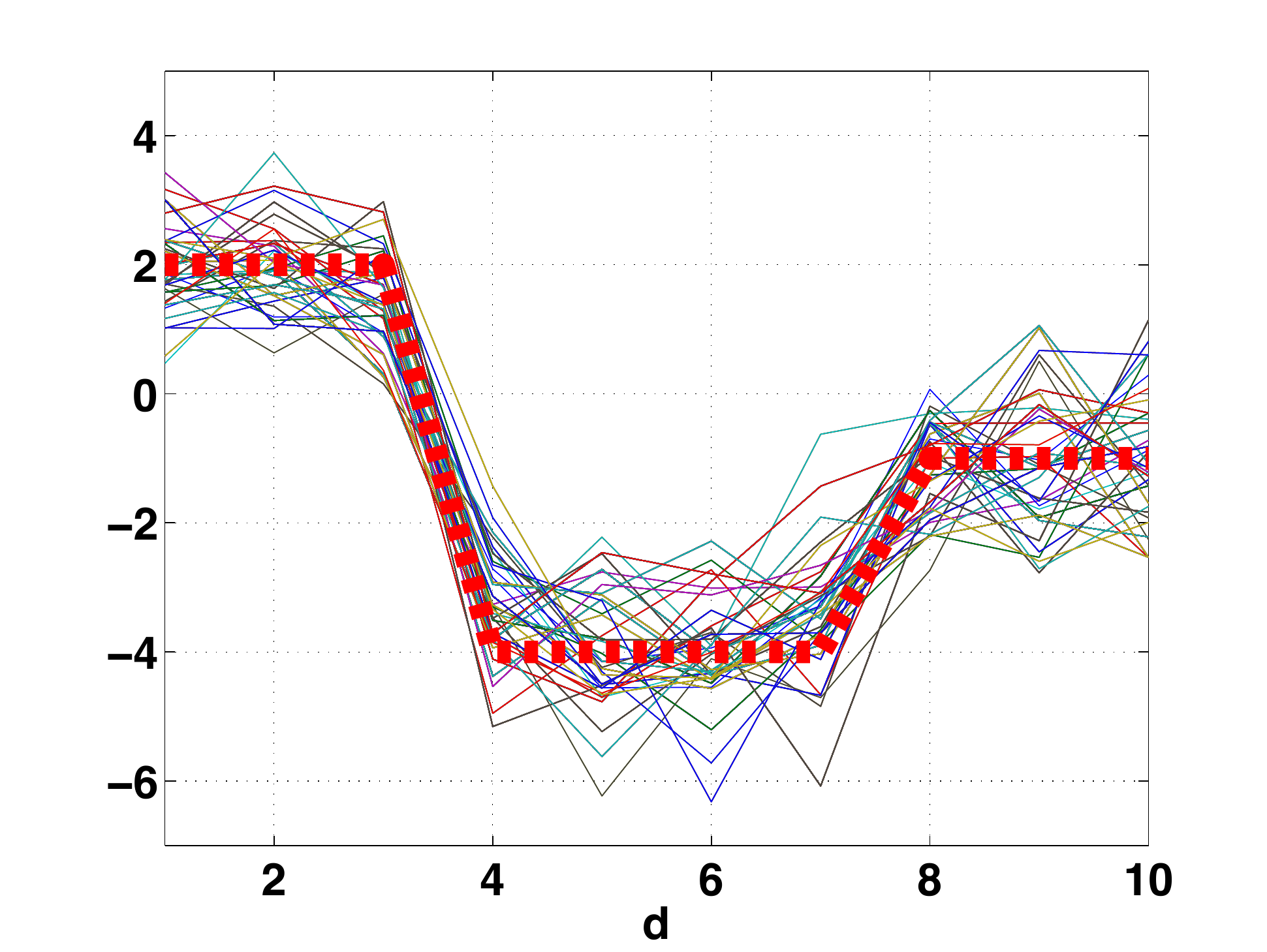}}
 }
\caption{ {\bf (a)}-{\bf (b)} MSE (semi-log scale) as function of  $T$  ($N=3$). {\bf (c)} MSE (semi-log scale) as function of $N$ ($T=2000$). {\bf (d)} Different states ${\bm \theta}_t={\theta}_{1:10,t}$ at different iterations $t$, obtained with var-PMH ($N=1000$ and $T=1000$). The values ${\bm \mu}=\mu_{1:10}$ are shown in dashed line ($\mu_{1:3}=2$,  $\mu_{4:7}=4$ and $\mu_{8:10}=-1$).}
\label{FigEX2_pmh}
\end{center}
\end{figure*}

In this section,  in order to clarify the differences between batch and particle schemes, we consider again a multidimensional Gaussian target density, that can be express as
\begin{equation}
\label{EQTAR}
\bar{\pi}({\bm \theta})=\bar{\pi}(\theta_1,\ldots, \theta_D)=\prod_{d=1}^D \mathcal{N}({\theta}_d|\mu_d,\sigma^2),
\end{equation}
with ${\bm \theta}=\theta_{1:D}\in \mathbb{R}^D$, $D=10$, with $\mu_{1:3}=2$,  $\mu_{4:7}=4$, $\mu_{8:10}=-1$ (shown dashed line in Figure \ref{FigEX2_pmh}(d)), and $\sigma=\frac{1}{2}$.
The target pdf $\bar{\pi}({\bm \theta})=\mathcal{N}({\bm \theta}|{\bm \mu},\sigma^2{\bf I}_D)$ (with ${\bm \mu}=\mu_{1:10}$) is formed by independent Gaussian pieces, and can factored as in Eq. \eqref{EQTAR}.  The factorization allows the use of a sequential proposal procedure and hence particle schemes can be employed, even if there is not correlation between ${\theta}_d$ and ${\theta}_{d-1}$.
\footnote{For the particle schemes, we consider ${\bm \theta}={\bf x}$ as a dynamic parameter (see Section \ref{PS_section}).}

We apply I-MTM, I-MTM2, PMH, and a variant of PMH, denote as var-PMH which uses the corresponding acceptance probability of I-MTM in Eq. \eqref{eq:alphaI_MTM} instead of the acceptance function of the classical PMH in Eq. \eqref{A1pmh}. The goal is to estimate the vector ${\bm \mu}$. We compute the MSE in estimating the vector ${\bm \mu}=\mu_{1:10}$, averaging the results over $500$ independent simulations. The components of ${\bm \mu}$ are shown in Figure \ref{FigEX2_pmh}(d) with a dashed line.

For all the techniques, we employ a sequential construction of the $N$ candidates  (using the chain rule, see below): in PMH and var-PMH the resampling is applied at each iteration whereas in I-MTM and I-MTM2 no resampling is applied. More specifically, the proposal density for all the methods is 
\begin{equation}
q({\bm \theta})=q_1(\theta_1)\prod_{d=2}^D q_d(\theta_d|\theta_{d-1}),
\end{equation}
where $q_1(\theta_1)=\mathcal{N}(\theta_1|-2, 4)$ and $q_d(\theta_d|\theta_{d-1})=\mathcal{N}(\theta_d|\theta_{d-1}, \sigma_p^2)$, but PMH and var-PMH employ resampling steps so that the generated tries are correlated (whereas in  I-MTM and I-MTM2 the generated candidates are independent).

 We test all the techniques considering different value of number of tries $N$ and number of iterations of the chain $T$. 
 Figures \ref{FigEX2_pmh}(a)-(b) show the MSE    
as function of number of iterations $T$, keeping fixed the number of tries $N=3$. Figure \ref{FigEX2_pmh}(a) reports the results of the MTM schemes whereas Figure \ref{FigEX2_pmh}(b) reports the results of the PMH schemes. Figure \ref{FigEX2_pmh}(c) depicts the MSE as function of $N$ (with $T=2000$), for the PMH methods. Note that the use of only $N=3$ particles and the application of the resampling at each iteration is clearly a disadvantage for the PMH schemes. If the resampling is applied very often (as in this case), a greater number of $N$ is advisable (such as $N=100$ or $N=1000$). Hence, the results confirm that applying a resampling step at each iteration is not optimal and that a smaller rate of resampling steps could improve the performance \cite{Djuric03,Doucet08tut}. The results also confirm that the use of an acceptance probability of type in Eq. \eqref{eq:alphaI_MTM} provides smaller MSE, i.e., I-MTM and var-PMH perform better than I-MTM2 and PMH, respectively. This is more evident for small number of candidates $N$. 
 When $N$ grows,  the performance of  PMH  and var-PMH methods becomes similar, since the acceptance probability approaches $1$, in both cases.  Figure \ref{FigEX2_pmh}(d) depicts 35 different states ${\bm \theta}_t=\theta_{1:10,t}$ at different iteration indices $t$, obtained with var-PMH ($N=1000$ and $T=1000$) and the values $\mu_{1:10}$ are given in dashed line.

\subsection{Hyperparameter tuning for Gaussian Process (GP) regression models}
\label{GPexample}

Let us assume observed data pairs $\{y_j,{\bf z}_j\}_{j=1}^{P}$, with $y_j\in \mathbb{R}$ and ${\bf z}_j\in \mathbb{R}^{L}$. We also denote the corresponding $P\times 1$ output vector as ${\bf  y}=[y_1,\ldots,y_P]^{\top}$ and the $L\times P$ input matrix as ${\bf  Z}=[{\bf z}_1,\ldots,{\bf z}_P]$. We address the regression problem of inferring the unknown function $f$ which links the variable $y$ and ${\bf z}$. Thus, the assumed model is $y=f({\bf z})+e$,   
where $e\sim N(e;0,\sigma^2)$, and that $f({\bf z})$ is a realization of a Gaussian Process (GP)~\cite{Bishop,rasmussen2006gaussian}. Hence $f({\bf z}) \sim \mathcal{GP}(\mu({\bf z}),\kappa({\bf z},{\bf r}))$ where $\mu({\bf z})=0$, ${\bf z},{\bf r} \in \mathbb{R}^{L}$, and we consider the kernel function 
\begin{equation}
\label{EqKernel}
\kappa({\bf z},{\bf r})=\exp\left(-\sum_{\ell=1}^{L}\frac{(z_\ell-r_\ell)^2}{2\delta^2}\right).
\end{equation}
Given these assumptions, the vector ${\bf  f}=[f({\bf z}_1),\ldots, f({\bf z}_P)]^\top$ is distributed as 
$p({\bf  f}|{\bf  Z},\delta, \kappa)=\mathcal{N}({\bf  f};{\bf  0},{\bf  K})$,
where ${\bf  0}$ is a $P\times 1$ null vector, and ${\bf  K}_{ij}:=\kappa({\bf z}_i,{\bf z}_j)$, for all $i,j=1,\ldots,P$, is a $P\times P$ matrix.
The vector containing all the hyperparameters of the model is
${\bm \theta}=[\delta, \sigma]$,
i.e., all the parameters of the kernel function in Eq.~\eqref{EqKernel} and standard deviation $\sigma$ of the observation noise. In this experiment, we focus on the marginal posterior density of the hyperparameters,
${\bar \pi}({\bm \theta}|{\bf  y}, {\bf  Z}, \kappa)\propto \pi({\bm \theta}|{\bf  y}, {\bf  Z}, \kappa)=p({\bf  y}|{\bm \theta}, {\bf  Z}, \kappa) p({\bm \theta})$,
which can be evaluated analytically, but we cannot compute integrals involving it \cite{rasmussen2006gaussian}. Considering a uniform prior within $[0,20]^2$, $p(\x)$ and since $p({\bf  y}|{\bm \theta}, {\bf  Z}, \kappa)=\mathcal{N}({\bf  y};{\bf  0},{\bf K}+\sigma^2 {\bf  I})$, we have 
\begin{gather}
\begin{split}
\nonumber
\log \left[\pi({\bm \theta}|{\bf  y}, {\bf  Z}, \kappa)\right] = -\frac{1}{2} {\bf  y}^{\top} ({\bf  K}+\sigma^2 {\bf  I})^{-1} {\bf  y} -\frac{1}{2} \log\left[\mbox{det} \left({\bf  K} +\sigma^2 {\bf  I}\right)\right]{\color{red}+C},  
\end{split}
\end{gather}
 where $C>0$, and clearly ${\bf  K}$ depends on $\delta$~\cite{rasmussen2006gaussian}. The moments of this marginal posterior cannot be computed analytically. Then, in order to compute the Minimum Mean Square Error (MMSE) estimator $\widehat{{\bm \theta}}=[\widehat{\delta},\widehat{\sigma}]$, i.e., the expected value $E[{\bm \Theta}]$ with 
${\bm \Theta} \sim {\bar \pi}({\bm \theta}|{\bf  y}, {\bf  Z}, \kappa)$, we approximate $E[{\bm \Theta}]$ via Monte Carlo quadrature. More specifically, we apply I-MTM2, GMS, a MH scheme with a longer chain and a static IS method. For all these methodologies, we consider the same number of target evaluations, denoted as $E$, in order to provide a fair comparison. 

\begin{figure*}[h!]
\begin{center}
\centerline{
\subfigure[]{\includegraphics[width=0.45\textwidth]{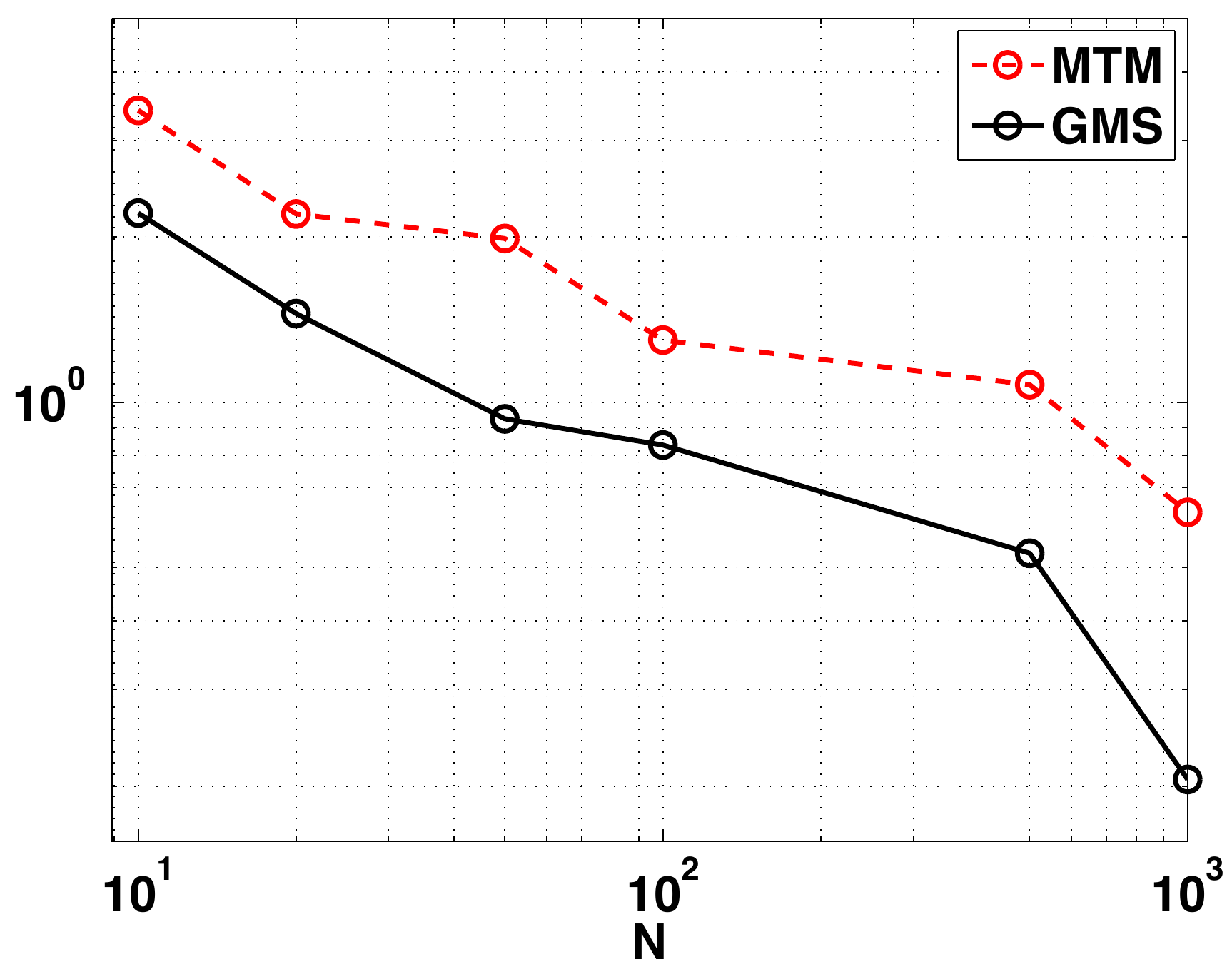}}
\subfigure[]{\includegraphics[width=0.45\textwidth]{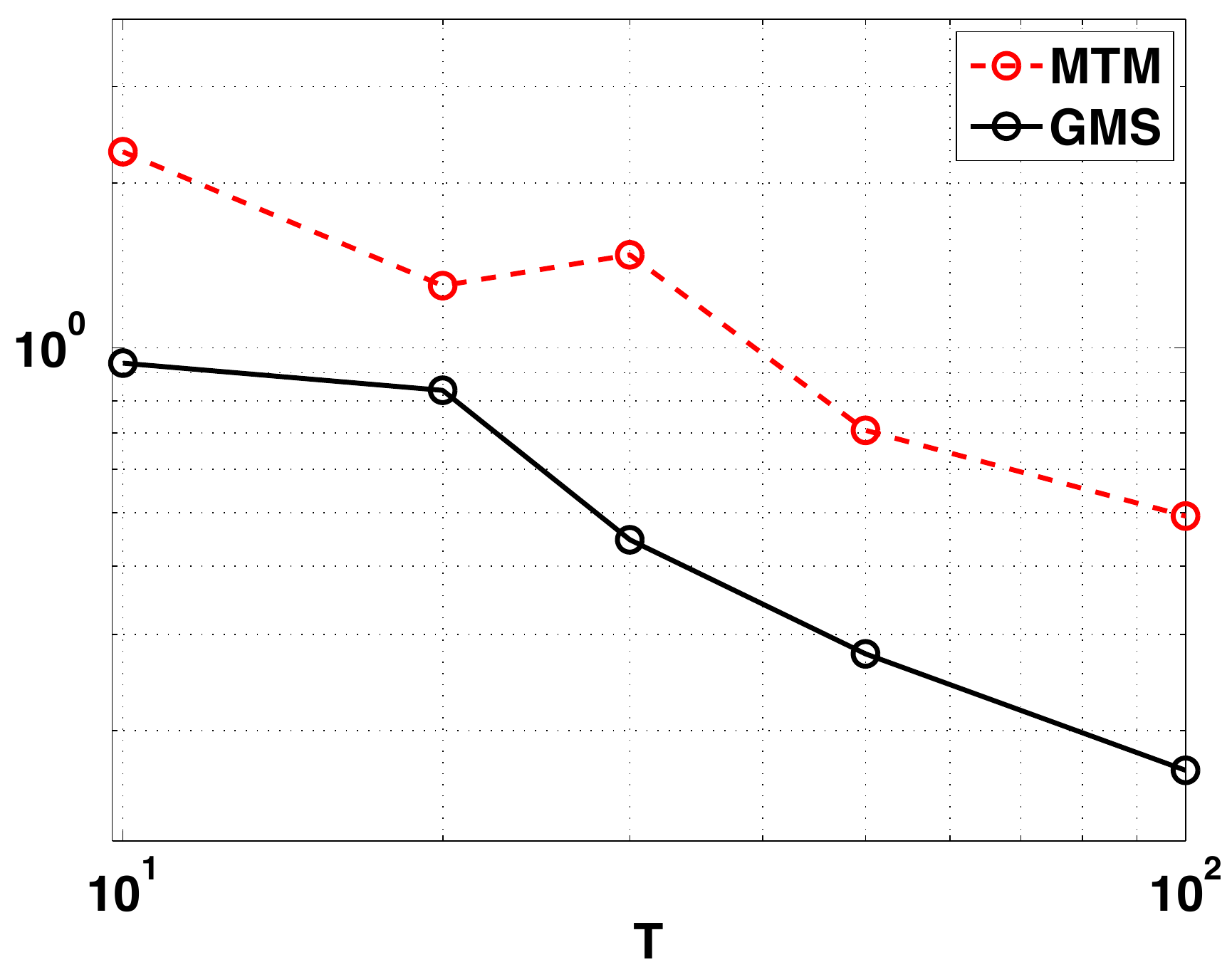}}
}
\vspace{-0.25cm}
\caption{MSE (loglog-scale; averaged over $10^3$ independent runs) obtained with the I-MTM2 and GMS algorithms (using the same proposal pdf and the same values of $N$ and $T$) {\bf (a)} as function of $N$ with $T=20$ and {\bf (b)} as function of $T$ with $N=100$. 
}
\label{FigSIMU}
\end{center}
\end{figure*}

\begin{figure*}[h!]
\begin{center}
\centerline{
\includegraphics[width=0.45\textwidth]{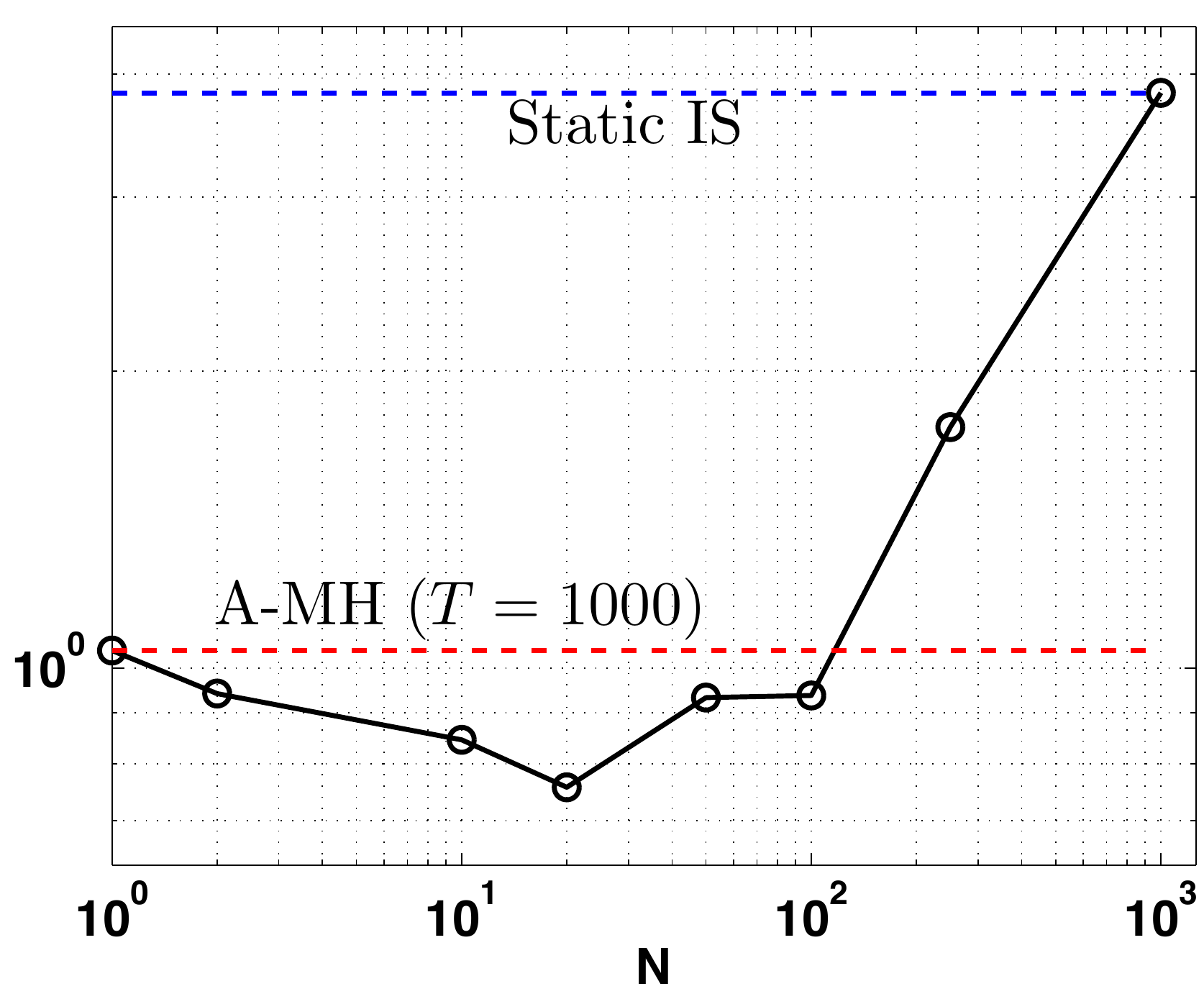}
}
\vspace{-0.25cm}
\caption{MSE (loglog-scale; averaged over $10^3$ independent runs) of GMS (circles) versus the number of candidates $N\in\{1,2,10,20,50,100,250,10^3\}$, but keeping fixed the total number of posterior evaluations $E=NT=1000$, so that $T\in\{1000,500,100,50,20,10,4,1\}$. The MSE values the extreme cases $N=1$, $T=1000$, and $N=1000$, $T=1$, are depicted with dashed lines. In first case, GMS coincides with an adaptive MH scheme (due the adaptation of the proposal, in this example) with a longer chain. The second one represents a static IS scheme (clearly, using the sample proposal than GMS). We can observe the benefit of the dynamic combination of IS estimators obtained by GMS. 
}
\label{FigSIMU2}
\end{center}
\end{figure*}

We generated $P=200$ pairs of data, $\{y_j,{\bf z}_j\}_{j=1}^{P}$, according to the GP model above setting $\delta^*=3$, $\sigma^*=10$, $L=1$, and drawing $z_j\sim\mathcal{U}([0,10])$. We keep fixed these data over the different runs. We computed the ground-truth ${\bm \theta}= [\delta= 3.5200,\sigma= 9.2811]$ using an exhaustive and costly grid approximation, in order to compare the different techniques. 
For I-MTM2, GMS, and MH schemes, we consider the same adaptive Gaussian proposal pdf $q_t({\bm \theta}|{\bm \mu}_t,\lambda^2 {\bf I})=\mathcal{N}({\bm \theta}|{\bm \mu}_t,\lambda^2 {\bf I})$, with $\lambda=5$ and ${\bm \mu}_t$ is adapted considering the arithmetic mean of the outputs after a training period, $t\geq 0.2 T$, in the same fashion of \cite{Luengo13,Haario01} (${\bm \mu}_0=[1,1]^{\top}$).  
First, we test both techniques fixing $T=20$ and varying the number of tries $N$. Then, we set $N=100$ and vary the number of iterations $T$. Figure \ref{FigSIMU} (log-log plot) shows the Mean Square Error (MSE) in the approximation of $\widehat{{\bm \theta}}$ averaged over $10^3$ independent runs. Observe that GMS always outperforms the corresponding I-MTM2 scheme. These results confirm the advantage of recycling the auxiliary samples drawn at each iteration during an I-MTM2 run.  In Figure \ref{FigSIMU2}, we show the MSE obtained by GMS keeping invariant the number of target evaluations $E=NT=10^3$ and varying $N\in\{1,2,10,20,50,100,250,10^3\}$. As a consequence, we have $T\in\{10^3,500,100,50,20,10,4,1\}$. Note that the case  $N=1$, $T=10^3$, corresponds to an adaptive MH (A-MH) method with a longer chain, whereas the case $N=10^3$, $T=1$, corresponds to a static IS scheme (both with the same posterior evaluations $E=NT=10^3$). We observe that the GMS always provides smaller MSE than the static IS approach. Moreover, GMS outperforms A-MH with the exception of two cases where $T\in\{1,4\}$.

\subsection{Localization of a target in a wireless sensor network}
\begin{figure*}[h!]
\begin{center}
\centerline{
\subfigure[]{\includegraphics[width=0.39\textwidth]{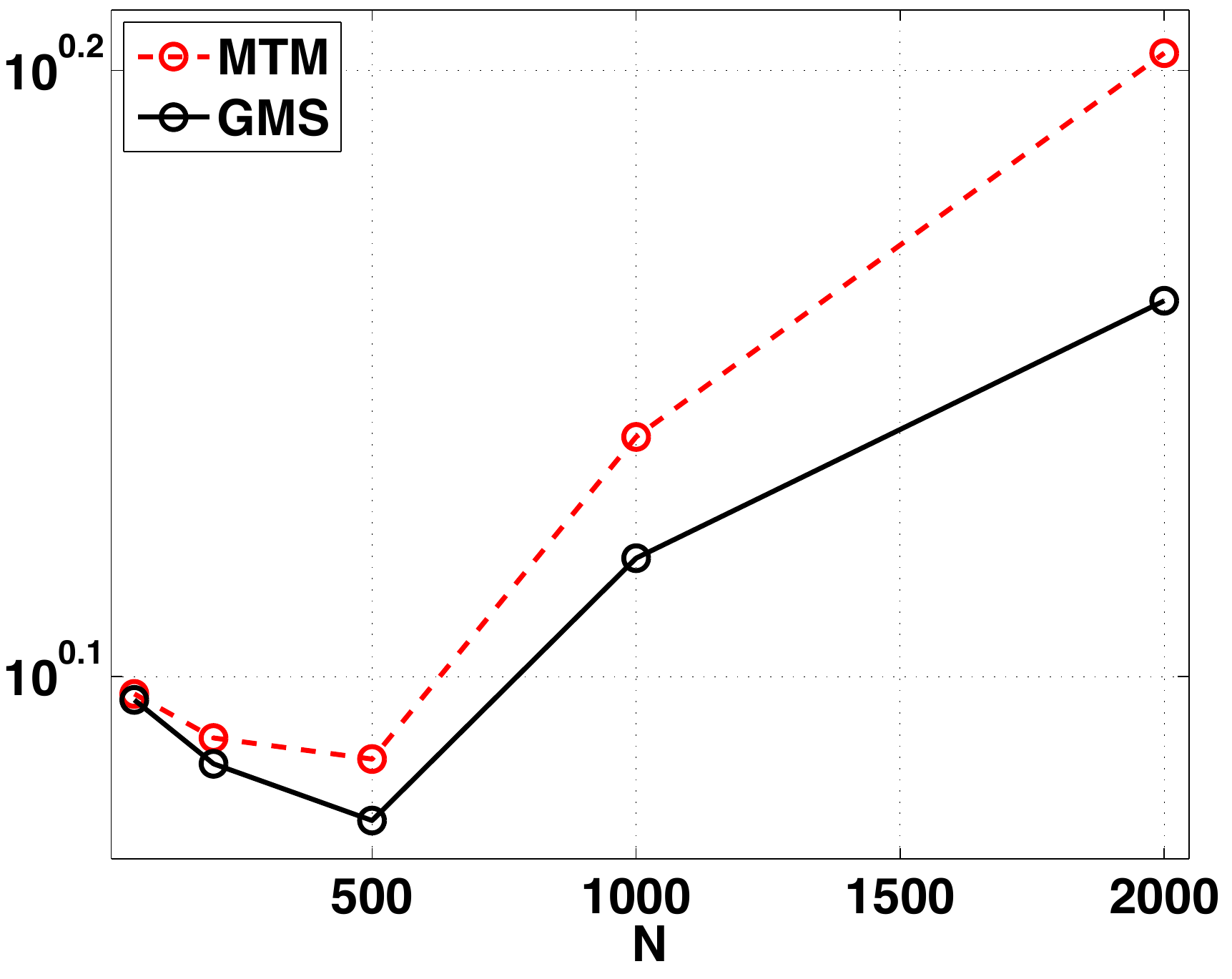}}
}
\vspace{-0.25cm}
\caption{MSE (log-scale) versus the number of candidates $N\in\{50,200, 500, 1000, 2000\}$ obtained by GMS and the corresponding I-MTM algorithm (without  using all the samples of the accepted sets, but only a resampled one), keeping fixed the total number of evaluations $E=NT=10^4$ of the posterior pdf, so that $T\in\{200,50,20,10,5\}$. 
}
\label{FigSIMU_Ex2}
\end{center}
\end{figure*}

We consider the problem of positioning a target in $\mathbb{R}^{2}$ using a range measurements in a wireless sensor network (WSN) \cite{Ali07,Ihler05}.  
We also assume that the measurements are contaminated by noise with different unknown power, one per each sensor. This situation is common in several practical scenarios. Indeed, even if the sensors have the same construction features, the noise perturbation of each the sensor can vary with the time and depends on the location of the sensor. This occurs owing to different causes:  manufacturing defects, obstacles in the reception, different physical environmental conditions (such as humidity and temperature) etc. Moreover, in general, these conditions change along time, hence it is necessary that the central node of the network is able to re-estimate the noise powers jointly with position of the target (and other parameters of the models if required) whenever a new block of observations is processed. More specifically, let us denote the target position with the random vector $\textbf{Z}=[Z_1,Z_2]^{\top}$. The position of the target is then a specific realization ${\bf Z}={\bf z}$. 
The range measurements are obtained from $N_S=6$ sensors located at $\textbf{h}_1=[3, -8]^{\top}$, $\textbf{h}_2=[8,10]^{\top}$, $\textbf{h}_3=[-4,-6]^{\top}$, $\textbf{h}_4=[-8,1]^{\top}$, $\textbf{h}_5=[10,0]^{\top}$ and $\textbf{h}_6=[0,10]^{\top}$ as shown in Figure \ref{FigSIMU_Ex2}(a). The observation models are given by
\begin{gather}
\label{IStemaejemplo}
\begin{split}
Y_{j}=20\log\left(||{\bf z}-{\bf h}_j ||\right)+B_{j}, \quad j=1,\ldots, N_S, \\
\end{split}   
\end{gather}   
where $B_{j}$ are independent Gaussian random variables with pdfs, $\mathcal{N}(b_j;0,\zeta_j^2)$, $j=1,\ldots, N_S$.  We denote 
${\bm \zeta}=[\zeta_1,\ldots,\zeta_{N_S}]$ the vector of standard deviations.
Given the position of the target ${\bf z}^*=[z_1^*=2.5,z_2^*=2.5]^{\top}$ and setting ${\bm \zeta}^*=[\zeta_1^*=1,\zeta_2^*=2,\zeta_3^*=1,\zeta_4^*=0.5,\zeta_5^*=3,\zeta_{6}^*=0.2]$ (since $N_S=6$ and $D=N_S+2=8$), we generate $N_O=20$ observations from each sensor according to the model in Eq. \eqref{IStemaejemplo}.
 Then, we finally obtain a measurement matrix ${\bf Y}=[y_{k,1},\ldots, y_{k,N_S}] \in \mathbb{R}^{d_Y}$, where $d_Y=N_ON_S=120$, $k=1,\ldots,N_O$.  
We consider uniform prior $\mathcal{U}(\mathcal{R}_z)$ over the position $[z_1,z_2]^{\top}$  with $\mathcal{R}_z=[-30\times 30]^2$, and a uniform prior over $\zeta_j$, so that ${\bm \zeta}$ has prior $\mathcal{U}(\mathcal{R}_\zeta)$ with $\mathcal{R}_\zeta=[0,20]^{N_S}$. Thus, the posterior pdf is 
\begin{eqnarray*}
&&{\bar \pi}({\bm \theta}|\textbf{Y})={\bar \pi}({\bf z},{\bm \zeta}|\textbf{Y})= \ell(\textbf{y}|z_1,z_2,\zeta_1,\ldots,\zeta_{N_S})\prod_{i=1}^2p(z_i)\prod_{j=1}^{N_S} p(\zeta_j), \\
&=&\left[\prod_{k=1}^{N_O}\prod_{j=1}^{N_S} \frac{1}{\sqrt{2\pi \zeta_j^2}}\exp\left(-\frac{1}{2\zeta_j^2}(y_{k,j}+10\log\left(||{\bf z}-{\bf h}_j ||\right)^2\right) \right]\mathbb{I}_{z}(\mathcal{R}_z){\bf I}_{\zeta}(\mathcal{R}_\zeta), 
\end{eqnarray*}
where ${\bm \theta}=[{\bf z},{\bm \zeta}]^{\top}$ is a vector of parameters of dimension $D=N_S+2=8$ that we desire to infer, and $\mathbb{I}_{c}(\mathcal{R})$ is an indicator variable that is $1$ if $c\in\mathcal{R}$, otherwise is $0$. 

Our goal is to compute the Minimum Mean Square Error (MMSE) estimator, i.e., the expected value of the posterior ${\bar \pi}({\bm \theta}|\textbf{Y})={\bar \pi}({\bf z},{\bm \zeta}|\textbf{Y})$ (recall that $D=8$). Since the MMSE estimator cannot be computed analytically, we apply Monte Carlo methods for approximating it.
We compare GMS, the corresponding MTM scheme,  the Adaptive Multiple Importance Sampling (AMIS) technique \cite{CORNUET12}, and $N$ parallel MH chains with a random walk proposal pdf. For all of them we consider Gaussian proposal densities. For GMS and MTM, we set $q_t({\bm \theta}|{\bm \mu}_{n,t},\sigma^2{\bf I})=\mathcal{N}({\bm \theta}|{\bm \mu}_{t},\sigma^2{\bf I})$ which is adapted considering the empirical mean of the generated samples after a training period, $t\geq 0.2 T$ \cite{Luengo13,Haario01}, ${\bm \mu}_{0}\sim\mathcal{U}([1,5]^{D})$ and $\sigma=1$. For AMIS, we have $q_{t}({\bm \theta}|{\bm \mu}_{t},{\bf C}_{t})=\mathcal{N}({\bm \theta}|{\bm \mu}_{t},{\bf C}_{t})$, where ${\bm \mu}_{t}$ is as previously described (with ${\bm \mu}_{0}\sim\mathcal{U}([1,5]^{D})$) and  ${\bf C}_{t}$ is also adapted using the empirical covariance matrix, starting ${\bf C}_0= 4  {\bf I}$. We also test the use of $N$ parallel Metropolis-Hastings (MH) chains (we also consider the case of $N=1$, i.e., a single chain), with a Gaussian random-walk proposal pdf, $q_n({\bm \mu}_{n,t}|{\bm \mu}_{n,t-1},\sigma^2 {\bf I})=\mathcal{N}({\bm \mu}_{n,t}|{\bm \mu}_{n,t-1},\sigma^2 {\bf I})$ with ${\bm \mu}_{n,0}\sim\mathcal{U}([1,5]^{D})$ for all $n$ and $\sigma=1$.

We fix the total number of evaluations of the posterior density as $E=NT=10^4$. Note that, generally, the evaluation of the posterior is the most costly step in MC algorithms (however, AMIS has the additional cost of re-weighting all the samples at each iteration according to the deterministic mixture procedure \cite{Bugallo15,CORNUET12}).  We recall that $T$ denotes the total number of iterations and $N$ the number of samples drawn from each proposal at each iteration.  We consider  ${\bm \theta}^*=[{\bf z}^*,{\bm \zeta}^*]^{\top}$ as the ground-truth and compute the Mean Square Error (MSE) in the estimation obtained with the different algorithms. The results are averaged over $500$ independent runs and they are provided in Tables \ref{GMS_localization}, \ref{AMIS_localization}, and \ref{MH_localization} and Figure \ref{FigSIMU_Ex2}(b). Note that GMS outperforms AMIS for each a pair $\{N,T\}$ (keeping fixed $E=NT=10^4$), and GMS also provides smaller MSE values than $N$ parallel MH chains (the case $N=1$ corresponds to a unique longer chain). Figure \ref{FigSIMU_Ex2}(b) shows the MSE versus $N$ maintaining $E=NT=10^4$ for GMS and the corresponding MTM method. This figure again confirms the advantage of recycling the samples in a MTM scheme.

\begin{table}[h!]
\begin{center}
\caption{{\bf Results GMS.}}
\begin{tabular}{|c|c|c|c|c|c|c|c|c|}
\hline 
{\bf MSE}  &  1.30 &  1.24 &    1.22   &  1.21 &    1.22 &    {\bf 1.19} &  1.31  &    {\bf 1.44}   \\ 
\hline
\hline
$N$   &  10 & 20  & 50  & 100 & 200  & 500   & 1000&  2000  \\ 
$T$   &  1000  & 500  & 200  & 100  & 50 & 20  & 10 & 5 \\ 
\hline
$E$  &  \multicolumn{8}{c|}{$NT=10^4$}\\
\hline
{\bf MSE range} &\multicolumn{8}{c|}{ {\bf Min MSE= 1.19}  \quad   --------- \quad    {\bf Max MSE= 1.44} }     \\
\hline
\end{tabular}
\label{GMS_localization}
\end{center}
\end{table}

\begin{table}[h!]
\begin{center}
\caption{{\bf Results AMIS \cite{CORNUET12}.}}
\begin{tabular}{|c|c|c|c|c|c|c|c|c|}
\hline 
{\bf MSE}   & 1.58 & 1.57 &  1.53   & 1.48 & 1.42  & {\bf 1.29}   &  1.48 &  {\bf 1.71}  \\  
\hline
$N$   &10 & 20 &50 & 100 & 200 & 500 & 1000 & 2000 \\
$T$   & 1000 & 500 & 200 &  100 & 50 & 20 & 10 & 5   \\ 
\hline
$E$  &  \multicolumn{8}{c|}{$NT=10^4$}\\
\hline
{\bf MSE range} &\multicolumn{8}{c|}{ {\bf Min MSE= 1.29}  \quad   --------- \quad    {\bf Max MSE= 1.71} }     \\
\hline
\end{tabular}
\label{AMIS_localization}
\end{center}
\end{table}

\begin{table}[h!]
\begin{center}
\caption{{\bf Results $N$ parallel MH chains with random-walk proposal pdf.}}
\begin{tabular}{|c|c|c|c|c|c|c|c|c|}
\hline 
{\bf MSE} & 1.42 &  {\bf 1.31} &    1.44 &  2.32 &   2.73 &    {\bf 3.21} &    3.18 &    3.15  \\ 
\hline
\hline
$N$  &1 & 5 & 10 &  50 & 100 & 500 & 1000 &2000 \\
$T$   & $10^4$  & 2000    & 1000 & 200  & 100     &  20   & 10    & 5 \\ 
\hline
$E$  &  \multicolumn{8}{c|}{$NT=10^4$}\\
\hline
{\bf MSE range} &\multicolumn{8}{c|}{ {\bf Min MSE= 1.31}  \quad   --------- \quad    {\bf Max MSE=3.21 } }     \\
\hline
\end{tabular}
\label{MH_localization}
\end{center}
\end{table}

\subsection{Localization with real data}
In this section, we describe a numerical experiment involving real data. More specifically, we consider a localization problem \cite{Ali07}. We have carried out an experiment with a network consisting of four nodes. Three of them are placed at fixed positions and play the role of sensors that measure the strength of the radio signals transmitted by the target. The other node plays the role of the target to be localized. All nodes are bluetooth devices (Conceptronic CBT200U2A) with a nominal maximum range of 200 m. 
We consider a square monitored area of $4 \times 4$ m and place the sensors at fixed positions $\textbf{h}_1=[0.5,1]$, $\textbf{h}_2=[3.5,1]$ and $\textbf{h}_3=[2,3]$, with all coordinates in meters. The target is located at $\textbf{z}=[z_1=2.5,z_2=2]$.  The measurement provided by the $i$-th sensor is denoted as a random variable $Y_i$, considering the following model 
\begin{equation}
\label{modeloobserv}
	Y_i=\kappa-20\log\left[||\textbf{z}-\textbf{h}_i||\right]+B_{i},
\end{equation}
where $B_{i}$ are again independent Gaussian random variables with pdfs $\mathcal{N}(b_i;0,\zeta^2)$, for all $i=1,2,3$. Differently from the previous section, we estimate in advance the following parameters of the model, ${\hat \kappa}\approx -26.58$ and ${\hat\zeta}\approx 4.73$, using a least square fitting. We obtain $N_O=5$ measurements from each sensor ($d_Y=3N_O=15$),
and we consider a uniform prior on the $4 \times 4$ m  area. Given these measurements, we approximate the expected value $E[{\bf Z}]$ of the corresponding posterior ${\bar \pi}({\bf z})$ (here ${\bm \theta}={\bf z}$) using a thin deterministic bivariate grid, obtaining the ground truth $\approx [3.17,2.62]^{\top}$. 
We test an MH method and MTM scheme using with a random walk Gaussian proposal pdf, $q({\bf z}|{\b z}_{t-1})= \mathcal{N}({\bf z}|{\bf z}_{t-1},\sigma^2{\bf I}_2)$, with $\sigma=1$, $T\in\{1000,5000\}$, and $N\in \{10, 100, 1000\}$ tries for MTM (clearly, $N=1$ for MH).  We also test a Metropolis-adjusted Langevin algorithm (MALA), where the proposal is Gaussian random walk density with mean ${\bf z}_{t-1}+\beta\nabla[\log \pi({\bf z})]$ and  $\nabla[\log \pi({\bf z})]$ denotes the gradient of $\log \pi({\bf z})$ \cite{Roberts98}.  The covariance matrix of MALA Gaussian proposal is  $\sigma^2{\bf I}_2$ (as the other techniques) and the drift parameter $\beta=\sigma^2/2$. We compute the MSE in estimating $E[{\bf Z}]\approx [3.17,2.62]^{\top}$ and averaged the results over $2000$ independent runs (at each run, we take the mean of the  square error values of each component). The results are shown in Table \ref{Ex_localization_realdata}. Recall that MALA uses the additional information of the gradient. Note that, a MALA-type proposal pdf can also be used in a MTM scheme. 
 The use of multiple tries improves the mixing of the Markov chain and speeds up the convergence.

\begin{table}[h!]
\begin{center}
\caption{{\bf MSE obtained in the localization problem with real data.}}
\begin{tabular}{|c|c|c|c|c|c|c|}
\hline 
{\bf Algorithm}  &  {\bf MH} &  {\bf MALA} &  {\bf MALA} & {\bf MTM}   &   {\bf MTM}   &  {\bf MTM}\\ 
\hline
$N$   &  1 &  1  & 1 & 10  & 100  & 1000   \\ 
$T$   &  1000 & 1000 & 5000 &  1000  & 1000  & 1000   \\ 
\hline
\hline
{\bf MSE}   & 0.2511 & 0.2081 & 0.1359  & 0.0944  & 0.0469  & 0.0030   \\ 
\hline
\end{tabular}
\label{Ex_localization_realdata}
\end{center}
\end{table}

\section{Conclusions}
\label{conclSect}

We have provided a thorough review of MCMC methods using multiple candidates in order to select the next state of the chain. We have presented and compared different Multiple Try Metropolis, Ensemble MCMC and Delayed Rejection Metropolis schemes. We have also described the Group Metropolis Sampling technique which generates a chain of set of weighted samples, so that some candidates are properly reused in the final estimators. Furthermore, we have shown how the Particle Metropolis-Hastings algorithm can be interpreted as an MTM scheme using a particle filter for generating the different weighted candidates. Several connections and differences have been pointed out. 
Finally, we have tested several techniques in different numerical experiments: two toy examples in order to provide an exhaustive comparison among the methods, a numerical example regarding the hyperparameter selection for a Gaussian Process (GP) regression model, and two localization problems, one of them involving a real data analysis.



\section*{Acknowledgements}

This work has been supported by the European Research Council (ERC) through the ERC Consolidator Grant SEDAL ERC-2014-CoG 647423.



\bibliography{bibliografia,bibliografia_GIS,fusion}


\begin{appendices}

\section{Distribution after resampling}
\label{AfterRES}
Let us also denote as ${\bm \theta} \in\{{\bm \theta}^{(1)}\ldots,{\bm \theta}^{(N)}\}$, a generic sample after applying one multinomial resampling step according to the normalized IS weights ${\bar w}_n$, $n=1,\ldots,N$.
The density of ${\bm \theta}$ is given by
\begin{eqnarray}
\label{CompleteProp_otra}
\widetilde{q}({\bm \theta})=\int_{\mathcal{D}^N} \widehat{\pi}({\bm \theta}|{\bm \theta}^{(1:N)}) \left[\prod_{i=1}^N q({\bm \theta}^{(i)})\right]d{\bm \theta}^{(1:N)},
\end{eqnarray}
where 
\begin{equation}
 \widehat{\pi}({\bm \theta}|{\bm \theta}^{(1:N)})=\sum_{j=1}^N {\bar w}_j \delta({\bm \theta}-{\bm \theta}^{(j)}).
\end{equation} 
We also define also the matrix
$
{\bf m}_{\neg n}=[{\bm \theta}^{(1)},\ldots, {\bm \theta}^{(n)},{\bm \theta}^{(n+1)},\ldots, {\bm \theta}^{(N)}],
$ 
containing all the samples except for the $n$-th.
After some straightforward rearrangements,  Eq. \eqref{CompleteProp_otra} can be rewritten as
\begin{equation}
	\widetilde{q}({\bm \theta})= \sum_{j=1}^{N}\left(\int_{\mathcal{D}^{N-1}} \frac{\pi({\bm \theta})}{\sum_{n=1}^{N}{\frac{\pi({\bm \theta}^{(n)})}{q({\bm \theta}^{(n)})}}}\left[\prod_{\substack{n=1 \\ n \neq j}}^{N}{q({\bm \theta}^{(n)})}\right]
		 d{\bf m}_{\neg j}\right).
\label{eq:phiIS0}
\end{equation}
Finally, we can write
\begin{equation}
	\widetilde{q}({\bm \theta}) = \pi({\bm \theta}) \sum_{j=1}^{N}{\int_{\mathcal{D}^{N-1}} \frac{1}{N\widehat{Z}}
		\left[\prod_{\substack{n=1 \\ n \neq j}}^N q({\bm \theta}^{(n)})\right]  d{\bf m}_{\neg j}},
\label{eq:phiIS2}
\end{equation}
where $\widehat{Z}= \frac{1}{N}\sum_{n=1}^N\frac{\pi({\bm \theta}^{(n)})}{q({\bm \theta}^{(n)})}$ that is that  IS estimator of $Z$. The equation above represents the density of a resampled particle ${\bm \theta} \in\{{\bm \theta}^{(1)}\ldots,{\bm \theta}^{(N)}\}$. Note that if $\widehat{Z}=Z$ then $\widetilde{q}({\bm \theta}) =\pi({\bm \theta})$. Clearly, for a finite value of $N$, there exists a discrepancy between $\widetilde{q}({\bm \theta}) $ and $\bar{\pi}({\bm \theta})$, but this discrepancy decreases as $N$ grows.

\section{Particle Filtering}
\label{AppPF}
In this appendix, we recall some basic concepts and more recent results about the Sequential Importance Sampling (SIS) and Sequential Importance Resampling (SIR) methods, important for the description of the techniques above. 
\newline
{\bf SIS procedure.} We assume  that the variable of interest is formed by only a dynamical variable, i.e., ${\bm \theta}={\bf x}=x_{1:D}=[x_1\dots,x_D]^{\top}$ (for simplicity, consider $x_d\in\mathbb{R}$) and  the target can be factorized as
\begin{eqnarray}
\bar{\pi}({\bf x})\propto \pi({\bf x})&=&\gamma_1(x_1) \prod_{d=2}^D  \gamma_d(x_d|x_{d-1}).
\end{eqnarray}
Given a proposal of type $q({\bf x})=q_1(x_1) \prod_{d=2}^D  q_d(x_d|x_{d-1})$, and a sample ${\bf x}^{(n)}=x_{1:D}^{(n)}\sim q({\bf x})$ with $x_{d}^{(n)}\sim q_d(x_d|x_{d-1})$, we assign the importance weight  
\begin{equation}
\label{EqFinRecW}
w({\bf x}^{(n)})=w_D^{(n)}=\frac{\pi({\bf x}^{(n)})}{q({\bf x}^{(n)})}=\frac{ \gamma_1(x_1^{(n)}) \gamma_2(x_2^{(n)}|x_1^{(n)}) \cdots  \gamma_D(x_D^{(n)}|x_{1:D-1}^{(n)})}{q_1(x_1^{(n)}) q_2(x_2^{(n)}|x_1^{(n)}) \cdots  q_D(x_D^{(n)}|x_{1:D-1}^{(n)})}.
\end{equation}
 The weight above can be compute with a recursive procedure for computing the importance weights: starting with $w_1^{(n)}=\frac{\pi(x_1^{(n)})}{q(x_1^{(n)})}$ and then
 \begin{gather}
 \label{RecWeights}
 \begin{split}
  w_d^{(n)}=w_{d-1}^{(n)} \beta_d^{(n)}=\prod_{j=1}^{d} \beta_j^{(n)},   \quad \quad d=1,\ldots,D,   
 \end{split}
 \end{gather}
 where we have set
 \begin{equation}
  \label{AumWeights}
 \beta_1^{(n)}=w_1^{(n)} \quad \mbox{and} \quad \beta_d^{(n)}=\frac{\gamma_d(x_d^{(n)}|x_{1:d-1}^{(n)})}{q_d(x_d^{(n)}|x_{1:d-1}^{(n)})},
\end{equation} 
for $d=2,\ldots,D$.  Let also define the partial target pdfs
\begin{equation}
\pi_d(x_{1:d})=\gamma_1(x_1) \prod_{i=2}^d  \gamma_i(x_i|x_{i-1})
\end{equation}


{\bf SIR procedure.} In SIR, a.k.a., standard particle filtering, resampling steps are incorporated during the recursion as shown of Table \ref{alg:SIRpartialRes} \cite{Djuric03,Doucet08tut}. In general, the resampling steps are applied only in certain iterations in order to avoid the path degeneration, taking into account an approximation $\widehat{ESS}$ of the Effective Sampling Size (ESS) \cite{ESSmartino}. If $\widehat{ESS}$ is smaller than a pre-established threshold, the particles are resampled. Two examples of ESS approximation are $\widehat{ESS}=\frac{1}{\sum_{n=1}^N (\bar{w}_d^{(n)})^2}$ and $\widehat{ESS}=\frac{1}{\max \bar{w}_d^{(n)}}$ where $\bar{w}_d^{(n)}=\frac{w_d^{(n)}}{\sum_{i=1}^N w_d^{(i)}}$ (note that $1 \leq \widehat{ESS} \leq N$).
Hence, the condition for the adaptive resampling can be expressed as $\widehat{ESS} < \eta N$ where $\eta\in [0,1]$. SIS is given when $\eta=0$ and SIR  for $\eta\in(0,1]$. When $\eta=1$, the resampling is applied at each iteration and in this case SIR is often called {\it bootstrap particle filter} \cite{Djuric03,Doucet08tut}. If $\eta= 0$, no resampling steps are applied, and we have the SIS method described above.

{\rem Note that in Table \ref{alg:SIRpartialRes}, we have employed a proper weighting for resampling particles \cite{GISssp16},
\begin{equation}
w_{d}^{(1)}=w_{d}^{(2)}=\ldots= w_{d}^{(n)}=\widehat{Z}_{d}.
\end{equation}
Generally, it is remarked that $w_{d}^{(1)}=w_{d}^{(2)}=\ldots= w_{d}^{(n)}$ but a specific value is not given. If a different value $c\neq \widehat{Z}_{d}$ is employed, i.e., $w_{d}^{(1)}=\ldots= w_{d}^{(n)}=c$, the algorithm is still valid but the weight recursion loses part of the statistical meaning. This is the reason why the marginal likelihood estimator $\widehat{Z}=\widehat{Z}_{D}=\frac{1}{N} \sum\limits_{n=1}^Nw_{D}^{(n)}$ is consistent only, if a proper weighting after resampling is used \cite{GISssp16,GIS17,Martino15PF}.  
}

\begin{table}[!h]
	\centering
	\caption{{\bf The SIR method with proper weighting after resampling.}}
	\vspace{0.3cm}
	    \begin{tabular}{|p{0.95\columnwidth}|}
		\hline
  \begin{enumerate}
\item Choose $N$ the number of particles, the initial particles $x_0^{(n)}$, $n=1,\ldots,N$,  an ESS approximation $\widehat{ESS}$ \cite{ESSmartino}  and a constant value $\eta\in[0,1]$.
    \item For $d = 1, \ldots, D$:
    	\begin{enumerate}
        \item {\bf Propagation:}\label{PropStep} Draw  $x_d^{(n)} \sim q_d(x_d|x_{d-1}^{(n)})$, for $n=1,\ldots,N$.
        \item  {\bf Weighting:}\label{WStep} Compute the weights       
                 \begin{equation}
                   w_d^{(n)}=w_{d-1}^{(n)} \beta_d^{(n)} =\prod_{j=1}^{d} \beta_j^{(n)},   \quad \quad n=1,\ldots,N,   
                \end{equation}       
                   where  $\beta_d^{(n)}=\frac{\gamma_d(x_d^{(n)}|x_{d-1}^{(n)})}{q_d(x_d^{(n)}|x_{d-1}^{(n)})}$. 
             \item\label{StepResampling}  if $\widehat{ESS} < \eta N$ then: 
             \begin{enumerate}
\item {\bf Resampling:}\label{StepResampling2} Resample $N$ times within the set $\{x_{d-1}^{(n)}\}_{n=1}^N$ according to the probabilities $\bar{w}_d^{(n)}=\frac{w_{d}^{(n)}}{ \sum_{j=1}^Nw_{d}^{(j)}}$, obtaining $N$ resampled particles $\{\bar{x}_d^{(n)}\}_{n=1}^N$. Then, set $x_d^{(n)}=\bar{x}_d^{(n)}$, for $n=1,\ldots,N$.
\item\label{StepProperGIS}  {\bf Proper weighting:} Compute $\widehat{Z}_{d}=\frac{1}{N} \sum\limits_{n=1}^Nw_{d}^{(n)}$ and set $w_{d}^{(n)}=\widehat{Z}_{d}$ for all $n=1,\ldots,N$.
\end{enumerate}             
         \end{enumerate}
         \item Return $\{\x_n=x_{1:D}^{(n)},  w_n=w_D^{(n)}\}_{n=1}^N$.
   \end{enumerate} 	
\\
	\hline
	\end{tabular}		
	\label{alg:SIRpartialRes}
\end{table}

\subsection{Marginal likelihood estimators in SIS}
{\rem 
 In SIS, there are two possible formulations of the estimator of the marginal likelihoods $Z_d =\int_{\mathbb{R}^{d}} \pi_d(x_{1:d}) d x_{1:d}$, 
\begin{eqnarray}
\widehat{Z}_d&=&\frac{1}{N} \sum_{n=1}^N w_d^{(n)}= \frac{1}{N} \sum_{n=1}^N w_{d-1}^{(n)}  \beta_d^{(n)},  \label{Est1}\\
\overline{Z}_d&=& \prod_{j=1}^d\left[\sum_{n=1}^N{\bar w}_{j-1}^{(n)}\beta_j^{(n)}\right]. \label{Est2}
\end{eqnarray}
In SIS, both estimators are equivalent $\overline{Z}_d\equiv \widehat{Z}_d$. }

Indeed, the classical IS estimator of the normalizing constant $Z_d$ at the $d$-th iteration is
\begin{eqnarray}
\widehat{Z}_d&=&\frac{1}{N} \sum_{n=1}^N w_d^{(n)}=\frac{1}{N} \sum_{n=1}^N w_{d-1}^{(n)}\beta_d^{(n)},  \label{Z_approx2}\\
&=&\frac{1}{N} \sum_{n=1}^N\left[\prod_{j=1}^d \beta_j^{(n)}\right].   \label{Z_approx3}
\end{eqnarray} 
An alternative formulation, denoted as $\overline{Z}_d$, is often used  
\begin{eqnarray}
 \label{EstZ2_now}
\overline{Z}_d&=& \prod_{j=1}^d\left[\sum_{n=1}^N{\bar w}_{j-1}^{(n)}\beta_j^{(n)}\right]  \\
&=&\prod_{j=1}^d\left[\frac{\sum_{n=1}^N w_{j}^{(n)}}{\sum_{n=1}^N w_{j-1}^{(n)}}\right]=
\widehat{Z}_1\prod_{j=2}^d\left[  \frac{ \widehat{Z}_j}{\widehat{Z}_{j-1}}\right]=\widehat{Z}_{d}.  \label{EstZ2_5}
\end{eqnarray} 
where we have employed ${\bar w}_{j-1}^{(n)}=\frac{w_{j-1}^{(n)}}{\sum_{i=1}^N w_{j-1}^{(i)}}$ and $w_{j}^{(n)}=w_{j-1}^{(n)}\beta_j^{(n)}$ \cite{Doucet08tut}. 

Furthermore, note that $\overline{Z}_d$ can be written in a recursive form as
\begin{eqnarray}
 \label{EstZ3}
\overline{Z}_d=\overline{Z}_{d-1}\left[\sum_{n=1}^N{\bar w}_{d-1}^{(n)}\beta_d^{(n)}\right].
\end{eqnarray}

\subsection{Marginal likelihood estimators in SIR}
\label{SIRsect}

{\rem 
If  a proper weighting after resampling is applied in SIR, both formulations $\widehat{Z}_d$  and $\overline{Z}_d$ in Eqs. \eqref{Est1}-\eqref{Est2} provide consistent estimator of $Z_d$ and they are equivalent, $\widehat{Z}_d\equiv \overline{Z}_d$ (as in SIS). 
}

If a proper weighting is not applied, only  
$$
\overline{Z}_d=\prod_{j=1}^d\left[\sum_{n=1}^N{\bar w}_{j-1}^{(n)}\beta_j^{(n)}\right].
$$ 
is a  consistent estimator of $Z_d$, in SIR. In this case, $\widehat{Z}_d=\frac{1}{N}\sum_{n=1}^N w_{d}^{(n)}$ is not a possible alternative (without using a proper weighting after resampling). However, considering the proper weighting of the resampled particles, then $\widehat{Z}_d$ is also a consistent estimator of $Z_d$ and it is equivalent to $\overline{Z}_d$. Below, we analyze three cases:
\begin{itemize}
\item {\bf No Resampling ($\eta=0$):} this scenario corresponds to SIS where $\widehat{Z}_d$, $\overline{Z}_d$ are equivalent as shown in Eq. \eqref{EstZ2_5}. 
\item {\bf Resampling at each iteration ($\eta=1$):}  using the proper weighting, $w_{d-1}^{(n)}=\widehat{Z}_{d-1}$ for all $n$ and for all $d$, and replacing in Eq. \eqref{Z_approx2}  we have  
\begin{eqnarray}
\label{ZtotalRes2}
\widehat{Z}_{d}&=&\widehat{Z}_{d-1}\left[\frac{1}{N} \sum_{n=1}^{N} \beta_{d}^{(n)}\right], \\
&=&\frac{1}{N}\prod_{j=1}^d \left[ \sum_{n=1}^{N}  \beta_{j}^{(n)}\right]. \label{ZtotalRes3}
\end{eqnarray}
Since after resampling all particles have the same weight, we have ${\bar w}_{d-1}^{(n)}=\frac{1}{N}$ for all $n$. Replacing it in the expression of $\overline{Z}_{d}$ in \eqref{EstZ3}, we obtain
\begin{equation}
\label{ZtotalRes}
\overline{Z}_{d}=\frac{1}{N}\prod_{j=1}^d \left[  \sum_{n=1}^{N}  \beta_{j}^{(n)}\right],
\end{equation}
that coincides with $\widehat{Z}_{d}$ in Eq. \eqref{ZtotalRes3}.
\item {\bf Adaptive resampling ($0<\eta<1$):} for the sake of simplicity, let us start considering a unique resampling step applied at the $k$-th iteation with $k<d$. We check if both estimators are equal at $d$-th iteration of the recursion. Due to Eq. \eqref{EstZ2_5}, we have $\overline{Z}_{k}\equiv \widehat{Z}_{k}$,\footnote{We consider to compute the estimators before the resampling.} since before the $k$-th iteration no resampling has been applied. 
With the proper weighting $w_{k}^{(n)}=\widehat{Z}_{k}$ for all $n$, at the next iteration we have
\begin{eqnarray*}
\widehat{Z}_{k+1}&=&\frac{1}{N} \sum_{n=1}^N w_{k}^{(n)}\beta_{k+1}^{(n)}=\widehat{Z}_{k} \left[\frac{1}{N} \sum_{n=1}^N \beta_{k+1}^{(n)}\right],
\end{eqnarray*}
and using Eq. \eqref{EstZ3},  we obtain 
\begin{eqnarray*}
\overline{Z}_{k+1}&=&\overline{Z}_{k}\left[\sum_{n=1}^N\frac{1}{N}\beta_{k+1}^{(n)}\right]=\widehat{Z}_{k} \left[\frac{1}{N} \sum_{n=1}^N \beta_{k+1}^{(n)}\right], 
 \end{eqnarray*}
so that the estimators are equivalent also at the $(k+1)$-th iteration, $\overline{Z}_{k+1}\equiv \widehat{Z}_{k+1}$. Since we are assuming no resampling steps after the $k$-th iteration and until the $d$-th iteration, we have that $\overline{Z}_{i}\equiv \widehat{Z}_{i}$ for $i=k+2,\ldots,d$ due to we are in a SIS scenario  for $i>k$ (see Eq.  \eqref{EstZ2_5}). This reasoning can be easily extended for different number of resampling steps.  
\end{itemize}

\section{Consistency of GMS estimators}
\label{ConGMS}
{\bf Dynamic of GMS.}
We have already seen that we can recover an I-MTM chain from the GMS outputs applying one resampling step for each $t$ when $\mathcal{S}_t\neq  \mathcal{S}_{t-1}$, i.e.,       
\begin{gather}
{\bm \theta}_{t}= \left\{
\begin{split}
\label{RecChain_APP}
&{\bm \theta}_{t} \sim \sum_{n=1}^N \frac{\rho_{n,t}}{\sum_{i=1}^N \rho_{i,t}}  \delta({\bm \theta}-{\bm \theta}_{n,t}),   \quad \mbox{ if }  \quad \mathcal{S}_t\neq  \mathcal{S}_{t-1}, \\
&{\bm \theta}_{t-1}, \quad\quad\quad\quad\quad\quad\quad\quad\quad\quad\quad\mbox{ }\mbox{ }\mbox{ } \mbox{ if } \quad \mathcal{S}_t=  \mathcal{S}_{t-1},
\end{split}
\right. 
\end{gather}
for $t=1,\ldots,T$. The sequence $\{{\bm \theta}_t\}_{t=1}^T$ is a chain obtained by one run of an I-MTM2 technique. Note that 
(a) the sample generation, (b) the acceptance probability function and hence (c) the dynamics of GMS exactly coincide with the corresponding steps of I-MTM2 (or PMH; depending on candidate generation procedure). Hence, the ergodicity of the recovered chain is ensured.
\newline
{\bf Parallel chains from GMS outputs.} 
As described in Section \ref{Par_MTM}, we can extend the consideration above for generation $C$ parallel I-MTM2 chains. Indeed, we resample $C$ times instead of only one, i.e.,  
\begin{gather}
{\bm \theta}_{t}^{(c)}= \left\{
\begin{split}
\label{RecChain_parallelChains}
&{\bm \theta}_{t} \sim \sum_{n=1}^N \frac{\rho_{n,t}}{\sum_{i=1}^N \rho_{i,t}}  \delta({\bm \theta}-{\bm \theta}_{n,t}),   \quad \mbox{ if }  \quad \mathcal{S}_t\neq  \mathcal{S}_{t-1}, \\
&{\bm \theta}_{t-1}, \quad\quad\quad\quad\quad\quad\quad\quad\quad\quad\quad\mbox{ }\mbox{ }\mbox{ } \mbox{ if } \quad \mathcal{S}_t=  \mathcal{S}_{t-1}, \nonumber
\end{split}
\right. 
\end{gather}
for $c=1,\ldots,C$, where the super-index denotes the $c$-th chain (similar procedures have been suggested in \cite{Calderhead14,OMCMC_DSP}).
Clearly, the resulting $C$ parallel chains are not independent, and there is  an evident loss of performance w.r.t. the case of independent chains. However, at each iteration, the number of target evaluations per iteration is only $N$ instead of $NC$. 
Note that that each chain in ergodic, so that each estimator $\widetilde{I}_T^{(c)}=\frac{1}{T} \sum_{t=1}^Tg({\bm \theta}_t^{(c)})$ is consistent (i.e., convergence to the true value for $T\rightarrow \infty$). As a consequence, the arithmetic mean of consistent estimators,
\begin{equation}
\label{AquiApp}
\widetilde{I}_{C,T}=\frac{1}{C} \sum_{c=1}^C\widetilde{I}_T^{(c)}=\frac{1}{CT} \sum_{t=1}^T  \sum_{c=1}^C g({\bm \theta}_t^{(c)}),
\end{equation}
is also consistent, for all values of $C\geq 1$. 
\newline
{\bf GMS as limit case.} 
Let us consider the case $\mathcal{S}_t\neq  \mathcal{S}_{t-1}$ (the other one is trivial), at some iteration $t$. In this scenario, the samples of the $C$ parallel I-MTM2 chains, ${\bm \theta}_{t}^{(1)}$,${\bm \theta}_{t}^{(2)}$,...,${\bm \theta}_{t}^{(C)}$, are obtained by resampled independently $C$ samples from the set $\{{\bm \theta}_{1,t},\ldots,{\bm \theta}_{N,t}\}$ according to the normalized weights  $\bar{\rho}_{n,t}=\frac{\rho_{n,t}}{\sum_{i=1}^N \rho_{i,t}}$, for $n=1,\ldots,N$. Recall that the samples ${\bm \theta}_{t}^{(1)}$,${\bm \theta}_{t}^{(2)}$,...,${\bm \theta}_{t}^{(C)}$, will be used in the final estimator $\widetilde{I}_{C,T}$ in Eq. \eqref{AquiApp}.

Let us denote as $\#j$ the number of times that a specific candidate ${\bm \theta}_{j,t}$ (contained in the set $\{{\bm \theta}_{n,t}\}_{n=1}^N$) has been selected as state of one of $C$ chains, at the $t$ iteration. As $C\rightarrow \infty$, The fraction  $\frac{\#j}{C}$ approaches exactly the corresponding weights $\bar{\rho}_{j,t}$. Then, for $C\rightarrow \infty$, we have that the estimator in Eq. \eqref{AquiApp} approaches the GMS estimator, i.e.,
\begin{equation}
\label{AquiApp2}
\lim_{C\rightarrow \infty} \widetilde{I}_{C,T}= \frac{1}{T} \sum_{t=1}^T \sum_{n=1}^N \bar{\rho}_{n,t} g({\bm \theta}_{n,t}).
\end{equation}
 Since $ \widetilde{I}_{C,T}$ as $T\rightarrow \infty$ is consistent for all values of $C$, then the GMS estimator is also consistent (and it can be obtained as $C\rightarrow \infty$).
 
\end{appendices}

\end{document}